\documentclass[longauth]{aa}  

\usepackage{natbib}
\bibpunct{(}{)}{;}{a}{}{,} 
\usepackage{booktabs}
\usepackage{pdflscape}
\usepackage{siunitx}
\usepackage[varg]{txfonts}
\usepackage[breaklinks=true, hidelinks]{hyperref}
\usepackage{graphicx}
\usepackage{sidecap}
\DeclareUnicodeCharacter{0301}{*************************************}

\makeatletter
\renewcommand*\aa@pageof{, page \thepage{} of \pageref*{LastPage}}
\makeatother

\usepackage{amsmath}	
\usepackage{natbib}
\graphicspath{{./}{Figures/}}
\usepackage{ulem}
\usepackage{soul}

\newcommand{\HeI}{\textrm{He~{\textsc{i}}}}
\newcommand{\HII}{\textrm{H~{\textsc{ii}}}}
\newcommand{\HI}{\textrm{H~{\textsc{i}}}}

\newcommand{\FeII}{\textrm{Fe~{\textsc{ii}}}}

\usepackage{xcolor}

\definecolor{cbpurple}{rgb}{0.47, 0.37, 0.94}

\definecolor{gray}{rgb}{0.5, 0.5, 0.5}

\definecolor{darkorange}{rgb}{1.0, 0.55, 0.0}

\definecolor{darkteal}{rgb}{0, .3, .3}

\definecolor{cbred}{rgb}{0.9, 0.14, 0.46}

\definecolor{ochre}{rgb}{0.8, 0.47, 0.13}

\makeatletter
\renewcommand*\maketitle{%
  \thispagestyle{firstpage}
\begingroup
    \if@wideboxfn
    \setlength\bibindent{1.4\parindent}
    \else
    \setlength\bibindent{\parindent}
    \fi
    \renewcommand*\thefootnote{\@fnsymbol\c@footnote}%
    \renewcommand\@makefntext[1]{%
    \ifaa@longfn\hsize\textwidth\fi
    \noindent
    \hb@xt@\bibindent{\hss\@makefnmark\enspace}##1}
  \ifaa@twocolumn
  \begin{aa@strip}
    \aa@maketitle
  \end{aa@strip}
  \@thanks 
  \else
    \begingroup
      \let\thanks\footnote
      \aa@maketitle
    \endgroup
  \fi
\endgroup
  \setcounter{footnote}{0}%
}
\makeatother

\begin{document} 

\title{PDRs4All II: JWST's NIR and MIR imaging view of the Orion Nebula}

\author{Emilie Habart\inst{1} \and
           Els Peeters \inst{2, 3, 4} \and
           Olivier Bern\'{e} \inst{5} \and
           Boris Trahin \inst{1} \and
           Am\'elie Canin \inst{5} \and
           Ryan Chown \inst{2,3}\and
           Ameek Sidhu \inst{2,3} \and
          Dries Van De Putte \inst{6} \and
          Felipe Alarc\'on \inst{7} \and
          Ilane	Schroetter \inst{5} \and
          Emmanuel Dartois \inst{8}
% ...
          }

\institute{
    Institut d'Astrophysique Spatiale, Universit\'e Paris-Saclay, CNRS,  B$\hat{a}$timent121, 91405 Orsay Cedex, France; \email{pis@pdrs4all.org}  
    \label{inst1} \and
    Department of Physics \& Astronomy, The University of Western Ontario, London ON N6A 3K7, Canada;    
    \label{inst2} \and 
    Institute for Earth and Space Exploration, The University of Western Ontario, London ON N6A 3K7, Canada    
    \label{inst3} \and
    Carl Sagan Center, SETI Institute, 339 Bernardo Avenue, Suite 200, Mountain View, CA 94043, USA  
    \label{inst4} \and  
    Institut de Recherche en Astrophysique et Plan\'etologie, Universit\'e Toulouse III - Paul Sabatier, CNRS, CNES, 9 Av. du colonel Roche, 31028 Toulouse Cedex 04, France 
    \label{inst5} \and 
    Space Telescope Science Institute, 3700 San Martin Drive, Baltimore, MD, 21218, USA  
    \label{inst6} \and
    Department of Astronomy, University of Michigan, 1085 South University Avenue, Ann Arbor, MI 48109, USA  
    \label{inst7} \and
    Institut des Sciences Mol\'eculaires d'Orsay, CNRS, Universit\'e Paris-Saclay,  B$\hat{a}$timent 520, 91405 Orsay Cedex, France 
    \label{inst8} \and
    Instituto de Astrof\'isica e Ci\^{e}ncias do Espa\c co, Tapada da Ajuda, Edif\'icio Leste, 2\,$^{\circ}$ Piso, P-1349-018 Lisboa, Portugal  \label{inst9} \and    %
    ACRI-ST, Centre d’Etudes et de Recherche de Grasse (CERGA), 10 Av. Nicolas Copernic, F-06130 Grasse, France  
    \label{inst10} \and
    INCLASS Common Laboratory., 10 Av. Nicolas Copernic, 06130 Grasse, France \label{inst11} \and
    NASA Ames Research Center, MS 245-6, Moffett Field, CA 94035-1000, USA \label{inst12} \and
    LERMA, Observatoire de Paris, PSL Research University, CNRS, Sorbonne Universit\'es, F-92190 Meudon, France 
    \label{inst13} \and
    Instituto de F\'{\i}sica Fundamental  (CSIC),  Calle Serrano 121-123, 28006, Madrid, Spain 
    \label{inst14} \and
    UK Astronomy Technology Centre, Royal Observatory Edinburgh, Blackford Hill, Edinburgh EH9 3HJ, UK 
    \label{inst15} \and
    Observatorio Astron\'{o}mico Nacional (OAN,IGN), Alfonso XII, 3, E-28014 Madrid, Spain 
    \label{inst16} \and
    Sterrenkundig Observatorium, Universiteit Gent, Gent, Belgium \label{inst17} \and 
    Quantum Solid State Physics (QSP), Celestijnenlaan 200d - box 2414, 3001 Leuven, Belgium 
    \label{inst18} \and
    Institut de Plan\'etologie et d'Astrophysique de Grenoble (IPAG), Universit\'e Grenoble Alpes, CNRS, F-38000 Grenoble, France 
    \label{inst19} \and
    Institut de Radioastronomie Millim\'etrique (IRAM), 300 Rue de la Piscine, F-38406 Saint-Martin d'H\`{e}res, France 
    \label{inst20} \and 
    I. Physikalisches Institut der Universit\"{a}t zu K\"{o}ln, Z\"{u}lpicher Stra{\ss}e 77, 50937 K\"{o}ln, Germany 
    \label{inst21} \and
    Department of Astronomy, Graduate School of Science, The University of Tokyo, 7-3-1 Bunkyo-ku, Tokyo 113-0033, Japan 
    \label{inst22} \and
    Department of Physics, Faculty of Science and Engineering, Meisei University, 2-1-1 Hodokubo, Hino, Tokyo 191-8506, Japan 
    \label{inst22} \and
    Johns Hopkins University, 3400 N. Charles Street, Baltimore, MD, 21218, USA 
    \label{inst24} \and
    Physikalischer Verein - Gesellschaft für Bildung und Wissenschaft, Robert-Mayer-Straße 2, 60325 Frankfurt am Main, Germany 
    label{inst25} \and
    Institut für Angewandte Physik. Goethe-Universit{\"a}t Frankfurt, Max-von-Laue-Str. 1, 60438 Frankfurt am Main, Germany 
    \label{inst26} \and
    Department of Space, Earth and Environment, Chalmers University of Technology, Onsala Space Observatory, SE-439 92 Onsala, Sweden \label{inst27} \and
    Leiden Observatory, Leiden University, P.O. Box 9513, 2300 RA Leiden, The Netherlands  
    \label{inst28} \and
    Astronomy Department, University of Maryland, College Park, MD 20742, USA  \label{inst29} \and
    AIM, CEA, CNRS, Universit\'e Paris-Saclay, Universit\'e Paris Diderot, Sorbonne Paris Cit\'e, 91191 Gif-sur-Yvette, France 
    \label{inst30} \and
    Instituto de Física e Química, Universidade Federal de Itajubá, Av. BPS 1303, Pinheirinho, 37500-903, Itajubá, MG, Brazil 
    \label{inst31} \and
    Bay Area Environmental Research Institute, Moffett Field, CA 94035, USA \label{inst32} \and
    Australian Synchrotron, Australian Nuclear Science and Technology Organisation (ANSTO), Victoria, Australia 
    \label{inst33} \and
    INAF - Osservatorio Astrofisico di Catania, Via Santa Sofia 78, 95123 Catania, Italy  
    \label{inst34} \and
    Bay Area Environmental Research Institute, Moffett Field, CA 94035, USA \label{inst35} \and
    Laboratoire de Physique de l'\'Ecole Normale Sup\'erieure, ENS, Universit\'e PSL, CNRS, Sorbonne Universit\'e, Universit\'e Paris Cit\'e, F-75005, Paris, France
    \label{inst36} \and
    Laboratory for Atmospheric and Space Physics, University of Colorado, Boulder, CO 80303, USA 
    \label{inst37} \and
    Department of Chemistry, University of Colorado, Boulder, CO 80309, USA \label{inst38} \and
    Institute for Modeling Plasma, Atmospheres, and Cosmic Dust (IMPACT), University of Colorado, Boulder, CO 80303, USA 
    \label{inst39} \and 
    Faculty of Aerospace Engineering, Delft University of Technology, Kluyverweg 1, 2629 HS Delft, The Netherlands 
    \label{inst40} \and
    Radboud University, Institute for Molecules and Materials, FELIX Laboratory, Toernooiveld 7, 6525 ED Nijmegen, the Netherlands \label{inst41} \and
    DST INSPIRE School of Physics, University of Hyderabad, Hyderabad, Telangana 500046, India  
    \label{inst42} \and
    Department of Physics, Wellesley College, 106 Central Street, Wellesley, MA 02481, USA 
    \label{inst43} \and 
    Anton Pannekoek Institute for Astronomy (API), University of Amsterdam, Science Park 904, 1098 XH Amsterdam, The Netherlands 
    \label{inst44} \and 
    Delft University of Technology, Delft, The Netherlands 
    \label{inst45} \and 
    Laboratoire de Physique des deux infinis Ir\`ene Joliot-Curie, Universit\'e Paris-Saclay, CNRS/IN2P3,  B$\hat{a}$timent 104, 91405 Orsay Cedex, France 
    \label{inst46} \and
    Department of Chemistry, GITAM school of Science, GITAM Deemed to be University, Bangalore, India. 
    \label{inst47} \and
    Institut de Physique de Rennes, UMR CNRS 6251, Universit{\'e} de Rennes 1, Campus de Beaulieu, 35042 Rennes Cedex, France 
    \label{inst48} \and
    Department of Chemistry, The University of British Columbia, Vancouver, British Columbia, Canada 
    \label{inst49} \and 
    National Radio Astronomy Observatory (NRAO), 520 Edgemont Road, Charlottesville, VA 22903, USA 
    \label{inst50} \and
    European Space Astronomy Centre (ESAC/ESA), Villanueva de la Ca\~nada, E-28692 Madrid, Spain 
    \label{inst51} \and
    Observatoire de Paris, PSL University, Sorbonne Universit\'e, LERMA, 75014, Paris, France 
    \label{inst52} \and
    Harvard-Smithsonian Center for Astrophysics, 60 Garden Street, Cambridge MA 02138, USA 
    \label{inst53} \and
    Sorbonne Universit\'{e}, CNRS, UMR 7095, Institut d'Astrophysique de Paris, 98bis bd Arago, 75014 Paris, France 
    \label{inst54} \and
    Institut Universitaire de France, Minist\`{e}re de l'Enseignement Sup\'erieur et de la Recherche, 1 rue Descartes, 75231 Paris Cedex 05, France 
    \label{inst55} \and
    Department of Physics and Astronomy, Rice University, Houston TX, 77005-1892, USA 
    \label{inst56} \and
    Yunnan Observatories, Chinese Academy of Sciences, 396 Yangfangwang, Guandu District, Kunming, 650216, China 
    \label{inst57} \and
    Chinese Academy of Sciences South America Center for Astronomy, National Astronomical Observatories, CAS, Beijing 100101, China 
    \label{inst58} \and 
    Departamento de Astronom\'{i}a, Universidad de Chile, Casilla 36-D, Santiago, Chile
    \label{inst59} \and 
    Departments of Chemistry and Astronomy, University of Virginia, Charlottesville, Virginia 22904, USA 
    \label{inst60} \and
    InterCat and Dept. Physics and Astron., Aarhus University, Ny Munkegade 120, 8000 Aarhus C, Denmark 
    \label{inst61} \and
    Laboratory Astrophysics Group of the Max Planck Institute for Astronomy at the Friedrich Schiller University Jena, Institute of Solid State Physics, Helmholtzweg 3, 07743 Jena, Germany 
    \label{inst62} \and
    Instituto de Astronomia, Geof\'isica e Ci\^encias Atmosf\'ericas, Universidade de S\~ao Paulo, 05509-090 S\~ao Paulo, SP, Brazil 
    \label{inst63} \and
    Department of Physics and Astronomy, San Jos\'e State University, San Jose, CA 95192, USA 
    \label{inst64} \and
    Institut de Ciencies de l’Espai (ICE, CSIC), Can Magrans, s/n, E-08193 Bellaterra, Barcelona, Spain  
    \label{inst65} \and
    ICREA, Pg. Llu ́ıs Companys 23, E-08010 Barcelona, Spain 
    \label{inst66} \and
    Institut d’Estudis Espacials de Catalunya (IEEC), E-08034 Barcelona, Spain 
    \label{inst67} \and
    European Space Agency, Space Telescope Science Institute, 3700 San Martin Drive, Baltimore MD 21218, USA
    \label{inst68} \and
    Institute of Astronomy, Russian Academy of Sciences, 119017, Pyatnitskaya str., 48 , Moscow, Russia 
    \label{inst69} \and
    Department of Earth, Ocean, \& Atmospheric Sciences, University of British Columbia, Canada V6T 1Z4 
    \label{inst70} \and
    Telespazio UK for ESA, ESAC, E-28692 Villanueva de la Ca\~nada, Madrid, Spain 
    \label{inst71} \and
    IPAC, California Institute of Technology, Pasadena, CA, USA 
    \label{inst72} \and
    Department of Physics and Astronomy, University of Missouri, Columbia, MO 65211, USA 
    \label{inst73} \and
    Max Planck Institute for Astronomy, K\"onigstuhl 17, 69117 Heidelberg, Germany 
    \label{inst74} \and
    Chemical Sciences Division, Lawrence Berkeley National Laboratory, Berkeley, California, USA 
    \label{inst75} \and
    Kenneth S.~Pitzer Center for Theoretical Chemistry, Department of Chemistry, University of California -- Berkeley, Berkeley, California, USA 
    \label{inst76} \and
    Institut des Sciences Mol\'eculaires, CNRS, Universit\'e de Bordeaux, 33405 Talence, France 
    \label{inst77} \and
    Department of Chemistry, Massachusetts Institute of Technology, Cambridge, MA 02139, USA 
    \label{inst78} \and
    Instituto de Ciencia de Materiales de Madrid (CSIC), Sor Juana Ines de la Cruz 3, E28049, Madrid, Spain 
    \label{inst79} \and
    Department of Physics, PO Box 64, 00014 University of Helsinki, Finland 
    \label{inst80} \and
    Steward Observatory, University of Arizona, Tucson, AZ 85721-0065, USA 
    \label{inst81} \and
    AstronetX PBC, 55 Post Rd W FL 2, Westport, CT 06880  USA 
    \label{inst82} \and
    Department of Physics, College of Science, United Arab Emirates University (UAEU), Al-Ain, 15551, UAE 
    \label{inst83} \and
    National Astronomical Observatory of Japan, National Institutes of Natural Science, 2-21-1 Osawa, Mitaka, Tokyo 181-8588, Japan 
    \label{inst84} \and
    California Institute of Technology, IPAC, 770, S. Wilson Ave., Pasadena, CA 91125, USA 
    \label{inst85} \and
    Department of Physics, Institute of Science, Banaras Hindu University, Varanasi 221005, India 
    \label{inst86} \and
    University of Central Florida, Orlando, FL 32765 
    \label{inst87} \and
    Van't Hoff Institute for Molecular Sciences, University of Amsterdam, PO Box 94157, 1090 GD, Amsterdam, The Netherlands  
    \label{inst88} \and
    Laboratoire de Chimie et Physique Quantiques LCPQ/IRSAMC, UMR5626, Universit\'e de Toulouse (UPS) and CNRS, Toulouse, France 
    \label{inst89} \and
    Instituto de Matem\'atica, Estat\'istica e F\'isica, Universidade Federal do Rio Grande, 96201-900, Rio Grande, RS, Brazil 
    \label{inst90} \and
    Center for Astrophysics and Space Sciences, Department of Physics, University of California, San Diego, 9500 Gilman Drive, La Jolla, CA 92093, USA 
    \label{inst91} \and
    School of Chemistry, The University of Nottingham, University Park, Nottingham, NG7 2RD, United Kingdom 
    \label{inst92} \and
    Astronomy Department, Ohio State University, Columbus, OH 43210 USA 
    \label{inst93} \and
    Space Science Institute, 4765 Walnut St., R203, Boulder, CO 80301
    \label{inst94} \and
    Department of Physics, Stockholm University, SE-10691 Stockholm, Sweden
    \label{inst95} \and
    Department of Physics, Texas State University, San Marcos, TX 78666 USA 
    \label{inst96} \and
    Ritter Astrophysical Research Center, University of Toledo, Toledo, OH 43606, USA 
    \label{inst97} \and
    Department of Physics, Stockholm University, SE-10691 Stockholm, Sweden 
    \label{inst98} \and
    School of Physics and Astronomy, Sun Yat-sen University, 2 Da Xue Road, Tangjia, Zhuhai 519000,  Guangdong Province, China 
    \label{inst99} \and
    Star and Planet Formation Laboratory, RIKEN Cluster for Pioneering Research, Hirosawa 2-1, Wako, Saitama 351-0198, Japan 
    \label{inst100}  \and
    Institute of Deep Space Sciences, Deep Space Exploration Laboratory, Hefei 230026, China 
    \label{inst101}
}

\date{Received 26 April 2023; accepted 25 July 2023}

\abstract
   {The James Webb Space Telescope (JWST) has captured the most detailed and sharpest infrared (IR) images ever taken of the inner region of the Orion Nebula, the nearest massive star formation region, and a prototypical highly irradiated dense photo-dissociation region (PDR). 
   }
   {We investigate the fundamental interaction of far-ultraviolet (FUV) photons with molecular clouds. 
   The transitions across the ionization front (IF), dissociation front (DF), and the molecular cloud are studied at high-angular resolution. These transitions are  relevant to understanding the effects of radiative feedback from massive stars and the dominant physical and chemical processes that lead to the IR emission that JWST will detect in many Galactic and extragalactic environments.}
   {We utilized NIRCam and MIRI to obtain sub-arcsecond images over $\sim 150\arcsec$ and 42$\arcsec$ in key gas phase lines (e.g., Pa\,$\alpha$, Br\,$\alpha$, [FeII] 1.64~$\mu$m, H$_2$ 1-0 S(1) 2.12~$\mu$m, 0-0 S(9) 4.69~$\mu$m), aromatic and aliphatic infrared bands (aromatic infrared bands at  3.3-3.4~$\mu$m, 7.7, and 11.3~$\mu$m), dust emission, and scattered light. Their emission  
   are powerful tracers of the IF and DF, FUV radiation field and density distribution. 
Using NIRSpec observations the fractional contributions of lines, AIBs, and continuum emission to our NIRCam images were estimated. A very good agreement is found for the distribution and intensity of lines and AIBs between the NIRCam and 
NIRSpec observations.}
   {Due to the proximity of the Orion Nebula and the unprecedented angular resolution of JWST, these data reveal that the molecular cloud borders are hyper structured at small angular scales of $\sim 0.1-1\arcsec$ ($\sim$0.0002-0.002 pc or $\sim$40-400 au at 414 pc). 
A diverse set of features are observed such as ridges, waves, globules and photoevaporated protoplanetary disks. At the PDR atomic to molecular
transition, several bright features are detected that are associated with the highly irradiated surroundings of the
dense molecular condensations and embedded young star.
Toward the Orion Bar PDR, a highly sculpted interface is detected with sharp edges and density increases near the IF and DF. 
This was predicted by previous modeling studies, but the fronts were unresolved in most tracers. 
The spatial distribution of the AIBs reveals that the PDR edge is steep and is followed by an extensive warm atomic layer up to the DF with multiple ridges. A complex, structured, and folded H$^0$/H$_2$ DF surface was traced by the H$_2$ lines. 
This dataset was used to revisit the commonly adopted 2D PDR structure of the Orion Bar as our observations show that a 3D ``terraced'' geometry is required to explain the JWST observations. 
JWST provides us with a complete view of the PDR, all the way from the PDR edge to the substructured dense region, and this allowed us to determine, in detail, where the emission of the atomic and molecular lines, aromatic bands, and dust originate.
}
{This study offers an unprecedented dataset to 
benchmark and transform PDR physico-chemical and dynamical models for the JWST era. A fundamental step forward in our understanding of the interaction of FUV photons with molecular clouds and the role of FUV irradiation along the star formation sequence is provided. 
}
   

\maketitle

%


\begin{figure*}[h!]
\begin{center}
\includegraphics[width=0.9\textwidth]{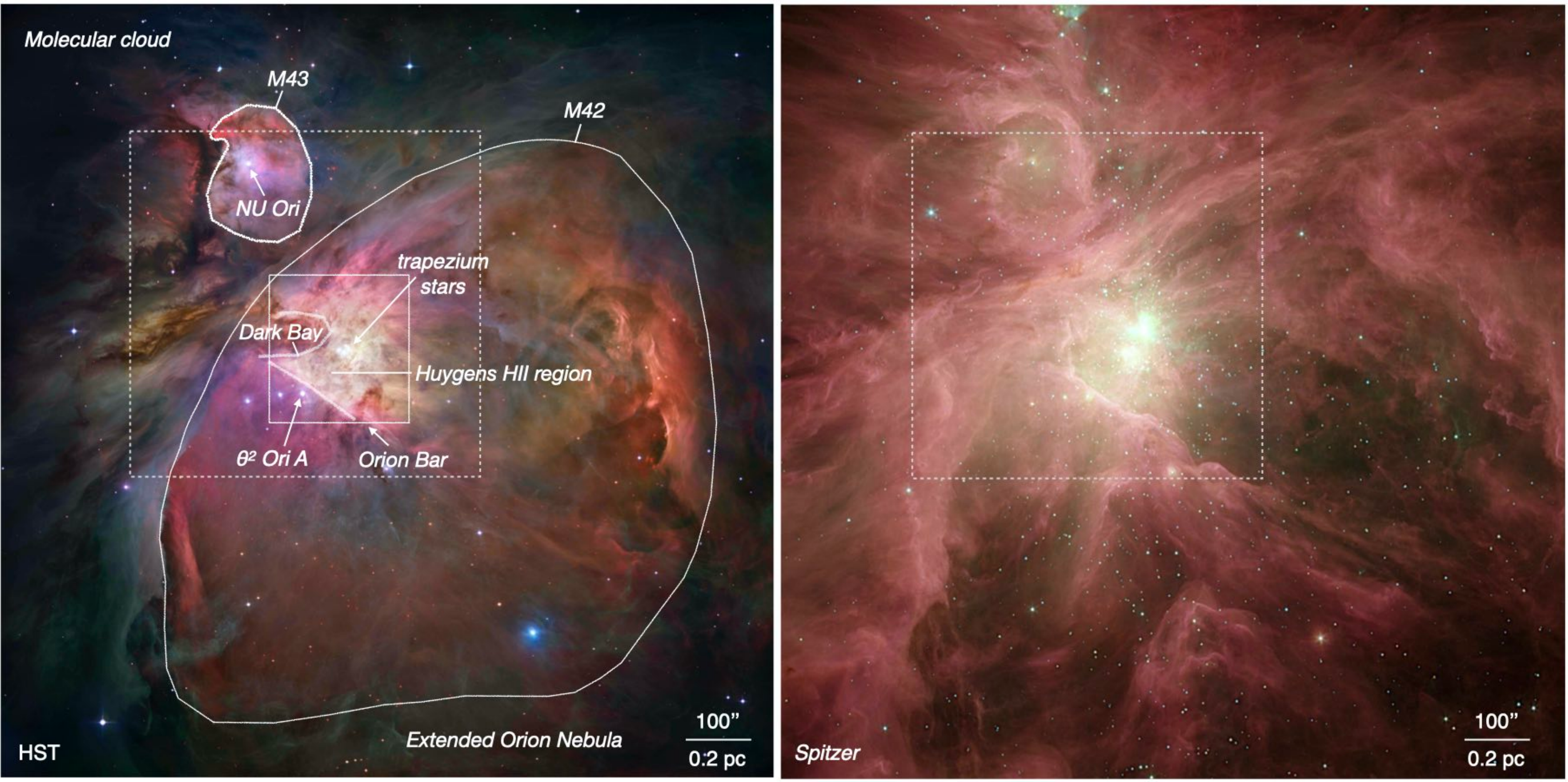}
\vspace*{0cm}
\caption{Composite color Hubble (left) and \textit{Spitzer} (right) views of M42 and M43 in the Orion Nebula complex. 
The solid white lines delineate the shells of M42 and M43, the Bar, the Huygens HII region, and the Dark Bay. The Huygens Region corresponds to the brightest parts of the Orion Nebula. The Dark Bay shows that there are regions of high optical depth between the observer and the ionized gas. 
Foreground depressions of visible light such as the Dark Bay are globally known as the Veil \citep{abel19,pabst2019}. 
Arrows indicate the positions of the most massive stars within each region: Trapezium stars in M42 and NU\,Ori in M43. 
The position of the $\theta ^2$ Ori A star near the center of the Bar is also indicated.
The areas where the JWST observations were obtained are the Bar, north of the Dark Bay, and north of M42 and M43 as presented in Fig.~\ref{fig:overlay_imaging}.  The large dashed rectangle shows the limits of the upper panel of Fig.~\ref{fig:overlay_imaging}.
The image credits for \textit{Hubble} are as follows:
NASA, ESA, M. Robberto (Space Telescope Science Institute/ESA) and the Hubble Space Telescope Orion Treasury Project Team. 
The image credits for \textit{Spitzer} are as follows: NASA/JPL-Caltech/Univ. of Toledo. The wavelengths of the composite color images correspond to 435, 555, 658, 775 and 850~nm for Hubble and 3.6, 4.5, 5.8 and 8~$\mu$m for \textit{Spitzer}.}
\label{fig:hst_orion_nebula}
\end{center}
\end{figure*}

\section{Introduction}
\label{sec:intro}

Massive stars dominate the evolution of the galaxy through the injection of radiative and mechanical energy into their natal molecular cloud and surrounding interstellar medium (ISM). This feedback stirs up and heats the gas and limits molecular cloud lifetimes through photo-ionization and photo-evaporation, inhibiting future star formation. Feedback can also trigger star formation as gas is swept up in dense and massive shells. Hence, feedback is closely tied to the star formation efficiency of molecular gas \citep[e.g.,][]{elmegreen2011,hopkins2014}.

Our understanding of stellar feedback is directly linked to studies of photo-dissociation regions (PDRs).  PDRs are the regions where far-ultraviolet
(FUV; $6$ eV $< h\nu < 13.6$ eV) radiation from massive stars dominates the
thermal processes and chemistry \citep[see reviews by][]{hollenbach1997,Wolfire2022}. PDRs separate the gas ionized by a star from the molecular cloud in which the star was born. Hence, stellar radiative energy is mainly deposited in a PDR, while the mechanical energy is transmitted through the PDR layer to the interiors of molecular clouds in the form of shock waves and/or turbulence.
The ``classic'' PDRs are at the interface between the HII region and the molecular cloud, extending into the deeper molecular layers \citep[e.g.,][]{tielens:85,tielens:85b}. PDRs however,
appear in many other environments including reflection nebulae
\citep{sheffer2011,peeters2017}, planetary
nebulae \citep{bernard2005physical}, surfaces of pillars and globules \citep{goicoechea2020,schneider2021}, the diffuse ISM \citep{wol95,wolfire_neutral_2003},
and in protostellar and protoplanetary disks (\citealt{gorti,vicente2013polycyclic,champion2017}; see reviews by \citealt{oberg2021,winter2022}).
In fact, most of the non stellar baryons in galaxies are in PDRs \citep{Hollenbach99}.
Thus, understanding the physics and chemistry of PDRs 
is critical
for understanding the star formation history of the Universe. 

With their wavelength coverage extending well into the mid-infrared, the instruments on JWST are well suited to study the physical and chemical characteristics of PDRs. The warm ($T\simeq 100-1000$ K) and dense ($n\gtrsim 10^3$ cm$^{-3}$) PDR gas -- mainly heated  through
photoelectric emission of electrons from small grains and molecules \citep{bakesandtielens94,weingartner_photoelectric_2001, berne2022contribution}
 -- is bright in the pure rotational transitions of H$_2$, the mid- and far-infrared fine-structure transitions of atomic ions and neutrals (e.g., Si$^+$, Fe$^+$, C$^+$, O), rotational transitions of CO and its isotopes, and rotational transitions of small radicals.  Carbon is singly ionized in the PDR surface layers and the cascade generated by electron recombinations will produce a rich spectrum of C$^0$ lines \citep{cesarsky1982,walmsley2000structure}. In addition, the strong FUV field produces bright fluorescence in the near-infrared ro-vibrational transitions of H$_2$, atomic transitions of O and N, and the Aromatic Infrared Bands (AIBs), generally attributed to the emission of vibrationally excited  polycyclic aromatic hydrocarbon molecules (PAHs) \citep[e.g.,][]{black1987,Tielens93,marconi1998,martini1999,Peeters04,tielens_interstellar_2008,habart2011}. The adjoining ionized gas will show bright IR line emission produced by collisional excitation of fine-structure levels (e.g., Fe$^+$, Ar$^+$, Ar$^{2+}$, Ne$^+$, Ne$^2+$, S$^{2+}$, S$^{3+}$)
and by recombination lines from HI and HeI \citep{martin-hernandez2002,rubin2007}. The spatial resolution of JWST ($\sim 0.1 - 1\arcsec$) exceeds that of all other space telescopes over the same wavelength range\footnote{The ground based Keck telescope using adaptive optics can
achieve a similar spatial resolution but with a restricted IR filter set
and wavelength range \citep[e.g.,][]{Habart2023}.} and is similar in spatial resolution to that
of the Atacama Large Millimeter Array (ALMA) at submillimeter wavelengths. 
JWST has many IR filters centered on gas lines, molecular spectroscopic patterns as well as the continuum due to interstellar dust (emission and scattering).
JWST emission line and continuum images of a  PDR thus carry key information relevant  to our understanding of the morphological impacts of stellar feedback, and JWST observations enable us to probe  at unprecedented
resolution how a molecular cloud is being disrupted by
strong stellar UV radiation, winds, outflows and jets.

The focus of this article is on JWST NIRCam  and  MIRI images of the Orion Nebula complex carried
out as part of the PDRs4All Early Release Science (ERS) program  \citep{pdrs4all}. 
The Orion Nebula complex is a nearby site of active star formation
exhibiting many feedback processes and PDR interfaces 
\citep[e.g.,][]{pabst2019}. The prototypical highly irradiated dense PDR in this nebula is usually referred to as the "Bright Bar" or "Orion Bar" \citep[e.g.,][]{Elliott1974,tielens1993anatomy,ODell2000}. In the following, we refer to it as the "Bar."
The ionizing and FUV radiation from the Orion Trapezium Cluster shines directly on the
face of the Bar.
At the outer layers, the ionized gas recombines
at the ionization front (IF) and the gas becomes neutral hydrogen.
This corresponds to the edge of the neutral PDR. 
The gas remains atomic \cite[e.g.,][]{vanderWerf13,Henney2021} until the H$_2$ 
dissociation front (DF), where the molecular hydrogen abundance increases rapidly. 
Over nearly 40 years, the Bar has been the target of many studies to elucidate the physical and chemical characteristics of PDRs \citep{Parmar91,Tielens93,Tauber94,YoungOwl00,Bernard-Salas2012,goicoechea_compression_2016,Parikka2018,kaplan2021} and provides therefore a widely-used template for the observational signature of the interaction of stars with their environment, both in the Milky Way and galaxies out to high redshifts \citep{Stacey2010,Vallini2018,Wolfire2022}. With its wide wavelength coverage, high sensitivity, multiple filters and high spectral resolution resulting in large line-to-continuum ratios, JWST has the potential to provide a  coherent vision of the structure of the Bar. Its structure includes the extended atomic layers (often called the ``inter-clump'' medium) and the thin emission layers of dense warm gas associated with the DF as well as the illuminated surfaces of large dense clumps. 

The article is organized as follows. 
In Sect.~\ref{sec:taget_M42_Orion_Bar}, we describe the  main physical characteristics of our target inferred from previous studies.
The observations, data reduction and the fractional contributions of line, AIB, and continuum emission to our NIRCam images are described in Sect.~\ref{sec:obs-reduction}.
In Sect.~\ref{sec:inner-nebula-morphology}, we describe the  structures observed by NIRCam and MIRI within the inner region of the OrionNebula. 
In Sect.~\ref{sec:Bar}, we focus on the Bar as a template region to understand the structure
 and morphology of a strongly irradiated PDR. The complex transition from the IF, the PDR DF to the molecular cloud is studied, and we determine in detail the origin of 
 the atomic and molecular lines, aromatic bands, and dust emission.
In Sect.~\ref{sec:proplyds}, we describe the photoevaporating proto-planetary disks observed in the whole NIRCam fields.
A summary and conclusions are given in Section~\ref{sec:conclusions}. 
In Appendix~\ref{Appendix:contribution_lines}, we show the template NIRSpec spectra presented in \citet{Peeters-spectro}  in the wavelength domain of NIRCam filters, illustrating the variation of the contribution of different lines into each imaging band. 
Appendix~\ref{Appendix:Images_all_filters_Orion_Bar} provides NIRCam and 
ground based images of the Bar.
In Appendix~\ref{Appendix:north-nebula-morphology}, we describe the  structures observed within the NIRCam fields north of the Dark Bay, north of M42 and in M43 (Fig.~\ref{fig:hst_orion_nebula}).

\section{M42 and the Bar}
\label{sec:taget_M42_Orion_Bar}

The Orion HII region, M42, -- the nearest site of massive star formation --  is illuminated by the Trapezium stars for which the \mbox{O7-type} star, \mbox{$\theta^1$ Ori C} dominates \citep{sota2011}, being the most massive and luminous member of the Trapezium  cluster at the heart of the Orion Nebula \citep[e.g.,][]{Odell01}. 
\mbox{$\theta^1$ Ori C} has created a concave blister of ionized gas on the surface of the underlying Orion Molecular Core 1 (OMC-1) \citep[see][and references therein]{wen1995three,Odell01}, with the brightest portion called the Huygens Region (Fig.~\ref{fig:hst_orion_nebula}). The electron density varies across this region from a central high of almost $10^4$ cm$^{-3}$ to 3$\times 10^3$ cm$^{-3}$ in the outer regions, while the electron temperature is usually about 9000 K \citep{Weilbacher15}. Between the ionized gas and OMC-1 lies a PDR that is face-on to our line of sight. 
In the region of the Bar, the ionized atomic layer and its PDR are tilted almost along the line of sight \citep[$\sim$4 (1-8) degrees, ][]{walmsley2000structure,Pellegrini09,Salgado16,Peeters-spectro}, forming an escarpment in the Main Ionization Front (MIF) and due to projection effects, producing one of the optically brightest features of the Huygens Region. This also provides the ability 
to probe without overlapping the multiple layers of the PDR 
\citep[][c.f. Fig.~\ref{fig:Geometry}]{Tielens93,hogerheijde1995millimeter}. The Bar is a strongly UV irradiated PDR viewed nearly edge-on.

The gas density ($n_H$) in the ambient molecular cloud is estimated to be $n_H = 0.5 - 1.0\times 10^5$ cm$^{-3}$ from a variety of IR and submillimeter gaseous emission lines \citep{tielens:85b,Hoger95,Bernard-Salas2012,goicoechea2017}. In addition, much denser  small structures and molecular condensations ("clumps") are embedded in the Bar \citep[$n_H \gtrsim 10^6$ cm$^{-3}$;][]{Lis03,goicoechea_compression_2016,Joblin2018,Cuadrado19}. The Far-UV (FUV) radiation field incident on the PDR of the Bar is G$_0= 2.2-7.1\times10^4$ in Habing units \citep[$1.6\times 10^{-3}$ erg cm$^{-2}$ s$^{-1}$;][]{Habing68} as derived from UV-pumped IR-fluorescent lines of OI by \citet{Marconi98} and \citet{Peeters-spectro}. Given the stellar characteristics of \mbox{$\theta^1$ Ori C} and the far-IR surface brightness of the Bar, this places \mbox{$\theta^1$ Ori C} at a physical distance of 0.33 pc from the far-IR dust emission \citep{Salgado16}. 
The projected distance between the star and the ionization front (IF) is about 0.2 pc (c.f., Fig.~\ref{fig:Geometry}).
Beyond the IF, where the hydrogen gas 
converts from ionized to neutral, only FUV photons with energies below 13.6\,eV penetrate the cloud. This corresponds to the edge of the neutral PDR but note that species with low ionization potentials (e.g., C, S, Fe) are still ionized in the surface layers of PDRs. Deeper in ($\Delta A_V\simeq 1-2$ mag), the radiation field has been sufficiently attenuated by dust extinction that hydrogen goes from atomic to molecular. At a depth of $\Delta A_V\simeq 2-4$ mag, the carbon balance shifts from C$^+$ to C$^0$ to CO \citep{tielens:85}. 
This chemical stratification has been verified by an important series of infrared to radio observations \cite[e.g.,][]{jansen1995millimeter,Tielens93,goicoechea_compression_2016}. 
Relative to the molecular gas in OMC-1, the PDR gas flows through the ionization front at $\simeq 1$ km/s \citep{pabst2019} and once ionized accelerates away at about 7$\pm$4~km/s for the [N II] emitting layer close to the ionization front and 12$\pm$4~km/s for the [O III] emitting layer further out \citep{odell2020}
as it joins the general expansion of the nebula. Close to  \mbox{$\theta^1$ Ori C}  there is a low density bubble of gas of diameter of 0.2 pc shaped by its stellar wind \citep{Odell2009,odell2020}.
This wind-blown cavity is open to the southwest and feeds the region of the Extended Orion Nebula (EON, Figure \ref{fig:hst_orion_nebula}) where hot shocked gas has been detected \citep{Gudel08}.  
A layer of ionized gas covers much of the Huygens Regions \citep{Garcia-Diaz2007,odell2020}
and outside of this is the atomic layer of gas known as the Veil, one portion of which is a hemispherical bubble best described as the Outer Shell, which was discovered by \cite{pabst2019} and expands at about 13 km/s away from the OMC. The Veil has a column density of $N_H=(2-6)\times 10^{21}$~cm$^{-2}$, depending on the direction, and obscures the Huygens Region by 1-3 magnitudes of visual extinction  \citep{odell1992,ODell2000,Weilbacher15}. 
The physical characteristics of the Bar are summarized in Table \ref{tab:param_OB}.

Assuming that the background PDR directly behind the Trapezium stars is a face-on PDR, the incident FUV field is estimated to be $10^5$ Habings from the observed far-IR surface brightness \citep{tielens:85b}; a factor of $\simeq 2-5$ higher than the FUV field incident on the Bar. The gas density in this PDR is estimated to be slightly higher ($10^5$ cm$^{-3}$) than in the Bar \citep{tielens:85b}, concomitant with the higher density of the ionized gas \citep{Weilbacher15}. 

The wide field of view of the JWST images include the nearby low-ionization HII region M43 (NGC 1982) powered by HD 37061 (also known as NU Ori), a B0.5V star (Fig. \ref{fig:hst_orion_nebula}). M43 lies to the northeast of M42 and this object has not been well studied but we include analysis of this JWST data in this article as well.
M43 is seen to be shielded from illumination by \mbox{$\theta^1$ Ori C} by the northeast portion of the wall bounding M42. 

M42 exhibits several high-velocity features, including microjets, large scale Herbig-Haro flows, and wind driven shocks \cite[e.g.,][]{bally2000,Odell01}.
Protostellar jets and outflows emanate from dust-enshrouded, nascent stars.
Shocks are formed where the collimated flows interact with the nebula's ambient ionized gas and the neutral foreground veil. 
Additionally, uncollimated flows from the low-mass accreting stars and the stellar wind from \mbox{$\theta^1$ Ori C} produce shocks. M42 further shows structures resulting from embedded sources of outflow in the BN-KL and Orion-S regions. 

As this summary demonstrates, many questions remain on: (i) the detailed geometry of this highly irradiated and very structured PDR; (ii) the best tracers of the different physical zones (H$^+$, H$^0$, H$_2$, C$^+$, C, CO); (iii) the physical and chemical conditions in these different zones, particularly at the ionization and dissociation fronts; (iv) the relationship of the various components (interclump, clumps, proplyds, winds \&\ jets) populating this region.
JWST, with its high spatial resolution, can uniquely address these issues and thereby provide valuable insight in the physical and chemical processes taking place in FUV irradiated, interstellar material.

\section{Observations and data reduction} 
\label{sec:obs-reduction}

\subsection{Observations}
\label{sec:obs-obs}

We provide in this section a summary of the main parameters of imaging observations obtained within the ERS project \#1288 "Radiative Feedback from Massive Stars as Traced by Multiband Imaging and Spectroscopic Mosaics," based on NIRCam and MIRI observations \citep{pdrs4all}.
The details of the observations can be retrieved from STScI using the Astronomer Proposal Toolkit (APT) under program ID 1288. 

\begin{figure*}[h!]
\begin{center}
\includegraphics[width=0.7\textwidth]{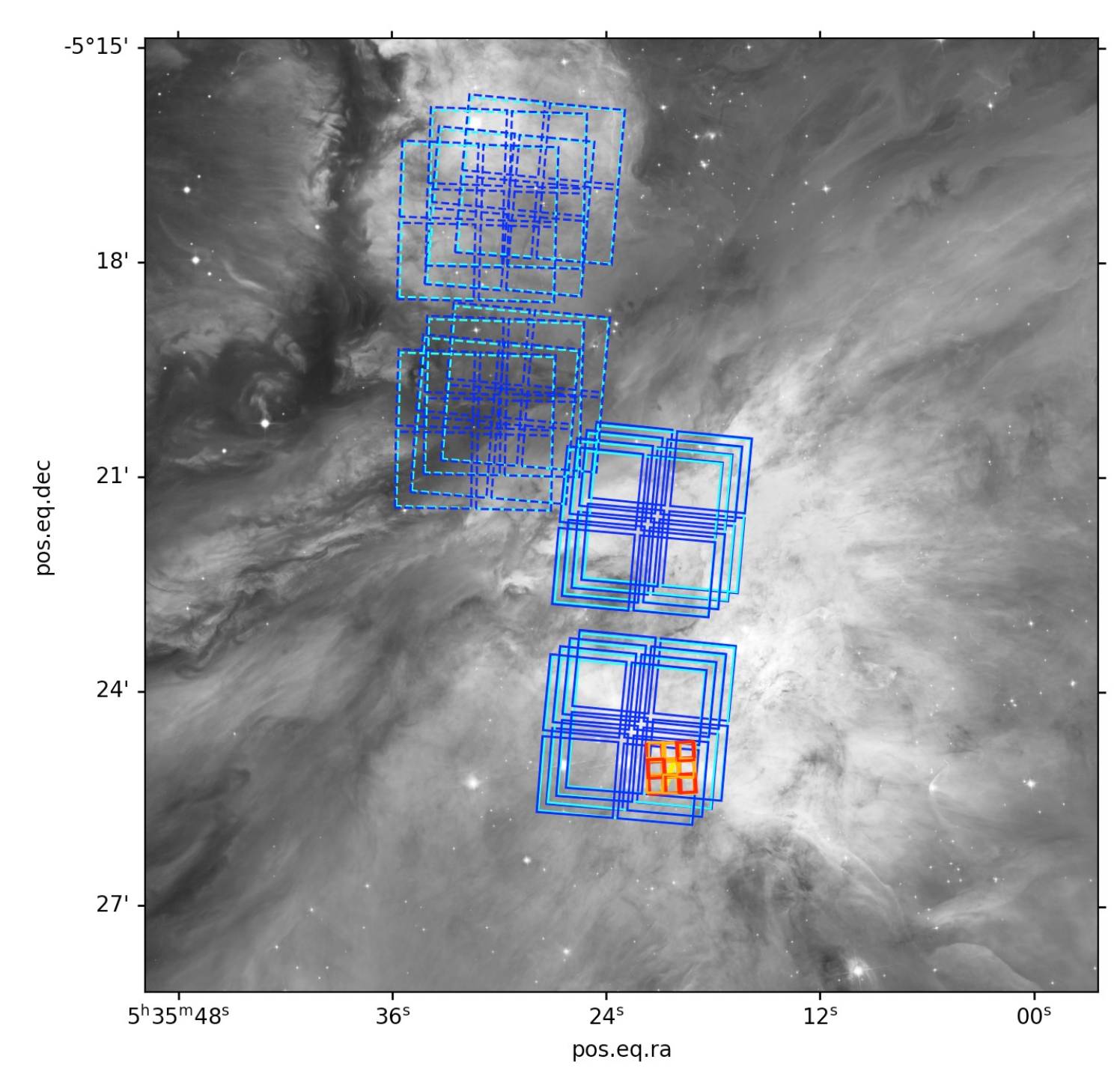}
\includegraphics[width=0.5\textwidth]{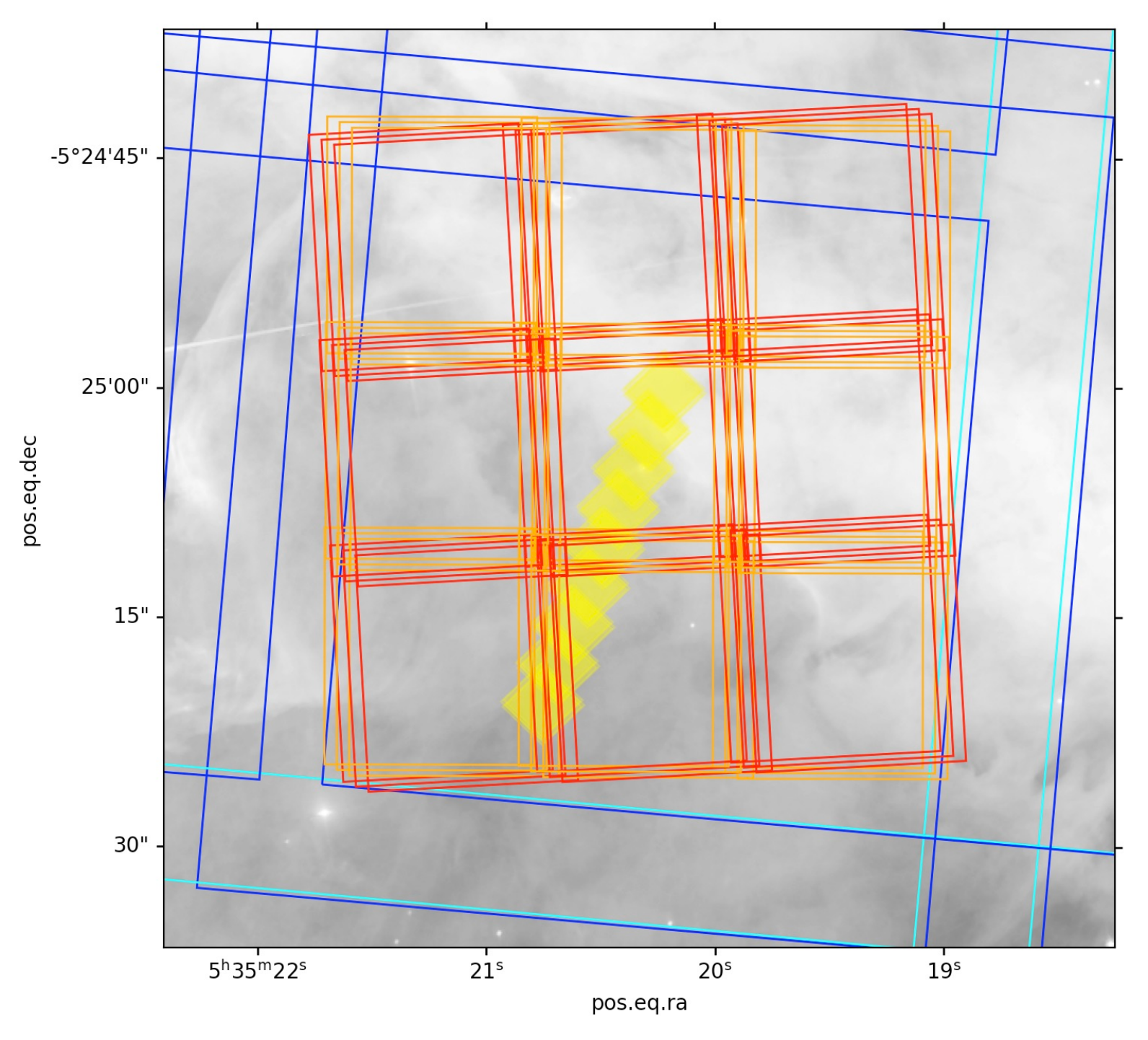}
\vspace*{0cm}
\caption{Overlay of the JWST NIRCam (blue) and MIRI (red) imaging and NIRSpec (yellow) footprints on the Hubble Space Telescope image of the Bar at 1.3 $\mu$m \citep{rob20}. 
{\bf Upper}: All of the NIRCam and MIRI observations. For NIRCam, module B covering the Bar and module A in the North of the Dark Bay are shown in this overlay. The pointing position 
is situated in between these two modules at RA=05$^{\rm h}$35$^{\rm m}$20$\fs$1963, DEC=-05$\degr$23$\arcmin$10.45$\arcsec$.
Dotted FOVs correspond to the NIRCam parallel observations. {\bf Lower}: MIRI observations are centered on position RA=05$^{\rm h}$35$^{\rm m}$20$\fs$3448, DEC=-05$\degr$25$\arcmin$4.01$\arcsec$. The NIRSpec footprint (yellow) is a strip oriented perpendicularly to the ionization front.}
\label{fig:overlay_imaging}
\end{center}
\end{figure*}


The telescope orientation (V3 Position Angle) was left unconstrained between 260 and 270 degrees. At the time of observation, 10-11th of September 2022, the telescope was oriented at about 265 degrees, resulting in the pointing illustrated in Fig. \ref{fig:overlay_imaging}. 

For NIRCam observations, we refer in particular to the data obtained under Observation 1 "NIRCam Orion Bar Imaging,", whereas for MIRI we refer to Observations 15 and 16 "MIRI Orion Bar Imaging." These two MIRI observations were executed with NIRCam in parallel on September 11th for the F1500W and F2550W filters and September 18th for the F770W and F1130W filters, with a difference between these two dates for the telescope orientation on the ecliptic plane of about 3 degrees. Due to the V3 orientation, the parallel NIRCam field observed covers the northern part of the Orion Nebula, near the M43 (NGC 1982) region. A second set of background images, Observations 14 and 17, were taken at an offset field about 2 degrees west of the Bar using again MIRI and NIRCam in parallel with the same parameters adopted for Observation 15 and 16, respectively. However, Observations 14 could not be achieved due to Fine Guidance Sensor (FGS) loss of fine guidance control, the background observations in the corresponding MIRI and NIRCAM filters are therefore not available.
An overview of the imaging filter selection is given in Table~\ref{tab:filters}. 

\begin{table*}
    \caption{Summary of imaging observations obtained with NIRCam and MIRI and filters selection. The main targeted species and features with the chosen filters are given in the first and third column.}
    \label{tab:filters}
    \begin{center}
    \begin{tabular}{lcccccccccc}
    \hline
    \hline
    Species & $\lambda~^1$ & Filter & Cont. & B$^3$ & S/R$^4$ & M43$^5$ & Obs. ID & Background$^7$ & Pixel scale & PSF FWHM \\
    & ($\mu$m) & & filter$^2$ & & & &(OB,M43/BKG) & &(arcsec) & (arcsec)\\
    \hline
    \\[-5pt]
    \multicolumn{11}{c}{NIRCam}\\[5pt]
 $\left[\mathrm{FeII}\right]~$   &  1.644 & F164N &  & $\surd$ & $>$4 & & 1 & & 0.031 &0.056 \\
   & &  &  F162M & $\surd$ & $>$30 & & 1& & 0.031 &0.055  \\
      Pa~$\alpha$  & 1.876 & F187N &  & $\surd$ & $>54$ & $\surd$ & 1,15/14 &$\times$ & 0.031 & 0.064\\
       &  &  & F182M & $\surd$ & $>75$ & $\surd$ & 1,16/17 & $\surd$ & 0.031 & 0.062\\
      Br~$\alpha$ & 4.052 & F405N & & $\surd$ & $>$326 & $\surd$ & 1,16/17 & $\surd$ & 0.063& 0.136\\
      &  &  & F480M & $\surd$ & $>$198 & & 1 & & 0.063& 0.164\\
      &  &  & F410M &  &  & $\surd$ & 1,15/14 &$\times$ & 0.063& 0.137 \\
      H$_2$ & 2.120 & F212N &  & $\surd$ & $>$6 & $\surd$& 1,15/14 &$\times$ & 0.031 &0.072\\
       &  &  & F210M & $\surd$ & $>$38 & $\surd$& 1,16/17 & $\surd$ & 0.031&0.071\\
      AIB & 3.235 & F323N$^6$ &  & $\surd$ & $>$60 & & 1 & & 0.063&0.108\\
       &  &  & F300M & $\surd$ & $>$131 & & 1,15/14 &$\times$& 0.063&0.100\\
      H$_2$ & 4.694 & F470N &  & $\surd$ & $>$54 & & 1& & 0.063&0.160\\
       & &  & F480M & $\surd$ & $>$198 & & 1 & & 0.063&0.164\\
    AIBs & 3.3-3.4 & F335M &  & $\surd$ & $>$382 & $\surd$& 1,16/17 & $\surd$ & 0.063&0.111\\
     &  & & F300M & $\surd$ & $>$131 & $\surd$& 1,15/14& & 0.063&0.100\\
      & 1.405 & F140M & & $\surd$ & $>$19 & & 1 & & 0.031&0.048\\
      & 2.672 & F277W & & $\surd$ & $>$182 & & 1 & & 0.063&0.092\\
    \hline
    \\[-5pt]
     \multicolumn{11}{c}{MIRI}\\[5pt]
     AIBs & 7.7 & F770W &  & $\surd$ &  & &15/14 & $\times$ & 0.11&0.269\\
     AIBs & 11.3& F1130W & & $\surd$ & & &15/14 & $\times$& 0.11&0.375\\
     & 15.0& F1500W & & $\surd$ &  & & 16/17& $\surd$& 0.11&0.488\\
     & 25.5& F2550W & & $\surd$ & & & 16/17& $\surd$& 0.11&0.803\\[5pt]
    \hline
   \end{tabular}
    \end{center}
    $^1$ The wavelength of the transition or pivot wavelength of the filter for NIRCam broad band filters. 
    $^2$ Continuum filter.
    $^3$ B: Bar (on-target observations) and north of the Bar.
    $^4$ S/R: Minimum signal-to-noise ratio measured at the template positions (HII, atomic and molecular regions) in the Bar and covered by the NIRSpec field. 
    $^5$ M43 (NGC 1982): NIRCam parallel observations.
    $^6$ This filter contains an H$_2$ line at 3.23 $\mu$m but is largely dominated by the 3.3 $\mu$m AIB.
    $^7$ A "$\times$" indicates that the observation has failed because of FGS loss.
\end{table*}

\subsubsection{NIRCam imaging} 

The selected NIRCam filters cover i) the 3.3-3.4 $\mu$m Aromatic Infrared Bands (AIBs), ii) the ro-vibrationally and rotationally excited lines of H$_2$ 1-0 S(1) at 2.12 and H$_2$ 0-0 S(9) at 4.69 $\mu$m, tracing the dissociation front, iii) the [FeII] line at 1.64 $\mu$m, tracing the ionization front, and iv) the Paschen Pa~$\alpha$ and Brackett Br~$\alpha$ atomic hydrogen lines, tracing the \HII\, region. Each filter was paired with a reference filter centered on an adjacent wavelength for subtraction of the underlying continuum emission.
We mapped the PDR region with a single pointing using a 4 point primary dither (Fig.~\ref{fig:overlay_imaging}). 
To avoid saturation, we used the RAPID readout mode with two groups per integration and two integrations per exposure. With 4 dithered pointings, this corresponds to 8 total integrations. The integration time per pixel from the reset to the second sample was about 21.47s, corresponding to a total exposure time of 171.788~s.
The minimum resulting signal-to-noise ratio on the extended emission 
 toward the Bar template regions covered by NIRSpec
is given  for all filters in Table \ref{tab:filters}.

\subsubsection{MIRI imaging} 

Data were obtained in  i) the 7.7 and 11.3 $\mu$m filters including AIBs which, when combined, could provide a proxy for PAH ionization  \citep[e.g.,][]{Joblin:96, galliano_variations_2008}
and ii) the 15 and 25 $\mu$m filters dominated by continuum emission tracing warm dust in the \HII\, and neutral region, similarly to the corresponding WISE, \textit{Spitzer}, and IRAS filters.
We obtain a $3\times3$ mosaic using a three point dither pattern (3-POINT-MIRI-F770W-WITH-NIRCam) (Fig. \ref{fig:overlay_imaging}). To prevent saturation given the brightness of the Bar, we use the FASTR1 readout pattern and the SUB128 imaging subarray. 

\subsubsection{NIRCam parallel observations} 

We obtained parallel NIRCam observations with the on-source MIRI imaging.  The adopted filters cover i) the 3.3 and 3.4 $\mu$m AIB, 
ii) the vibrationally excited line of H$_2$ 1-0 S(1) at 2.12 $\mu$m, and iii) the Pa~$\alpha$ and Br~$\alpha$ lines.
Also in this case, each filter was paired with a reference filter to estimate and be able to subtract the underlying continuum emission. 
The pointings, number of dithers, 
and dither pattern were set by those of the primary observations (on-source MIRI imaging). 
To accommodate the brightness of the Bar, we used the BRIGHT2 readout mode
with two groups per integration, one integration per exposure, and the three dithered positions. The effective exposure time 
in this case results is 1159.569 s, corresponding to 
580 s with correlated double sampling.

\subsection{Data reduction}

\subsubsection{NIRCam} 
Given the evolving nature of the automatic pipeline producing the data available on MAST, 
we have chosen to reduce our observations starting from the original {\tt \_uncal} files, that is those produced by the preliminary Stage 0, adopting the latest available development version of the pipeline at that time 1.7.3. 
Stage 1 corrects instrument signatures that need to be treated at the level of individual groups before ramp fitting, such as for dark current, nonlinearity detector response, and cosmic ray (CR) events. Given the intensity of the signal, we turned off the 
{\tt suppress\_one\_group} 
option of the ramp fitting step to recover signal saturated after the very first sample.  
Stage 2 operates on the count rate images produced by Stage 1 removing the background -if dedicated files are provided-, calibrating each exposure individually to produce images in physical units of MJy/sr and rectifying the images for final combination with other images. The last stage combines the rectified exposures from multiple exposures, performing outlier pixel removal and astrometric alignment to produce the final products, i.e. drizzled and mosaicked images with their associated data catalogs and segmentation maps. We have chosen to independently combine the different dithers for each module and each filter. 
We have also tested the impact of different parameters. First, we have tried bypassing the outlier detection step. This step allows finding some bad pixels and cosmic-rays not corrected in the previous stage. Without this step, the final images  show a strong "salt and pepper" pixeling effect.

On the short wavelength filters (F162M, F164N, and F212N), we detected some artefacts identified as wisps. They are due to straylight and are located at the same positions in each detector. The wisps are mainly on the B4 and A3 detectors and their positions depend on the filter, the detector, and the observation.
To correct them, we selected polygons encompassing the wisps by hand. Then, we flagged these polygons as DO\_NOT\_USE in the DQ array. The wisps were removed because they were 
not considered further but as a consequence
the noise was higher in these areas in the mosaic, because there is less information.
We corrected for the 1/f noise by subtracting the median value of each row and then each column before the JWST pipeline stage 3 process. We first took care to apply a mask on all the bright sources  when computing the median values for row and column subtractions and carefully inspected the results for any unintended consequences.

Finally, we improved the world coordinate system (WCS) alignment between the different mosaics by detecting unsaturated point sources in all detector images and comparing their positions with the astrometry from the Gaia DR3 catalog. The astrometry correction was at most of 0.3$\arcsec$.
To get the flux-calibrated line maps continuum subtracted in erg cm$^{-2}$ s$^{-1}$ sr$^{-1}$, the maps (in MJy/sr) are multiplied by $10^6 \times 10^{-23} \times \Delta\,\lambda \times c / (\lambda_{c})^2$ with $\Delta\,\lambda$ and $\lambda_{c}$ the bandwidth and pivot wavelength of the filters from the NIRCam manual.

\subsubsection{MIRI} 
As for the NIRCam observations, we reduced our observations starting from the raw data using the latest version of the pipeline and reference files available at that time (February 2023). 
The inter-pixel capacitance (IPC) step, correcting for the interpixel capacitance, was skipped in stage 1 as recommended by the instrument team because of poorly defined deconvolution kernels. Similarly, the Reset Switch Charge Decay (RSCD) step correcting for some ideal detector and readout effects was skipped because of the low number of groups per integration in our observations.
Dedicated backgrounds were observed for the F1500W and F2550W filters and were subtracted from the data in stage 2 of the pipeline. As mentioned in Section \ref{sec:obs-obs}, background observations in the F770W and F1130W filters were not successful due to FGS loss of fine guidance control.
The tweakreg step in stage 3 used to improve the alignment of the different input images was skipped because of the lack of point-like sources in the different filters, as well as the outlier detection step as we observed a deterioration of the final mosaics when applying this correction.
For the F2550W filter, as most of the groups reached the saturation level for the pixels along the Bar, we skipped the jump step because of poor performance.

Moreover, we observed strong edge-brightening effects in the final mosaics attributed to straylight features in the SUB128 array flat fields (the SUB128 array is located at the top left of the detector), stronger at longer wavelengths (F1500W and F2550W filters). A first solution was to flag the affected left columns and top rows   as DO\_NOT\_USE (about 15 rows and columns) so that they are not used in the final mosaic. 
For the F2550W a region around the Lyot mask has also been flagged. The number of rows/columns to flag has been selected for each filter as a trade off between having sufficient pixels for the overlap and removing pixels with poor quality. 
A better solution, used in this article, is to use the mosaics for flat field estimations with backprojection (projection of the computed mosaic back to the detector at the central position). Then  the mosaics are recomputed using the estimated flat fields. Repeating this process with an additional deflagging of the affected columns/rows in each iteration, significantly improved signal recovery.

\subsection{Line, AIBs, and continuum emission contributions to imaging bands in the NIRCam filters}
\label{sec:line_AIB_continuum_Filter}

Here we summarize our results regarding the fractional contributions
of line, AIBs, and continuum emission to our NIRCam images \citep[for a complete description see the Science Enabling Product 4 article, ][]{chown23}.
We computed synthetic NIRCam images from the stitched and co-added NIRSpec mosaic from \citet{Peeters-spectro}. We did this by applying Equation 5 of \citet{Gordon2022} to each spaxel.
This approach is similar to what has been done with \textit{Spitzer}/IRS data\footnote{See, e.g., \url{https://irsa.ipac.caltech.edu/data/Spitzer/docs/dataanalysistools/cookbook/14/}}. We call these images ``synthetic images'' in order to distinguish them from images observed by NIRCam.

We then decomposed each spaxel into line and continuum components (based on local baseline fit around prominent lines), and then calculated synthetic images from each of the line and continuum components. The fractional contribution of emission component $i$ to imaging band $j$, is given by the ratio of the synthetic image of component $i$ in band $j$ to the synthetic image in band $j$ of the total spectrum.
Note that this measurement relies solely on NIRSpec data, and so it is not affected by any differences from NIRSpec and NIRCam data (such as flux calibration, resolution, etc.) -- we investigate such differences in \cite{chown23}.

Figures~\ref{fig:template_FeII} to \ref{fig:template_AIB} show the three template spectra (HII region, Atomic PDR, Dissociation Front DF3) presented in \citet{Peeters-spectro}, illustrating the variation of the contribution of different lines into each imaging band.  Figures in \citet{chown23} show the fractional contribution of continuum and the line of interest into corresponding imaging bands. We thus derive the following for the different lines, bands and the continuum:

\begin{itemize}
    \item 
    {\bf \FeII\ 1.64~$\mu$m:} The F162M filter is dominated by continuum emission ($>80$\%), while F164N captures continuum,  \FeII\ 1.64~$\mu$m line, and the \HI\ (12-4) line (Fig.~\ref{fig:template_FeII}). As a result, F164N--F162M primarily traces \FeII\ 1.64~$\mu$m, except where the \HI\ (12-4) line contributes significantly. 
    
    \item 
    {\bf Pa $\alpha$:} The Pa $\alpha$ line contributes to 80--90\% of the emission in F187N (Fig.~\ref{fig:template_Pa_a}), without subtracting continuum using the F182M band. In F182M, continuum emission contributes 30--50\%, while Pa $\alpha$ contributes 40--60\%. Both F187N and F187N--F182M are well-correlated with the true Pa $\alpha$ line intensity, although some Pa $\alpha$ emission is self-subtracted in F187N--F182M.  
    \item 
    {\bf Br $\alpha$:} This line contributes 40--50\% of the F405N flux, which is well-correlated with the true Br $\alpha$ line intensity. In the Bar (on-target) observations, the F480M filter can be used to trace the continuum underlying F405N. 
    F480M traces the continuum well with a small contribution from an \HI\ line or an H$_2$ line (Fig.~\ref{fig:template_Br_a}). 
    F405N--F480M flux is overall a good measure of Br $\alpha$ intensity, but suffers from more scatter than the F405N vs. Br $\alpha$ correlation.  
    \item 
    {\bf H$_2$ 1-0 S(1) 2.12~$\mu$m:} 
    The fractional contribution of continuum emission in F212N is higher than that in F210M over a large area. This is due to bright \HI\ and \HeI\ emission lines that fall in the F210M filter but fall  outside of the F212N filter (Fig.~\ref{fig:template_H2_212}). This results in a negative flux in the F212N--F210M image in a significant fraction of area closer to the exciting sources. 
    Note also that the He lines that are close to H$_2$ 2.12~$\mu$m contribute to F212N at a comparable degree to the H$_2$ line except in regions that are sufficiently far away from the exciting sources. 
    \item 
    {\bf H$_2$ 1-0 O(5) 3.24~$\mu$m:} This line is located on a shoulder of the 3.3~$\mu$m AIB (Fig.~\ref{fig:template_H2_324}). The contribution of this line to F323N is 5--15\% across the observed area. The correlation of the F323N synthetic image with the actual line intensity is poor, and 
    subtracting continuum as traced by F300M does not improve this correlation.
    \item 
    {\bf H$_2$ 0-0 S(9) 4.69~$\mu$m:} In the northern half of the observed area with NIRSpec, \HI\ lines contribute to the observed F480M flux but not to F470N (Fig.~\ref{fig:template_H2_469}). Similar to the situation with H$_2$ 1-0 S(1) 2.12~$\mu$m in the F212N filter, this results in a negative flux in the F470N--F480M image. However, in the southern half (far from the exciting sources), F480M works reasonably well at subtracting continuum in F470N to estimate the intensity of the H$_2$ 4.69~$\mu$m line, although rising continuum towards longer wavelengths underestimates the line flux probed by F470N--F480M.
    \item 
    {\bf The 3.3~$\mu$m AIB:} This AIB is the dominant component of the emission in the F335M filter. This AIB contributes 50--60\%  of the F335M flux near the exciting sources, and 70--80\% at remaining positions. F300M traces the continuum on the shorter wavelength side of F335M reasonably well (Fig.~\ref{fig:template_AIB}). F335M--F300M is thus a good tracer of the integrated strength of the  3.3~$\mu$m AIB. 
\end{itemize}
A detailed analysis of a full set of lines will be presented in the science enabling product and the associated article \citep{chown23}.

\section{Morphology of the Orion Nebula inner region}
\label{sec:inner-nebula-morphology}

With their high angular resolution and unparalleled sensitivity, NIRCam and MIRI unveil the structures at very small scales of the Orion Nebula (0.1 to 1$\arcsec$ from 2 to 25 $\mu$m, equivalent to $\rm 2\times10^{-4}$ to $\rm 2\times10^{-3}$ pc at the Orion distance of 414~pc). It displays an incredible richness of substructures, 
 as well as previously hidden stars and even background galaxies.
In this section, we present several prominent features arising in the 
images within the inner region of the Orion Nebula (M42), i.e. the fields centered on the Bar. 

\begin{figure*}[h!]
\begin{center}
\vspace*{0cm}
\includegraphics[width=0.85\textwidth]{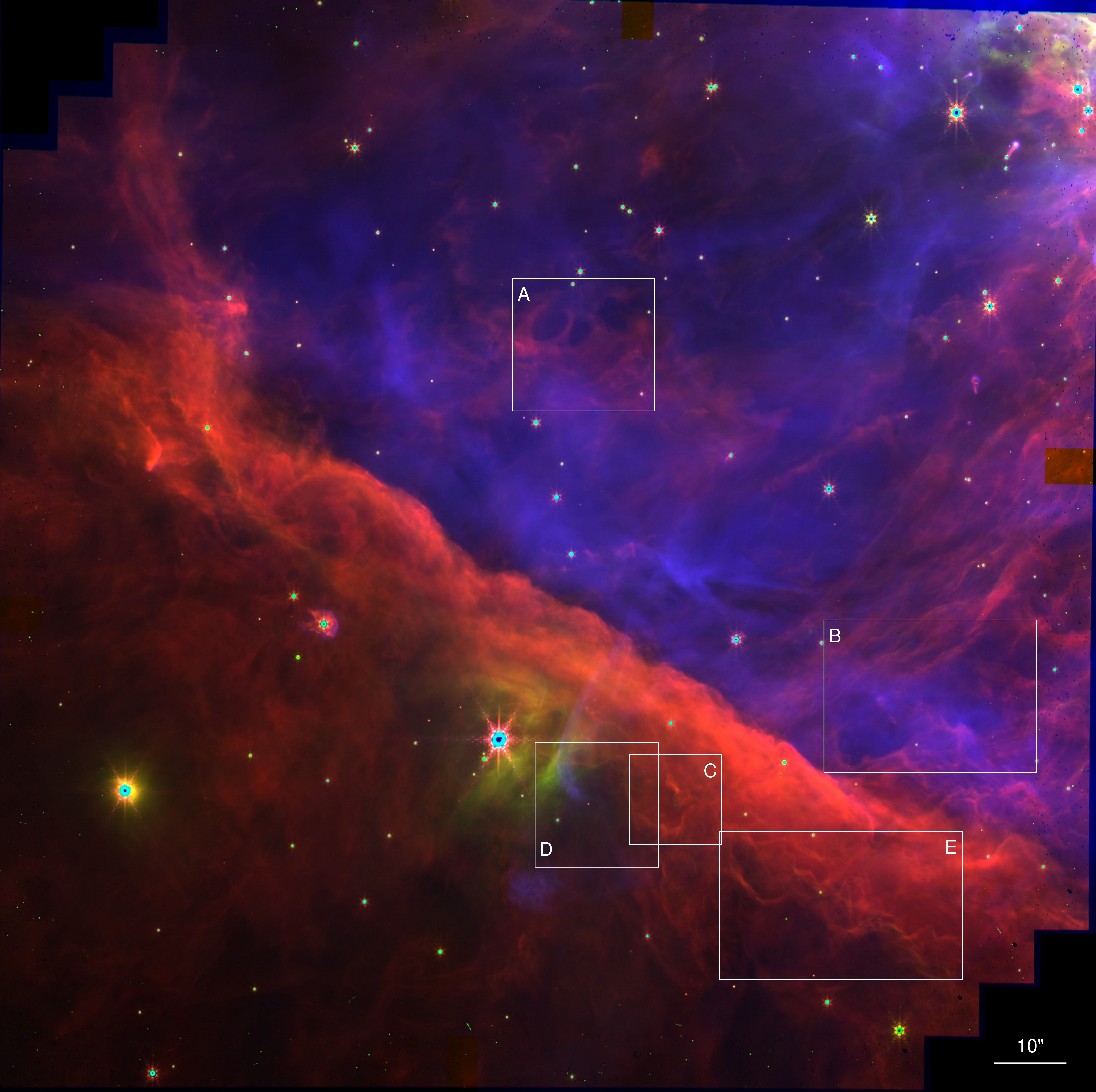}
\includegraphics[width=0.3\textwidth]{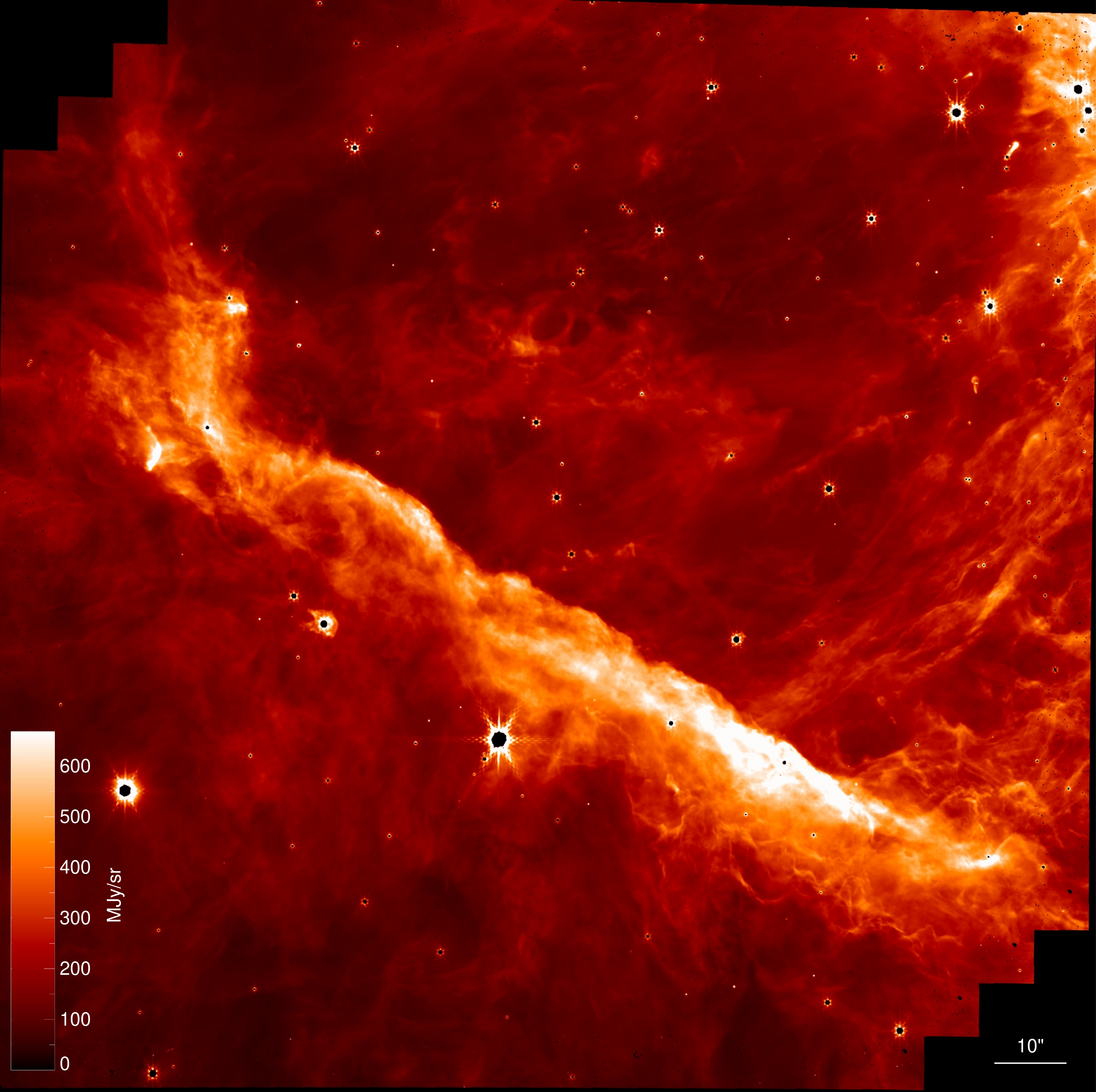}
\includegraphics[width=0.3\textwidth]{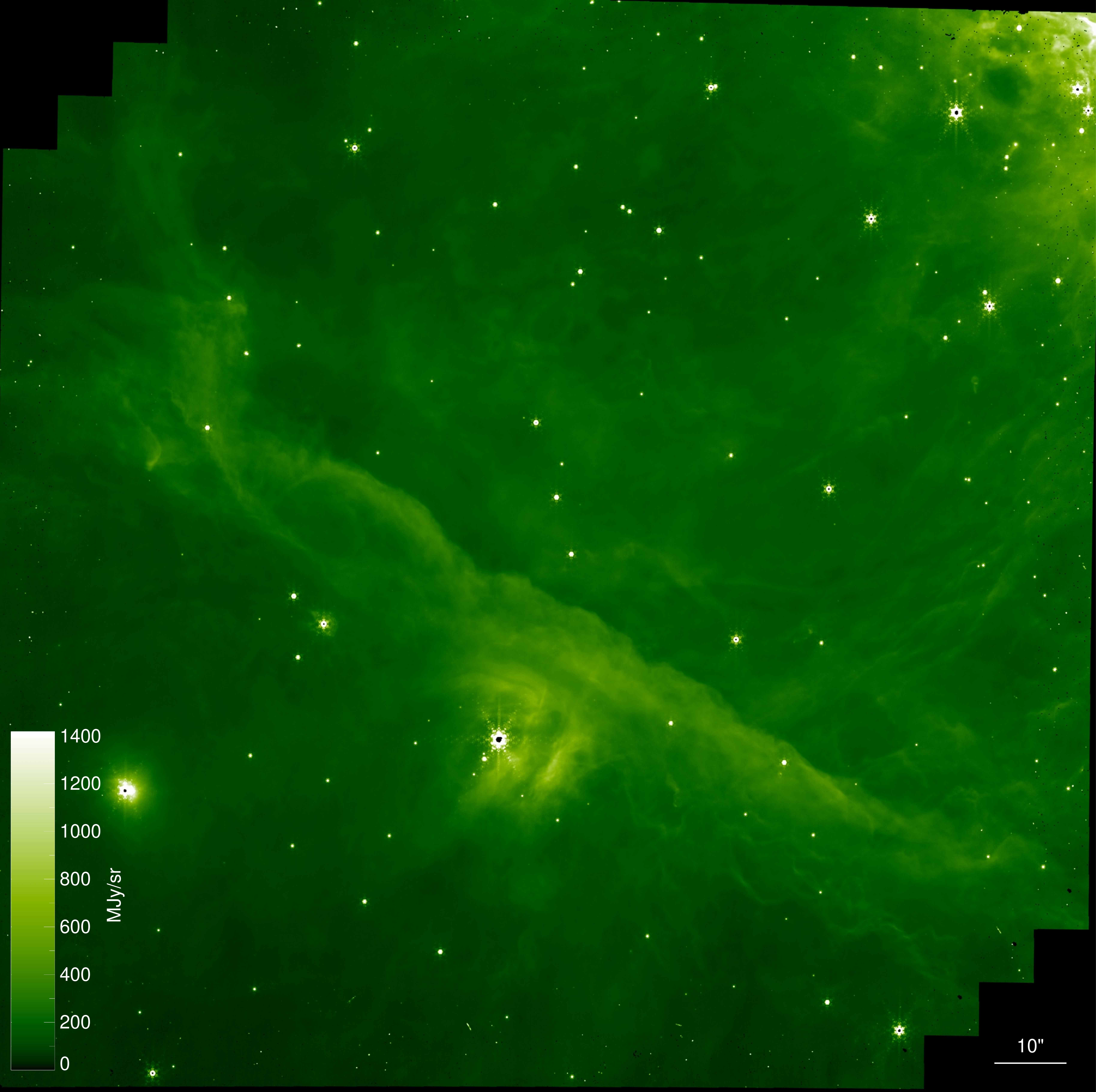}
\includegraphics[width=0.3\textwidth]{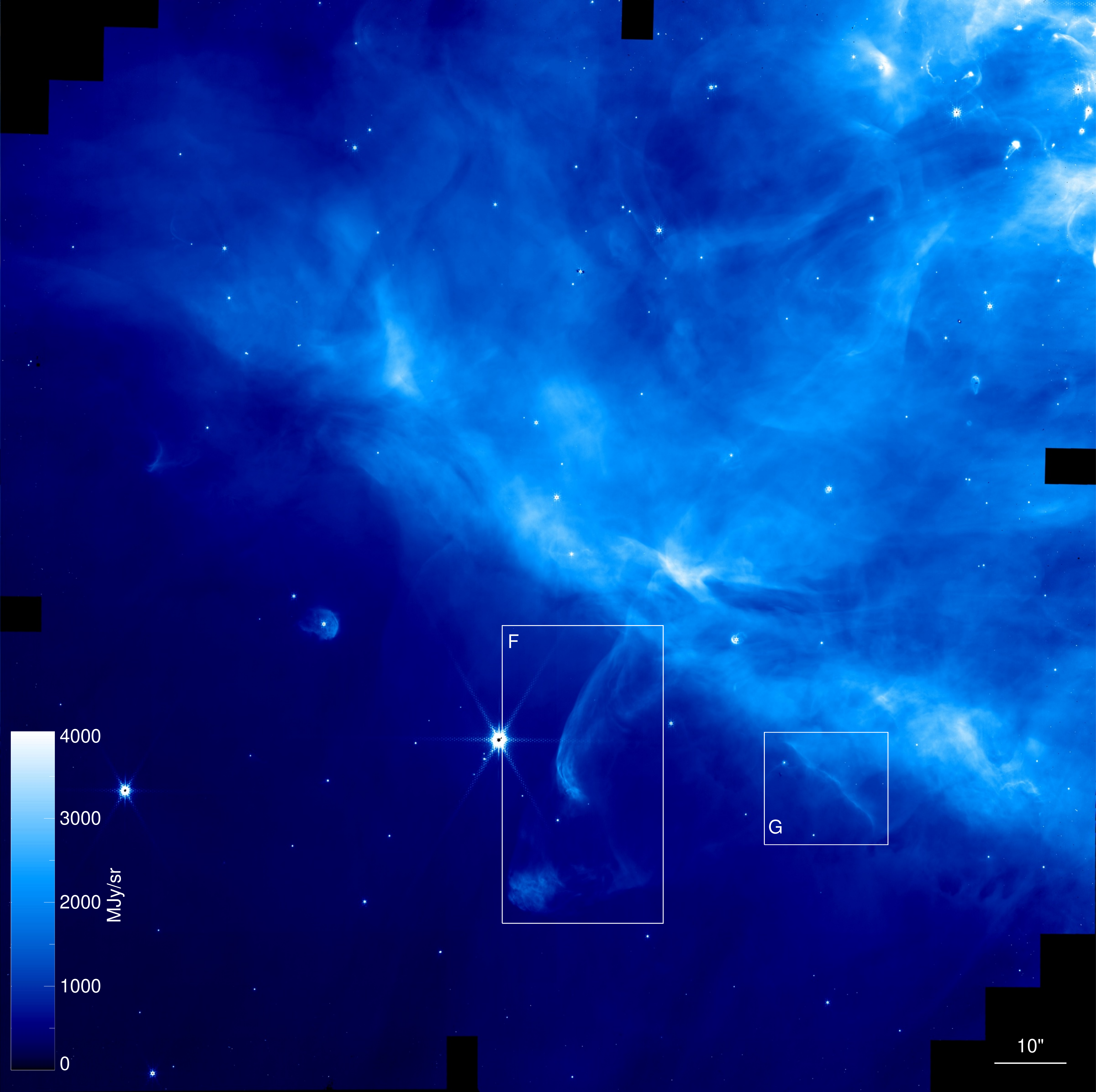}
\caption{Inner region of the Orion Nebula as seen by the JWST’s NIRCam instrument with north up and east left. 
In the upper panel, three images in different filters were combined to produce an RGB composite image: F335M (red), F470N (green) and F187N (blue)
that trace emission from hydrocarbons (AIBs), dust and molecular gas (H$_2$ 0-0 S(9) line), and  
ionized gas (Pa $\alpha$ line), respectively. The individual images used to make the RGB composite are shown in the lower panels. No continuum was subtracted. The size of the images is $\sim$150$\arcsec \times 150 \arcsec$. 
Most prominent is the   Bar, a wall of dense gas and dust that runs from the top left to the bottom right in these images, and that contains the bright star $\theta ^2$ Ori A \mbox{09.5 Vp} which shines in the center of the Bar.
The Bar and the surface of the Orion Molecular Cloud 1 (seen in the background in the front of the Bar) are illuminated by a group of hot, young massive stars forming the Trapezium Cluster, located just off the top right of the image. 
White boxes (labeled A, B, C, D, E, F and G) delineate regions of particular interest zoomed in and shown in Fig. \ref{fig:zoom-image-inside-Orion-nebula}.  
}
\label{fig:composite-image-inside-Orion-nebula}
\end{center}
\end{figure*}

Fig.~\ref{fig:composite-image-inside-Orion-nebula} 
shows composite NIRCam images in three selected filters (F187N, F335M, and F470N). 
 The F187N filter captures the distribution of ionized gas via the bright Pa~$\alpha$, F335M traces mostly emission from the AIB 3.3-3.4 $\mu$m aromatic and aliphatic CH stretching mode bands and F470N traces the dust continuum and the high excited  H$_2$ 0-0 S(9) pure rotational line.
Fig.~\ref{fig:composite-image-MIRI} shows composite MIRI images in the three selected filters F770W, F1130W, F1500W. 
These filters image the 7.7 and 11.3 $\mu$m aromatic bands and the continuum emission from hot/warm dust at thermal equilibrium. The emission at 15~$\mu$m is mainly produced by very small carbonaceous grains whereas at 25~$\mu$m slightly larger grains can contribute \cite[e.g.,][]{compiegne}.  
In Appendix~\ref{Appendix:Images_all_filters_Orion_Bar}, 
all the images obtained in the filters listed in Table~\ref{tab:filters}, and for some gas lines continuum subtracted, are presented. 

A schematic view of the Bar inferred from both these JWST observations and the literature on previous observations from visible to millimeter is presented in Fig.~\ref{fig:Geometry}.
We are viewing the main ionization front almost edge-on along the Bar.

\begin{figure*}[h!]
\begin{center}
\vspace*{0cm}
\includegraphics[width=0.2825\textwidth]{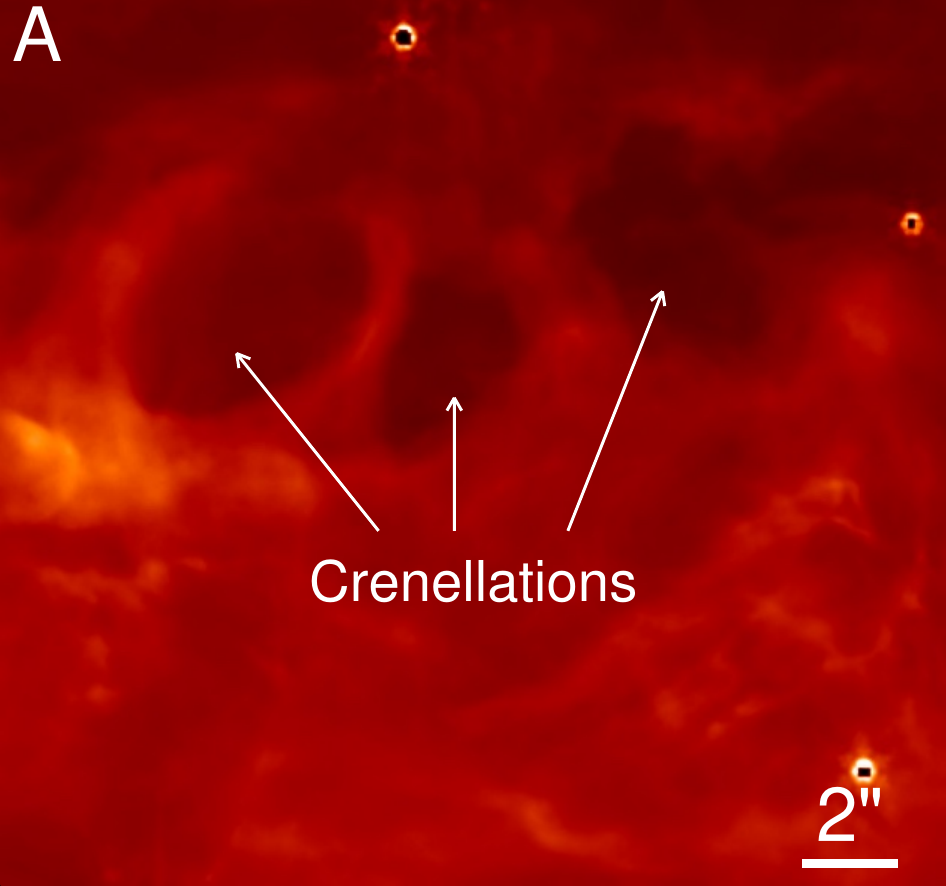}
\includegraphics[width=0.37\textwidth]{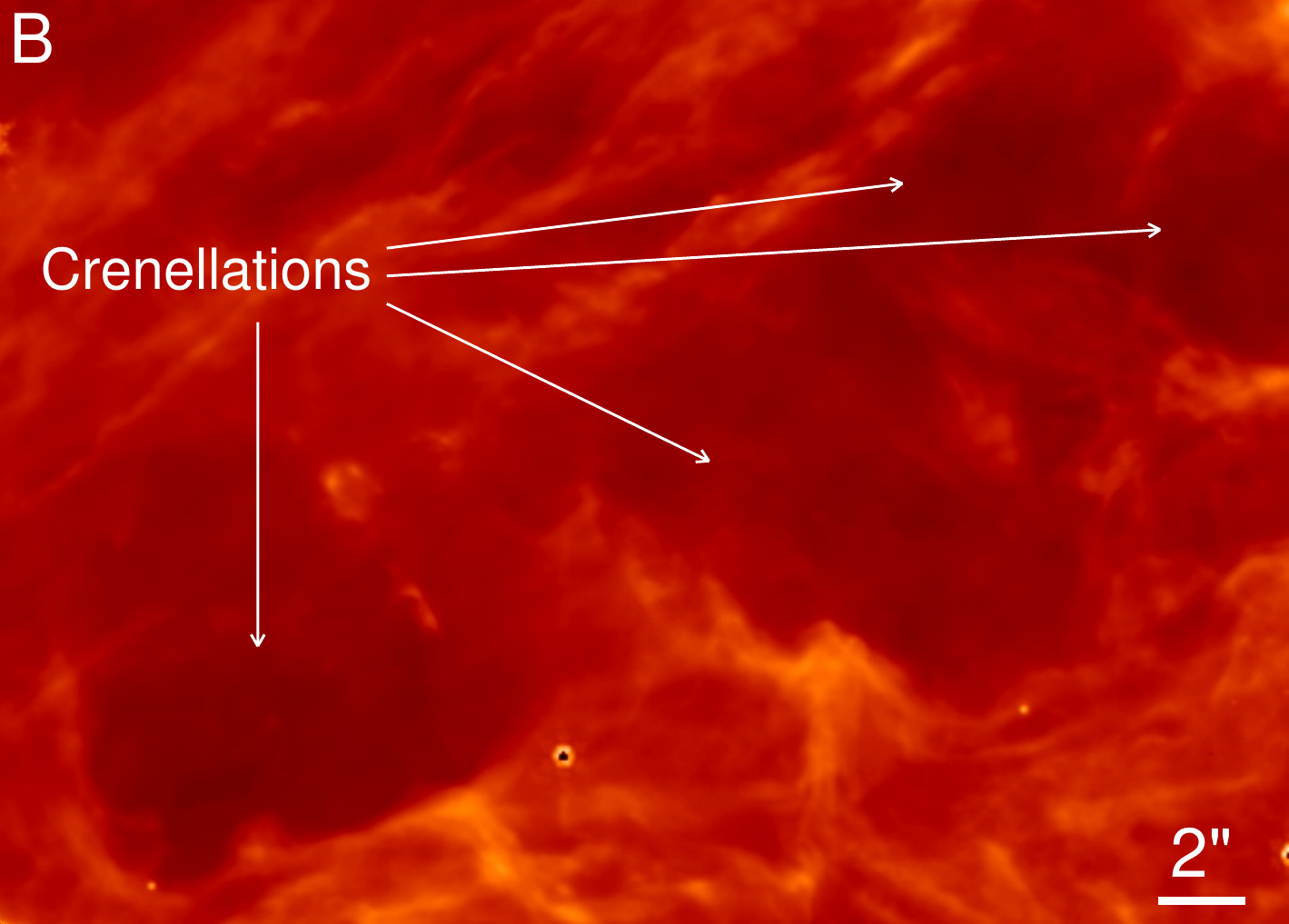}
\includegraphics[width=0.2725\textwidth]{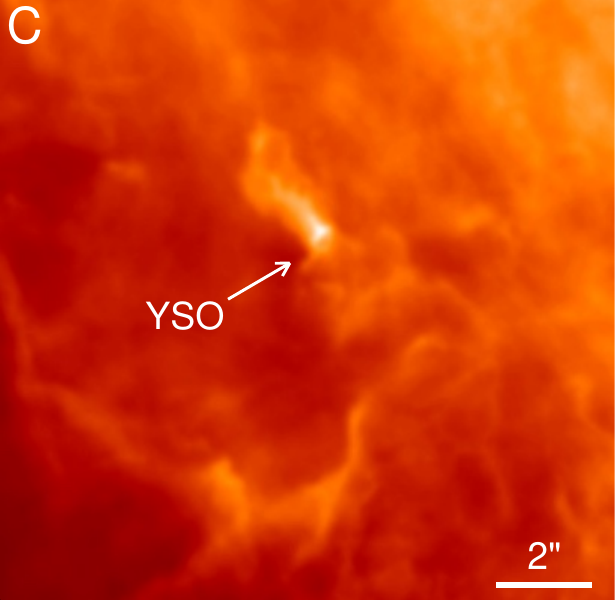}
\includegraphics[width=0.35\textwidth]{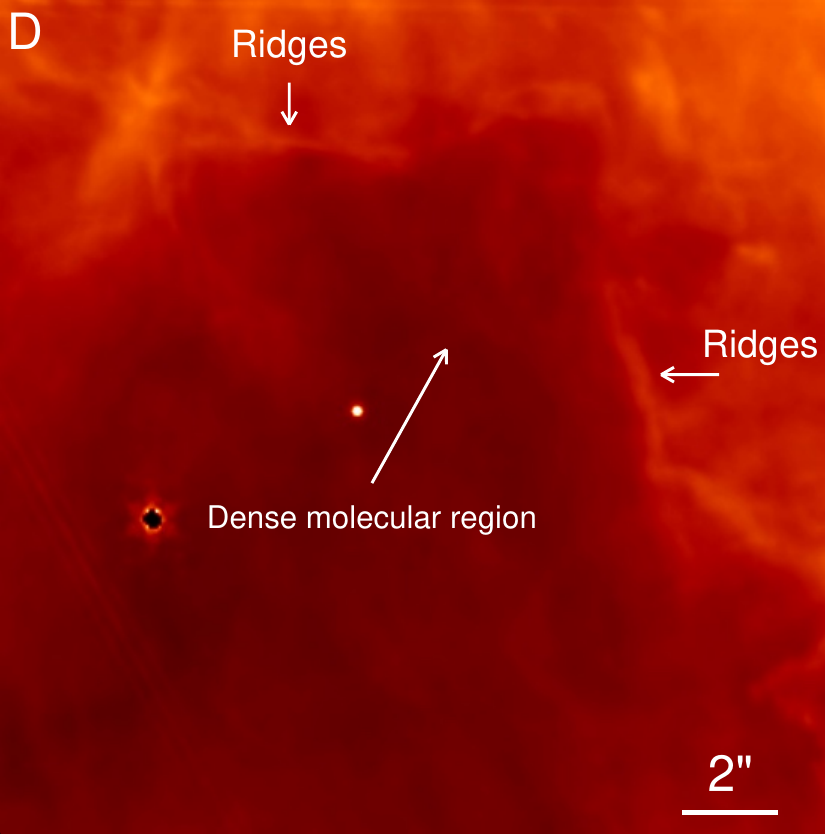}
\includegraphics[width=0.58\textwidth]{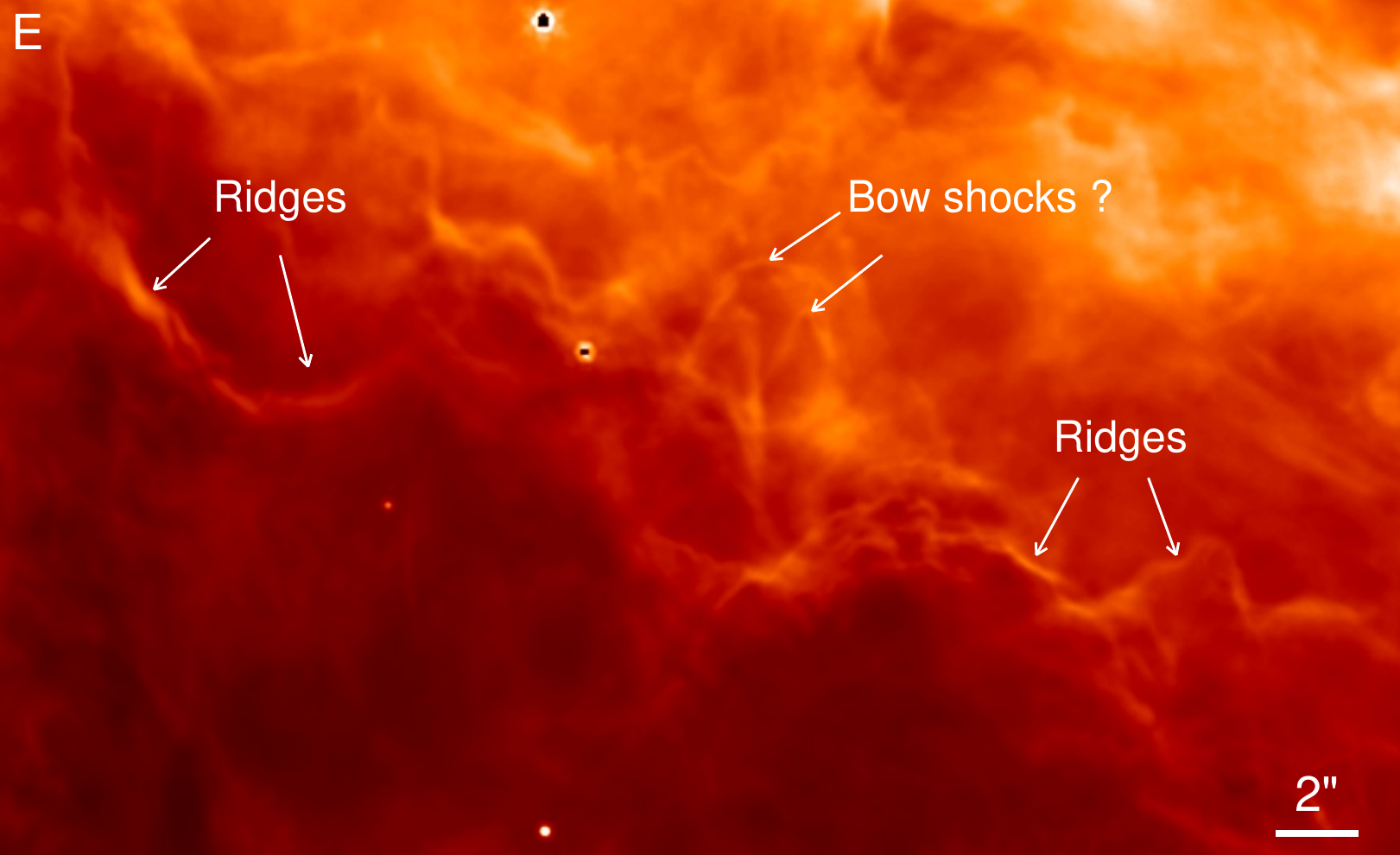}
\includegraphics[width=0.305\textwidth]{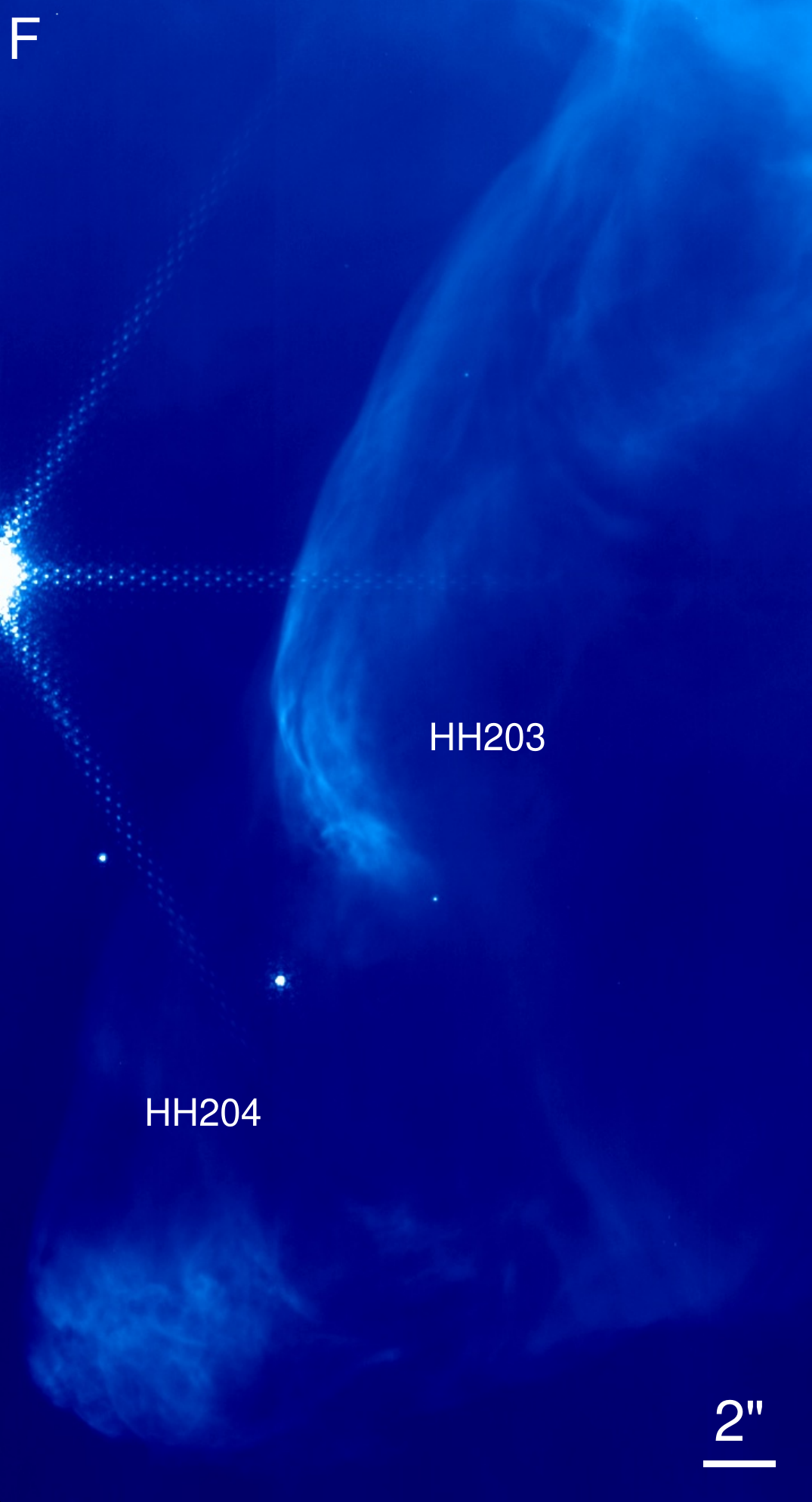}
\includegraphics[width=0.62\textwidth]{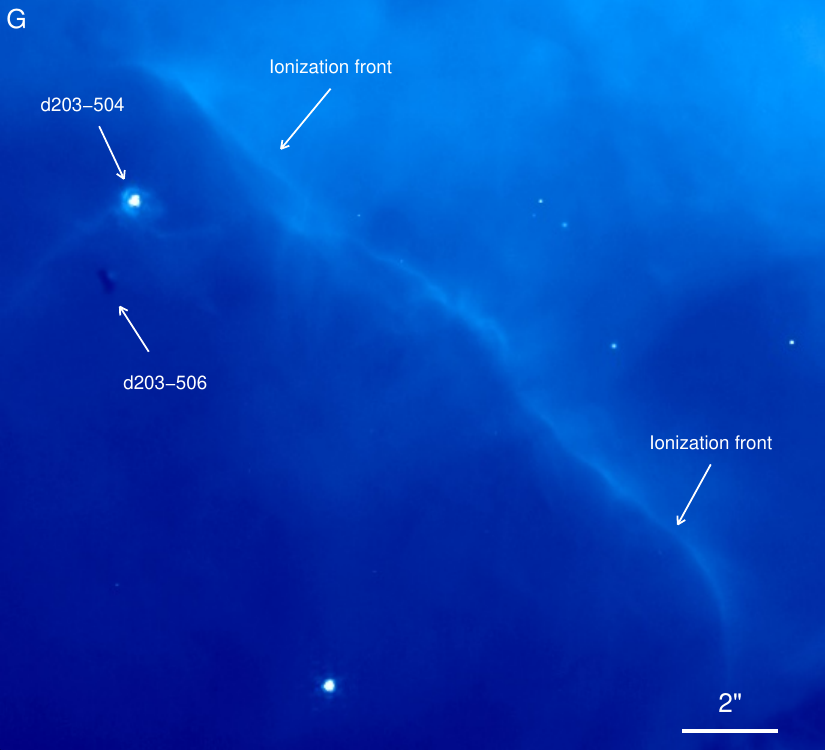}
\caption{Zoom into the F335M (red) and F187N (blue) areas shown in Fig. \ref{fig:composite-image-inside-Orion-nebula}. {\bf A, B} Several crenellations 
that may result from the interaction of collimated jets and outflows from protostars inside the molecular cloud. 
{\bf C} Bright substructure, likely a YSO candidate, detected with ALMA (Fig. \ref{fig:cut_Bar_H2_HCO_CO}). {\bf D, E} Irradiated molecular edges showing many patterns such as ridges. {\bf F} Features known as HH 203 and 204 driven by large outflows. {\bf G} Ionisation front of the Bar with the two irradiated proto-planetary disks 203-504 and 203-506.}
\label{fig:zoom-image-inside-Orion-nebula}
\end{center}
\end{figure*}

\begin{figure*}[h!]
\begin{center}
\vspace*{0cm}
\includegraphics[width=0.95\textwidth]{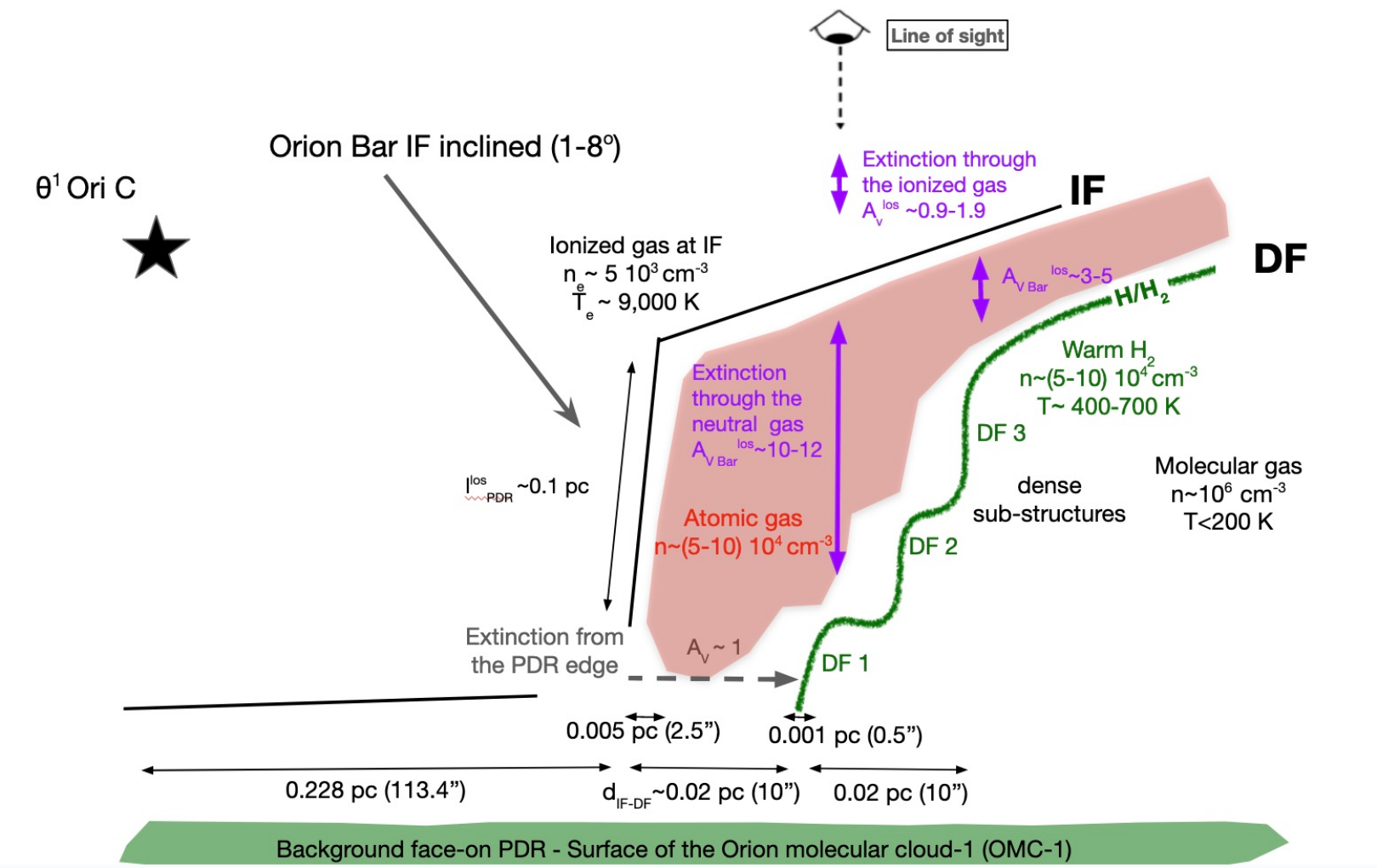}
\caption{Schematic view of the   Bar inferred from JWST and previous observations from visible to mm \cite[e.g.,][]{jansen1995millimeter,wen1995three,Odell01,Pellegrini09}.
It displays the main features discussed in detail in the core of this article, and inferred from both the imaging observations (this work) and NIRSpec spectroscopic observations \citep{Peeters-spectro}.
We note that for clarity the dimensions perpendicular to the Bar are not to scale,
the true spatial scales being explicitly given in the annotations. The adopted distance to the Bar is 414~pc.
In addition, the sketch does not include foreground material, which includes a layer of ionized gas \citep{odell2020} and, closer to the observer, layers that are grouped together under the designation as the Veil \citep[e.g.,][]{vanderWerf13,pabst2019,Pabst20}.}
\label{fig:Geometry}
\end{center}
\end{figure*}

\begin{figure}[h!]
\includegraphics[width=0.5\textwidth]{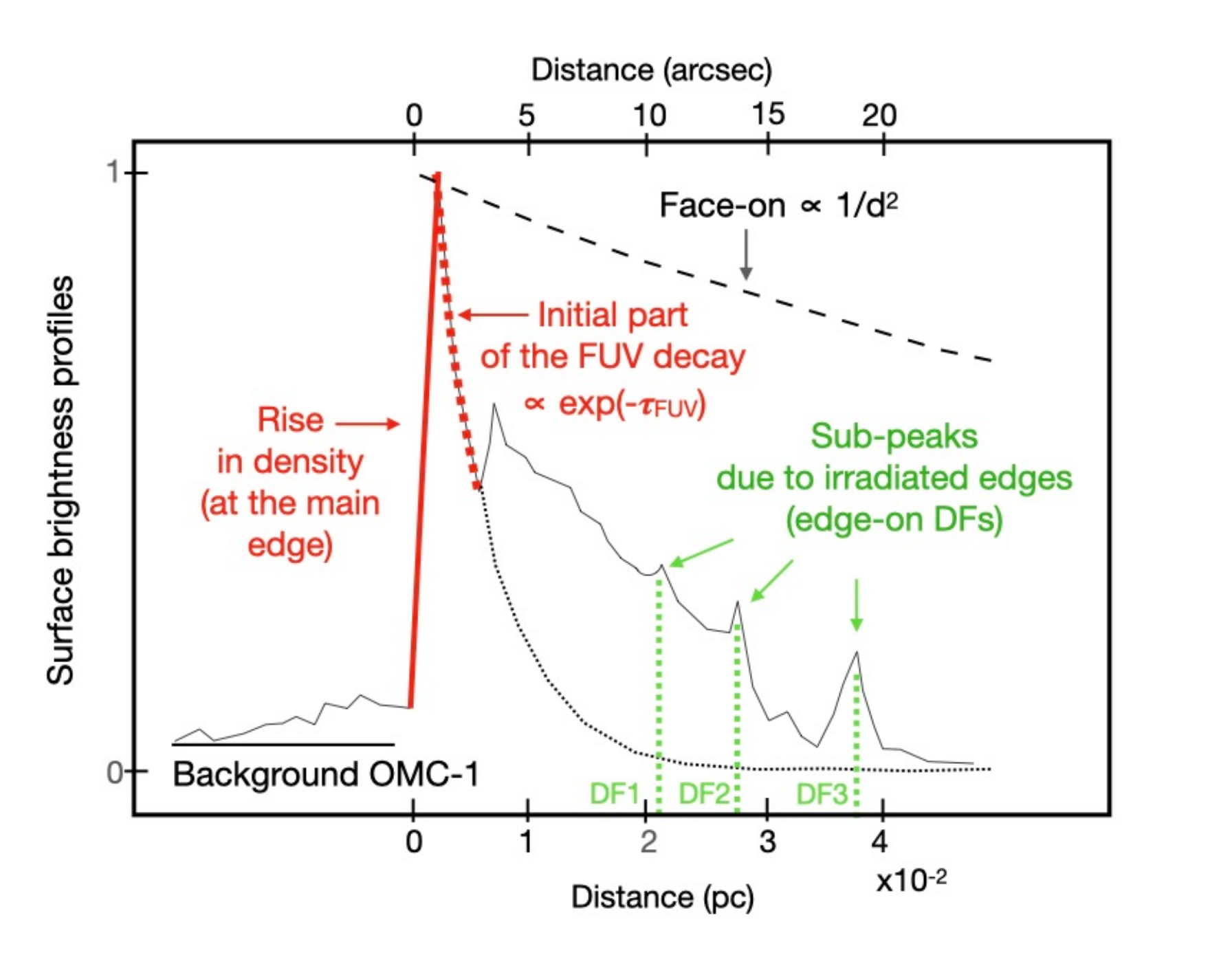}
\caption{Annotation of the different components seen in the AIB surface brightness profile perpendicular to the   Bar. These are shown for cut 2 in filter F333M from Fig.~\ref{fig:cut_Bar_NIRCAM_AIB}, where several other cuts through the Bar are shown.}
\label{fig:sketck-profile}
\end{figure}

\begin{table*}
     \caption{Parameters employed or derived for the   Bar.}
    \label{tab:param_OB}
   \begin{center}
    \begin{tabular}{lll}
        \hline \\[-10pt]
    \hline \\[-10pt]
    Parameter & Value & Reference\\
    \hline \\[-10pt]
    distance & 414$\pm$7~pc~\tablefootmark{a} & \citet{Menten07}\\
     & 1\arcsec = 0.002~pc & \\
    projected distance, $d_{proj}$, between $\theta ^1$ Ori C and the IF  & 0.228~pc& \\
    transverse size $l_{PDR}^{los}$ & $\sim$0.1~pc & \citet{Peeters-spectro}\\
     G$_0$ at IF & $\sim(2.2-7.1) \times 10^4$ & \citet{Peeters-spectro}\\
    FUV dust cross-section $\sigma _H$ & $6.5 \times 10^{-22}$  cm$^{2}$ / H & \citet{Cardelli89,blagrave2007}\\
    &  & \citet{Schirmer2022}\\
    $R_V =A_V /E(B-V)$ & 5.5 & \citet{Cardelli89,blagrave2007}\\
    $A_V /N_H$ & $3.5 \times 10^{-22}$~mag/cm$^{-2}$ & \citet{Cardelli89,blagrave2007} \\
         Density in atomic PDR n$_H$ & (5-10)$\times 10^{4}$  cm$^{-3}$ & This article (Sect.~\ref{sec:spatial-distribution-Bar-AIBs})\\
        Density at the IF n$_e$ & 5$\times 10^{3}$  cm$^{-3}$ & \citet{Weilbacher15}\\
    Temperature at the IF T$_e$ & $\sim 9~10^3$ K & \citet{Weilbacher15}\\
           Distance between IF and DF $d_{IF-DF}$ & 0.02-0.04~pc& This article (Sect. \ref{sec:H2_DF})\\
            Average atomic density between IF and DF  & 4.6 $\times 10^{4}$ cm$^{-3}$ & This article (Sect. \ref{sec:H2_DF})\\
           Density from NIR H$_2$ n$_H$ & (3.5)$\times 10^{4}$ to $\times 10^{5}$ cm$^{-3}$ & \citet{Peeters-spectro}\\
           Temperature at the DF $T$ & $\sim$ 400-700 K & \citet{Allers05,lettergas}\\
\hline 
    \end{tabular}
   \end{center}
\tablefoot{
   \tablefoottext{a}{The adopted distance is based on the \cite{Menten07} study using radio very long baseline interferometry. We underline that more recent observations, including Gaia, point to slightly lower or similar values. \cite{Kounkel17} used the Very Large Baseline Array and measured  a distance of 388$\pm$5~pc toward the Orion Nebula Cluster (ONC). This distance corresponds to the weighted average distance of all the stars located toward the ONC, including the Trapezium. 
\cite{Kounkel2018} used Gaia DR2 and measured an average distance of 386$\pm$3~pc for the ONC. However, other studies based on Gaia DR2 parallaxes such as \cite{Binder2018,Grossschedl2018,Kuhn2019} quote distances for the ONC of 410~pc, 400-410~pc and 403$^{+7}_{-6}$~pc, respectively which are near the estimate from \cite{Menten07}. \cite{Grossschedl2018} suggested that the about 10 pc difference compared to the literature value can be seen as an estimate of remaining systematic uncertainties. In Appendix A of \cite{Kuhn2019}, they compare their distance estimate to one from \cite{Kounkel2018} and discuss effects of the three-dimensional structure of Orion A. 
Moreover, one should note that these distances correspond to the distances of the stars and not the one of the molecular cloud and the Bar. New Gaia releases (in combination with new trigonometric parallax) will provide even more accurate distance determinations. For simplicity, we choose to assume the distance of 414$\pm$7 pc from \cite{Menten07} in order to remain consistent with \cite{pdrs4all}. The difference in distance values has anyway no important implications for the results presented in this article. 
}}
\end{table*}

\subsection{ionized gas, crenellations, bow-shocks and YSOs}

In Fig.~\ref{fig:composite-image-inside-Orion-nebula}, the ionized gas (in blue) comes from the MIF extending from the Trapezium grouping of stars to the Bar. Beyond the Bar, the MIF is primarily photoionized by $\Theta ^2$ Ori A \citep{Odell2017a}.
Due to intense ultraviolet and ionizing radiation, hot and ionized gas is photoevaporating away from the MIF. 
\HI~and \HeI~lines observed toward the Huygens Region at visible-wavelengths are blue-shifted, by $\simeq$10~km~s$^{-1}$, with respect to the molecular gas emission \citep{O'Dell2023}. This velocity difference approximately agrees with that inferred from observations of H and He  radio recombination lines  \cite[e.g.,][]{Goicoechea15,Cuadrado19}. 

In the NIRCam images, ionized gas flows from  the IF of the Bar (see panel G in Figs.~\ref{fig:composite-image-inside-Orion-nebula} and~\ref{fig:zoom-image-inside-Orion-nebula}) are not easy to discern because 
they are seen in the foreground of the MIF emission.
The NIRCam and MIRI images do not show any AIBs and ${\rm H_2}$ emission that would be associated with the photoevaporating flows from the IF of the Bar.
The AIBs and H$_2$ emission in front of the Bar most likely originate from the surface of the OMC-1. This surface is perpendicular to the line of sight and is illuminated by the Trapezium cluster, making it a face-on PDR \footnote{The emission from this background face-on PDR was previously observed with \textit{Herschel} in other PDR tracers, especially in high-J CO, CH$^+$ lines \citep{Parikka2018} and [OI] 63 and 145 $\mu$m and [CII] 158 $\mu$m \citep{Bernard-Salas2012}, as well as, with \textit{Spitzer} in AIBs \citep{knight22}.}.

On the background OMC-1 surface, several structures as shown in panels A and B in Fig.~\ref{fig:zoom-image-inside-Orion-nebula} are spatially resolved.
These types of features are not the only ones and several of them were observed at visual wavelengths with the HST. They were called “Crenellations" by \cite{ODell2015}. 
The interaction of collimated jets and outflows from protostars inside the molecular cloud likely drives the shocks that create these structures at the surface of the cloud \citep[see also ][]{Kavak2022a,Kavak2022b}.
In order to study whether some of these structures are “apparent" structures produced by extinction variations or the edges of dense molecular gas structures, high angular resolution molecular line tracers are needed.

One of the largest outflows with highly blue-shifted features \citep[e.g.,][]{Odell1997,Odell2008} 
is HH 203 and 204, 
localized southwest of $\theta ^2$ Ori A, are well seen in the Pa $\alpha$, Br $\alpha$ and [FeII] line NIRCam images and in the continuum filters at shorter wavelengths (panel F in Figs.~\ref{fig:composite-image-inside-Orion-nebula} and~\ref{fig:zoom-image-inside-Orion-nebula} and Figs. in Appendix \ref{Appendix:Images_all_filters_Orion_Bar}).
HH 203 is a well defined bow-shock with low ionization characteristics at its end. It is driven by a high-velocity jet that emerges into the area ionized by $\theta ^1$ Ori C or the nearest (in the plane of the sky) star $\theta ^2$ Ori A. HH 204 is almost at the same position angle as HH 203 but shows a different structure. Its top is a flocculant structure (that includes low ionization characteristics) and no jet is visible. 
These structures are not detected in H$_2$ line emission (Fig.~\ref{fig:cut_Bar_NIRCAM_H2} shows the H$_2$ emission in the map and the spatial profile towards the cut 5 which indicate that the H$_2$ emission is associated with the Bar and not with the bow shock of HH 203).
This supports the suggestion that these flows interact with the ambient ionized gas of the nebula and not the molecular gas \citep[e.g.,][]{Odell2008}.

NIRCam images in the F470N and F480M filters reveal for the first time strong emission in the surroundings of the bright star $\theta ^2$ Ori A (at about $\sim 8\arcsec$ or 0.015~pc from the star for a distance of 414~pc). These previously unknown features are also visible in F277W, F300M, F323N and F335M filters. A stellar wind from the $\theta ^2$ Ori A star most likely forms these features. 
A bow wave around the $\theta ^2$ Ori A star is probably moving into the MIF. 
The dynamical and radiative impact of $\theta ^2$ Ori A is influencing and complicating its nearby environment. 
\citet{Odell2017a} showed that foreground objects near this location are illuminated by $\theta ^2$ Ori A.

A very bright substructure 
located in the southern part of the Bar shows a very particular structure  $\sim$1000 au in size for a distance of 414~pc (see panel C in Figs.~\ref{fig:composite-image-inside-Orion-nebula} and~\ref{fig:zoom-image-inside-Orion-nebula}). This structure is also visible with ALMA (see Fig.~\ref{fig:cut_Bar_H2_HCO_CO} where it appears as a globule) and might correspond to the surroundings of a protostellar source embedded in the highly irradiated environment of the   Bar (Goicoechea et al. in prep). 
No MIRI point source is detected at this position, probably because the extended emission from the Bar is very bright and dominates the continuum emission. This indicates that this YSO is most likely too cold to emit strongly at mid-infrared wavelengths. 
The structures around this YSO are very bright in the H$_2$ 0-0 S(9) and 1-0 S(1) lines  in the NIRCam and Keck \citep[]{Habart2023} maps (see Fig.~\ref{fig:cut_Bar_H2_HCO_CO}) but also in the AIB emission (see panel C in Fig.~\ref{fig:zoom-image-inside-Orion-nebula}). 
This emission likely arises from a combination of irradiated shocks from the outflow and PDR emission. 
Finally, several bright emission features associated with embedded proplyds are detected in the Bar and in front of it (e.g., panel G in Figs.~\ref{fig:composite-image-inside-Orion-nebula} and~\ref{fig:zoom-image-inside-Orion-nebula}). The proplyds detected within the inner region of the Orion Nebula are discussed in Sect.~\ref{sec:proplyds}.

\subsection{Crenellations on the Bar}
 
One of the most striking features from the NIRCam and MIRI images, and in particular in filters probing AIBs and H$_2$ emission, is that the molecular cloud border in the background and the   Bar appears hyper structured, most likely turbulent \cite[e.g.,][]{goicoechea_compression_2016}. In the   Bar, lots of patterns are apparent such as crenellated structures or ridges (e.g., panels D, E in Figs.~\ref{fig:composite-image-inside-Orion-nebula} and~\ref{fig:zoom-image-inside-Orion-nebula}). The upper west corner of panel E in Fig.~\ref{fig:zoom-image-inside-Orion-nebula} coincides with a region where the Bar has no sharp boundary, a region labeled "SW-Gap" in Figure 24 of \cite{ODell2015}. In this zone, a number of crenellated structures were seen with HST. These features are both detected in the NIRCam filters dominated by the ionized gas and continuum below 3$\mu$m.  
As suggested by \cite{ODell2015}, these structures 
are very likely bow shocks forming in the tilted portion of the Bar or the foreground Veil.
On the other hand, in the regions further inside the Bar (e.g., panel D or panel E from upper east corner to west corner), the structures seen in the NIRCam and MIRI filters probing AIBs and  H$_2$ lines emission likely correspond to the edges of dense molecular gas inside the Bar. Most of these structures are in fact well seen in the submm HCO$^+$ (J=4-3) emission (with ALMA, see Sect.\ref{sec:Bar}) which is sensitive to molecular gas density variations. 

This highlights a very intricate irradiated cloud surface and sub-dense structures at such small scales that they were inaccessible to previous observations. 
The surge of stellar energy sculpts this region exhibiting an incredible richness of substructures.
The F335M filter is essential for bringing out the high-textured structure of the UV irradiated molecular cloud surfaces. The aromatic emission traces a combination of cloud density and strength of the local FUV field (see Sect.~\ref{sec:spatial-distribution-Bar-AIBs}). 
The F335M filter provides one of the highest resolution views of the outer molecular layer of the PDRs available with JWST. 
Intricate fine details of how interstellar matter is structured at small scale is thus revealed. At the cloud edge, AIB emission is more restricted to the atomic H layers of the PDR than the H$_2$ line emission (see Sect.~\ref{sec:Bar}). 
With the high incident FUV radiation field on the Bar, molecules in general are expected to survive longer in the shielded environment offered by the dense Bar or OMC-1.  
However, emission in highly rotationally and ro-vibrationally excited H$_2$ lines require FUV photons to be pumped. These lines are therefore observed at
the photo-dissociation front more specifically at the H$^0$/H$_2$ transition where atomic hydrogen becomes molecular.
Consequently, the  subtracted continuum H$_2$ line images (e.g., the H$_2$ 0-0 S(9) and 1-0 S(1) lines at 4.69 and 2.12~$\mu$m) highlight the irradiated edges of dense molecular structures.  
A detailed comparison between the atomic and molecular phase tracers across the   Bar PDR is given below in Sect.~\ref{sec:Bar}.

\section{Transition from ionization front to H$^0$/H$_2$ dissociation front}
\label{sec:Bar}

\begin{figure*}[h!]
\begin{center}
\vspace*{0cm}
\includegraphics[width=0.95\textwidth]{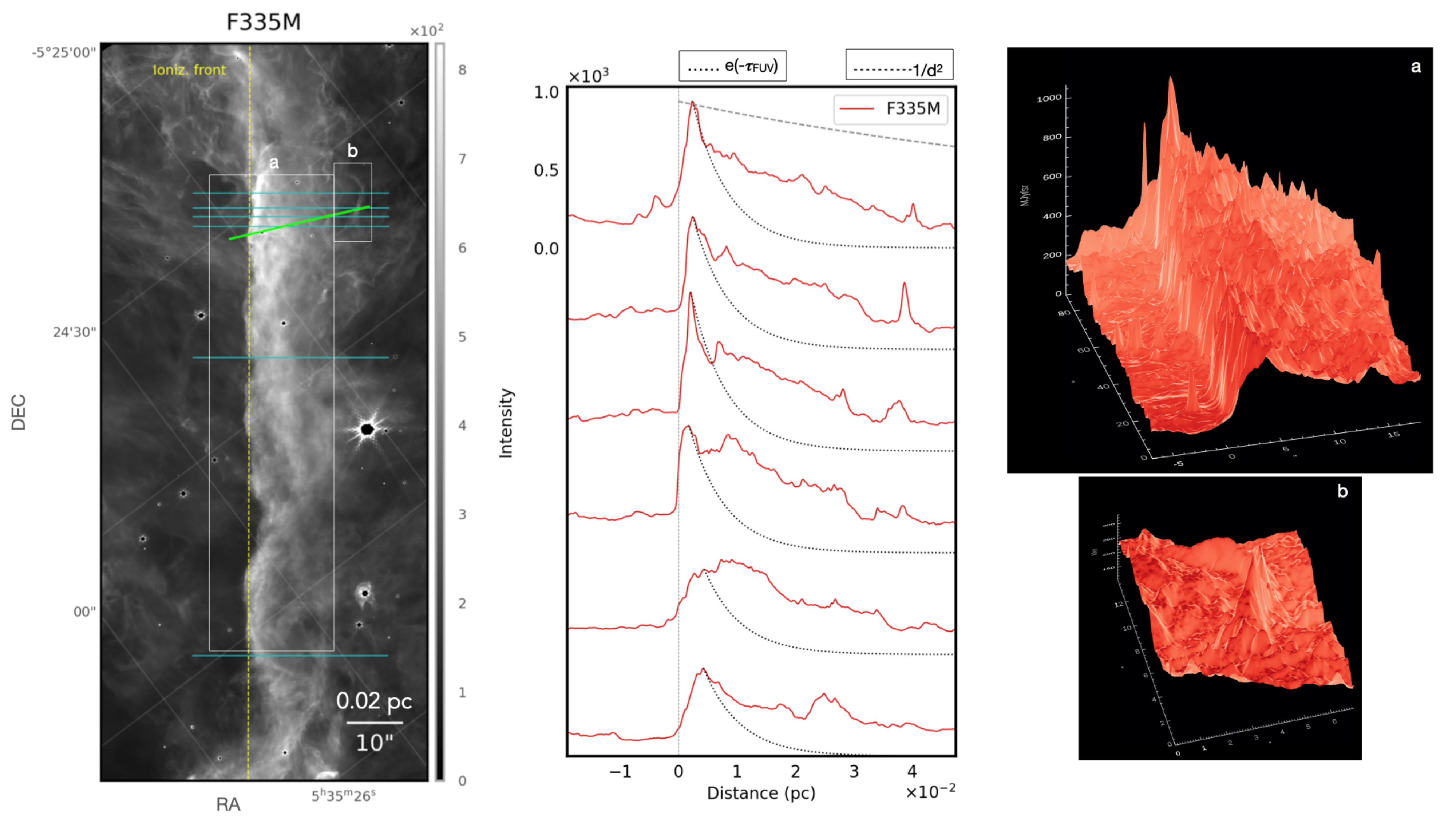}
\caption{Spatial distribution of
the F335M filter tracing mostly the AIBs CH stretch emission in the 3.17-3.54 $\mu$m range.
{\bf Left}: Map in the F335M filter centered on the Bar and rotated 
so that the ionizing radiation strikes the Bar from the left. 
Units are in MJy/sr. 
The vertical dashed line (in yellow) delimits the average position of the IF.
The horizontal lines (in cyan) give the position of 6 cuts, perpendicular to the Bar (the top cut toward the south), displayed in the middle panel.  The inclined line (in green) gives the position of the cut obtained from the NIRSpec field, shown in Fig.~\ref{fig:cut_Bar_NIRSpec}. The boxes (in white) delineate the regions of the 3D surface brightness maps shown on the right panels.
{\bf Middle}: Surface brightness profiles in the F335M filter shown as a function of the distance from the IF. 
The FUV extinction decrease in the Bar (dotted lines) and the expected geometric dilution factor of the incident FUV field intensity from the \mbox{O7-type} star \mbox{$\theta^1$ Ori C} (dashed lines in the upper cut) are both normalized to the peak of the emission in the F335M filter for comparison. The vertical dotted-dashed lines indicate the average position of the IF, also indicated in yellow in the left panel.  
{\bf Right}: 3D maps of the surface brightness profile in two regions defined in the left panel. For {\bf a}, the grid is $24" \times 85.2"$. It shows the sharp rise in brightness along the Bar. For {\bf b}, the grid is $6.6" \times 13.3"$. It shows bright fronts superimposed on the Bar emission. The orientations of the plots are optimized  for the best viewing angle.}
\label{fig:cut_Bar_NIRCAM_AIB}
\end{center}
\end{figure*}

\begin{figure*}[h!]
\begin{center}
\vspace*{0cm}
\includegraphics[width=0.95\textwidth]{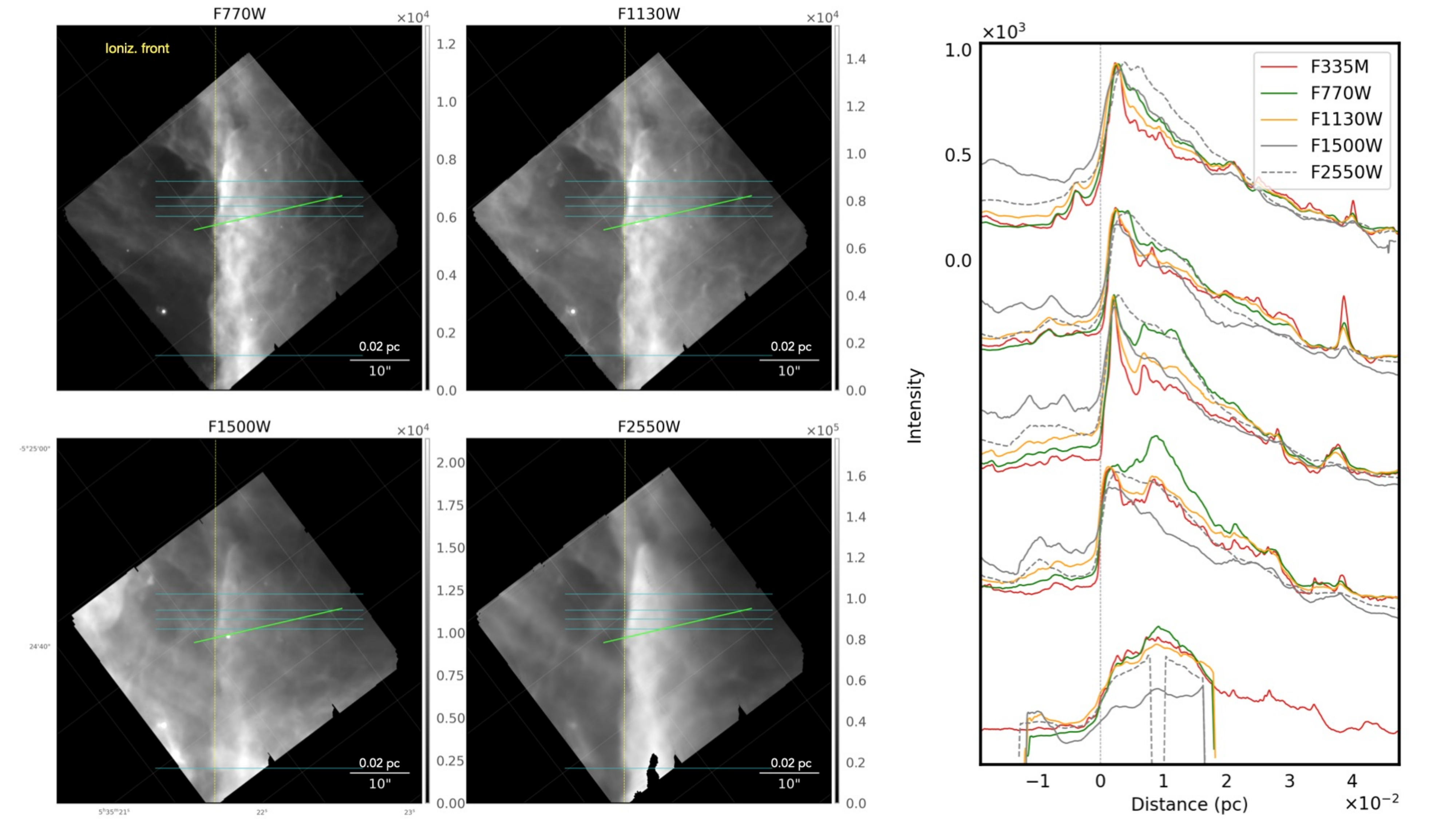}
\caption{Spatial distribution of
the MIRI filters tracing the AIBs at 7.7 and 11.3~$\mu$m and continuum dust emission. {\bf Left}: Maps in the MIRI filters. Same as Fig. \ref{fig:cut_Bar_NIRCAM_AIB} for the vertical and horizontal lines.
{\bf Right}: Surface brightness profiles of the F335M NIRCam filter (in red), and F770W (in green), F1130W (in yellow), F1500W (solid grey) and F2550W (dashed grey) MIRI filters as a function of the distance from the IF. In order to compare the brightness profiles along each cut in the Bar, the mean emission background in front of the Bar (face-on background PDR contribution) has been subtracted to each filter's cut and then scaled to the NIRCam F335M filter maximum amplitude. The background emission level subtracted to the profiles is about 160, 1700, 3200, 4900, 34000 MJy/sr for the F335M, F770W, F1130W, F1500W and F2500W filters, respectively. The vertical dotted-dashed line indicates the average position of the IF, also indicated by yellow line in the left panel.}
\label{fig:cut_Bar_NIRCAM_MIRI_AIB}
\end{center}
\end{figure*}

\begin{figure*}[h!]
\begin{center}
\vspace*{0cm}
\includegraphics[width=0.95\textwidth]{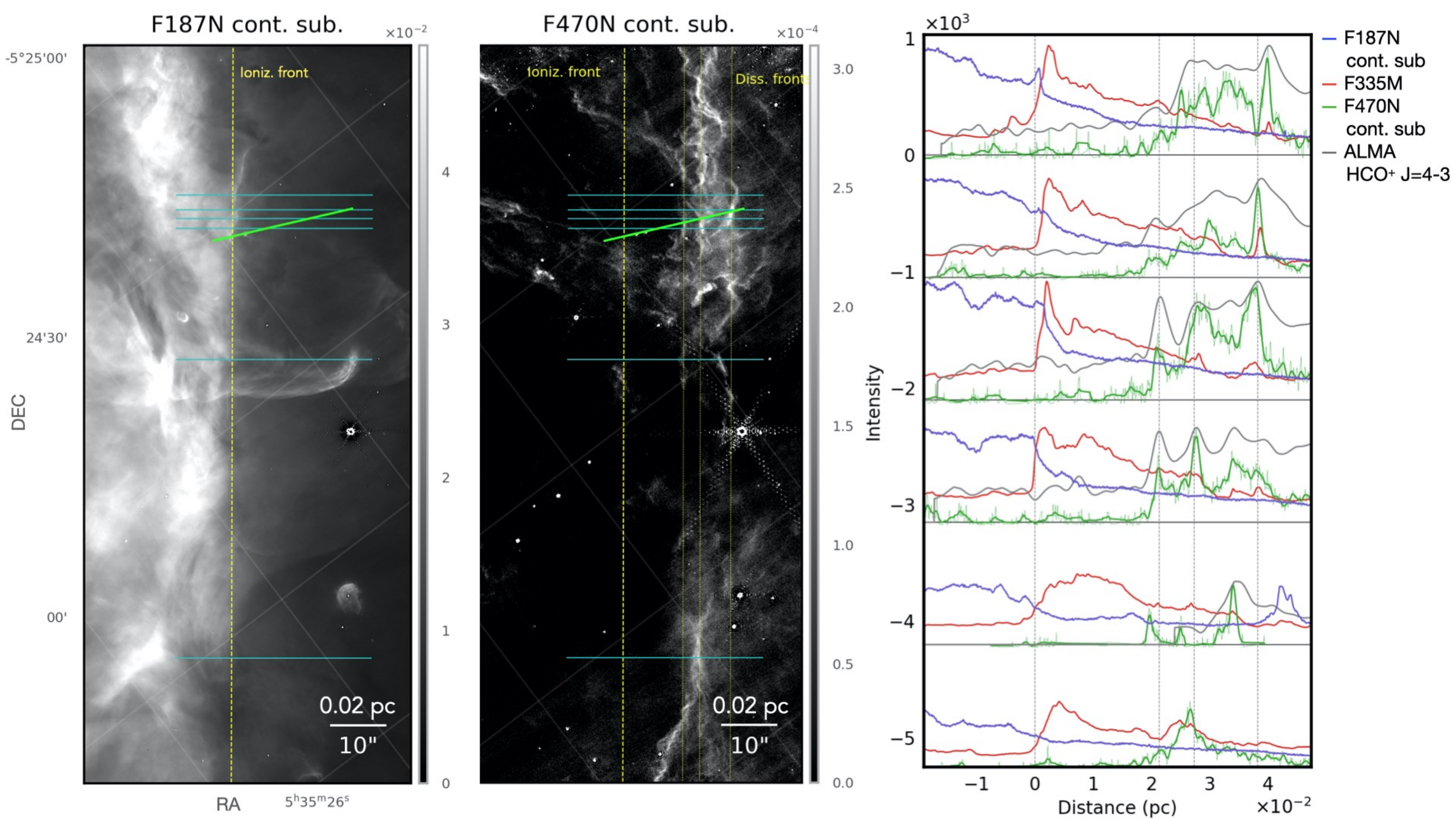}
\caption{Spatial distribution of
the ionized and excited dense molecular gas. {\bf Left}: Maps in the F187N filter (Pa $\alpha$ line) and in the F470N filter continuum subtracted (H$_2$ 0-0 S(9) line). Same as Fig.~\ref{fig:cut_Bar_NIRCAM_AIB} for the vertical and horizontal lines. Vertical dotted lines in the F470N filter continuum subtracted show the DFs position as determined in Sect.~\ref{sec:H2_DF}. Units are in erg cm$^{-2}$ s$^{-1}$ sr$^{-1}$. {\bf Right}: Surface brightness profiles of the Pa $\alpha$ (in blue) and H$_2$ 0-0 S(9) line (in green)  as a function of the distance from the IF.  The profiles of the lines are normalized to the  surface brightness profiles of the F335M filter (in red). The intensities have not been corrected for dust extinction. The ALMA HCO$^+$ J=4-3 line integrated intensity profile is shown in grey. The vertical dotted-dashed lines indicate the average position of the IF and the DFs, also indicated by yellow lines in the left panels.}
\label{fig:cut_Bar_NIRCAM_H2}
\end{center}
\end{figure*}

In this section, we focus on the  Bar, a prototypical highly irradiated dense PDR. 
Figs.~\ref{fig:cut_Bar_NIRCAM_AIB} to \ref{fig:cut_Bar_NIRSpec}
and \ref{fig:cut_Bar_H2_HCO_CO}
show close-up maps and surface brightness profiles of the   Bar viewed edge-on. A gradual structure is evident when moving away from the excitation source as the ionization front, the AIB and H$_2$ emission layers appear successively, in agreement with previous studies \citep[e.g.,][]{Tielens93}.  
However, instead of a smooth PDR transition,
multiple ridges in AIB and H$_2$ emission are spatially resolved for the first time.  In addition
JWST unambiguously reveals very sharp edges (on scales of $\sim$1$\arcsec$ or 0.002 pc) and rich small-scale structures (with typical widths of $\sim 0.5-1\arcsec$ or $\sim$0.001-0.002 pc).
This is in agreement with previous high spatial resolution ALMA HCO$^+$ emission maps \citep{goicoechea_compression_2016} which showed a highly sculpted interface. Along with 
analysis of
high-J CO and CH$^+$ \textit{Herschel} observations and H$_2$ pure rotational lines from ISO, 
\citet{Joblin2018} concluded  that the emission of these tracers arises from a thin (a few 10$^{-3}$\,pc), high thermal pressure (P$\sim$10$^8$\,K\,cm$^{-3}$) layer at the surface of the molecular region. 
However, the corresponding structures were unresolved in most tracers until now.

In the following, we use the notations of the scheme in Fig.~\ref{fig:Geometry} and Table~\ref{tab:param_OB} for the physical quantities related to the   Bar PDR. In particular, the assumed distance to the Bar is 414~pc \citep{Menten07}. This value is slightly higher or similar to more recent estimates using Gaia observations (see annotations in Table~\ref{tab:param_OB}.
We use the notation
$l_{PDR}^{los}$ the length of the PDR along the line of sight toward us and $d_{PDR}$ for the depth in the PDR from the IF.
$N_H^{los}$ and $A_V^{los}$ are the column density and visual extinction along the line of sight toward us. $N_H$ and $A_V$ are the column density and visual extinction in the UV illuminating direction (perpendicular to the Bar).

\subsection{Spatial distribution of the AIB emission}
\label{sec:spatial-distribution-Bar-AIBs}

We first analyze the surface brightness profiles of the filters probing the AIBs and continuum emission along the direction going away from the sources of UV illumination. 
In order to probe variations across the PDR, several cuts perpendicular to the Bar from southwest to northeast are shown in Figs.~\ref{fig:cut_Bar_NIRCAM_AIB}-\ref{fig:cut_Bar_NIRCAM_MIRI_AIB}. 
This allows us to probe the flux variations along the entire illuminated interface. 
 An approximate position of the IF has been marked in maps and profiles as a vertical line, which corresponds to the emission peak of the [FeII] 1.64 $\mu$m line
 and the rise of the AIBs and gas lines in the atomic zone (e.g., [OI] 63 and 145 $\mu$m).
The different components observed in the AIB emission profiles in the different PDR zones are annotated (Fig.~\ref{fig:sketck-profile}) and discussed next.
Our high spatial resolution JWST study permits filling in the near IF portion of the Bar's PDR that the study of \cite{Henney2021} based on Herschel data did not allow (c.f. his Figure 5).

\subsubsection{Steep density rise at the IF}
\label{sec:spatial-distribution-Bar-AIBs-rise-density}

At first glance, the spatial distribution of the AIB filter emission profiles follows the same trend as we move across the Bar, a  steep increase at the IF position,
followed by a slower decrease (with a typical scale of 10$\arcsec$ or $\sim$ 0.02~pc, see Figs.~\ref{fig:sketck-profile}, ~\ref{fig:cut_Bar_NIRCAM_AIB} and \ref{fig:cut_Bar_NIRCAM_MIRI_AIB}).
At first order, the AIBs surface brightness is proportional to the column density of the band carriers and to the local FUV flux strength, $I \propto N_{H}^{los} \times [C/H] \times G_0 \times e^{-\tau _{FUV}}$ with $N_{H}^{los} = n_H \times l_{PDR}^{los}$ and $[C/H]$ their carbon abundance. The column density is along the line of sight to us while the opacity $\tau _{FUV}$ is in the FUV illuminating direction, perpendicular to the Bar. 
The initial brightness increase is expected 
due to a large increase in dust column density at the IF, while the slower decrease is expected due to the extinction of the incident UV irradiation field.
What is noteworthy is that the emission spatial profiles show an extreme climb over few sub-arcsecs just after the IF. The emission peak arises $1\arcsec$ to $2.5\arcsec$ (0.002 to 0.005 pc) from the IF depending on the position in the Bar (see Fig. \ref{fig:cut_Bar_NIRCAM_AIB}). 
Such a sharp density rise is expected due to the sharp decrease in gas temperature at the IF
if the thermal pressures
in the ionized and neutral regions are of similar magnitude, as discussed below.
In addition, the extreme rise of the emission profile could put strong constraints on the tilt angle $\theta$ of the PDR between the plane of the irradiated surface and the line of sight. This is discussed in \citep{Peeters-spectro} and will be investigated in more detail in a future work.

\subsubsection{FUV dust extinction and density in the atomic PDR} 
\label{sec:spatial-distribution-Bar-AIBs-FUV-extinction}

Just after the rise, one can clearly observe in the AIB profiles 
a rapid decrease which must result from the FUV radiation field decreasing with depth inside the PDR as a consequence of the dust extinction.
Fluxes in filters centered on AIBs are most likely dominated by the emission produced by (sub)nanometric particles including large molecules (PAHs) stochastically heated, which is proportional to the FUV radiation field strength.
To visualize the decay of the FUV flux, the curve $\exp(-\tau_{FUV}$)=$\exp(-\sigma_H N_H$) with $\sigma_H$ the dust 
FUV extinction cross-section per proton, and $N_H$ the column density from AIB emission peak
(along the UV illuminating direction, perpendicular to the Bar), is plotted on Fig. \ref{fig:cut_Bar_NIRCAM_AIB} (dotted line). 
The parameters employed to reproduce the initial part of the observed decay are given in Table \ref{tab:param_OB}.
We assume $\sigma _H$, $R_V$ and $A_V/N_H$ in agreement with the extinction curve measured in the Orion Nebula by \citet{Cardelli89} and as refined by \citet{blagrave2007}. 
These values differ from that of the average measured in the ISM ($R_V$ = 3.1 and $A_V/N_H =5.3 \times 10^{-22}$ mag cm$^{-2}$) 
and lead to an increased penetration of FUV photons compared to the dust found in the diffuse ISM\footnote{This lower FUV extinction is in agreement with the findings of \citet{Schirmer2022} who fit \textit{Spitzer} and \textit{Herschel} observations with the THEMIS dust evolution model \citep{jones2017}. Despite the poor resolution, a radiative transfer model of the Bar including the evolution of dust grains (size, abundance, properties) allowed them to highlight a strong depletion of the subnanometric grain population and a size distribution shifted towards larger particles compared to the diffuse ISM. A more recent study with JWST data confirms this result (El Yajouri et al. in prep.).}. 
Then, in order to compute the column density $N_H$, we assumed after the initial rise in density (i.e., after the emission peak) a constant density $n_{H}$ in the atomic PDR as in \citet{Arab12} and \citet{Schirmer2022}. $N_H$ is then given by $n_H\times d_{PDR} $ with $d_{PDR}$ the distance from the emission peak. 
The density $n_H$ is adjusted to reproduce the initial part of the observed decay ($d_{PDR} \sim$0.002-0.01 pc). We derive n$_H$=(5-10)$\times 10^4$ cm$^{-3}$ (see Fig.~\ref{fig:cut_Bar_NIRCAM_AIB}). 
The density value range we derived in the atomic PDR is in agreement with the location of the H$^0$/H$_2$ transition obtained with NIRCam  (see Table \ref{tab:param_OB}) as discussed in Sect. \ref{sec:H2_density} as well as estimates from atomic gas FIR lines (e.g., [CII] and [OI] fine-structure lines, \citealt{Bernard-Salas2012}). It also agrees with
estimates from the Raman-scattered wings
of hydrogen H$\alpha$ lines in the Bar
\citep{Henney2021}.

The density we derived in the atomic PDR is significantly higher (factor 10-20) than the electron density $n_e$ derived at the IF which is about $\sim 5 \times 10^3$~cm$^{-3}$ \citep[see Figs. 26-27 in][]{Weilbacher15}. 
In a D-critical IF\footnote{In the usual classification of IF types (first established by \citealt{Kahn1954,Axford1961}, and summarized in usual textbooks, e.g.,  \citealt{spitzer_physical_2004,draine2011}), a D-critical front travels at subsonic speed with respect to the neutral gas, while the ionized gas is expelled at the speed of sound with respect to the front.} (that can be expected in a blister HII region), the pressure in the neutral gas at the close-back of the IF is expected to be a factor of 2 higher than in the ionized gas in the close-front of the IF. 
A strong rise in density is then expected to compensate for the much stronger temperature decrease between the ionized and neutral region.
In that case, a pressure of about $\sim 2 \times 10^8$ K~cm$^{-3}$ is expected in the neutral atomic region\footnote{The electron temperature $T_e$ derived at the IF is of the order of $\sim 9 \times 10^3$~K \citep[see Figs. 24-25 in][]{Weilbacher15} which gives a thermal pressure of $P_{th/k}=2 \times n_e \times T_e \sim 10^8$ K~cm$^{-3}$ in the neutral gas. }, and given our density estimate, this would mean a temperature of 
a few $10^3$ K.

\subsubsection{Extended emission and secondary peaks towards the molecular region}

Another important characteristic of the observed emission profiles is
that when entering the PDR, 
the emission decays to a non zero plateau that extends into the molecular region (Figs. \ref{fig:sketck-profile} and \ref{fig:cut_Bar_NIRCAM_AIB}). 
This extended emission is most likely due to irradiated atomic material along the line of sight located in front of the molecular region, in the foreground face-on PDR surface layer (as seen in geometry on Fig. \ref{fig:Geometry}). This emission may originate in the flattened region that is still illuminated directly by the ionizing stars. The MIF turns up at the Bar and then continues a flatter rise further away (as illustrated in the lower part of Figure 13 in \citealt{Odell2010}). A detailed \textit{Spitzer} study revealing what is behind the Bar and supporting this geometry is published by \cite{Rubin2011}.
For AIBs and dust continuum emission, the further inside the Bar, the greater the face-on PDR contribution
with the FUV radiation in the Bar being rapidly attenuated. 

Furthermore, in the decaying part of the profile,  several secondary peaks are visible (Fig.~\ref{fig:sketck-profile} and \ref{fig:cut_Bar_NIRCAM_AIB}). This might be associated with multiple irradiated ridges with varying densities. These ridges are located after the main edge in the FUV illuminating direction. 
The sub-peaks in the region where the hydrogen is mostly molecular ($d_{PDR}>0.02$~pc) spatially and individually coincide with strong H$_2$ line emission peaks (Fig. \ref{fig:cut_Bar_NIRCAM_H2}), hinting that these AIB emission peaks arise from the material at or close to the DF. 
Some AIBs sub-peaks are very pronounced (such as the one at $d_{PDR}=0.04$~pc, Fig. \ref{fig:cut_Bar_NIRCAM_AIB}, panel b). 
For those, the AIB emission sub-peak is observed slighlty shifted (by $\sim 0.2\arcsec$) from the H$_2$ emission peak (i.e. H$_2$ is observed closer to the Trapezium). 
In order to investigate this shift in
detail, radiative transfer calculations are required.

\subsection{Excited dense molecular gas}
\label{sec:spatial-distribution-Bar-molecular-gas}

In this section, we analyze the distribution of the excited dense molecular gas traced by H$_2$ emission in order to probe the gas physical structure and the location of the key chemical transitions occurring in the molecular PDR. With JWST, we are for the first time able to spatially resolve the emission profiles of both the 
high rotationally and vibrationally excited H$_2$ lines (see Figs. \ref{fig:cut_Bar_NIRCAM_H2}, \ref{fig:cut_Bar_H2_HCO_CO} and \ref{fig:cut_Bar_NIRSpec}). Our NIRCam observations show a very good agreement with Keck/NIRC2 observations \citep{Habart2023} in terms of vibrationally excited line distribution and intensity. 
In the following, we examine the highly structured H$^0$/H$_2$ dissociation fronts towards the Bar as well as the remarkably similar spatial distribution between the highly rotationally (0-0 S(9)) and vibrationally (1-0 S(1)) excited H$_2$ lines and the HCO$^+$ line J=4-3 emission observed with ALMA.
The spatial distribution of the emission lines as a function of the geometry
of the DF surface layer, density variation with depth into the PDR and
extinction along the line of sight is now discussed in detail.

\subsubsection{Highly structured H$^0$/H$_2$ dissociation fronts}
\label{sec:H2_DF}

\begin{figure*}[h!]
\begin{center}
\vspace*{0cm}
\includegraphics[width=0.95\textwidth]{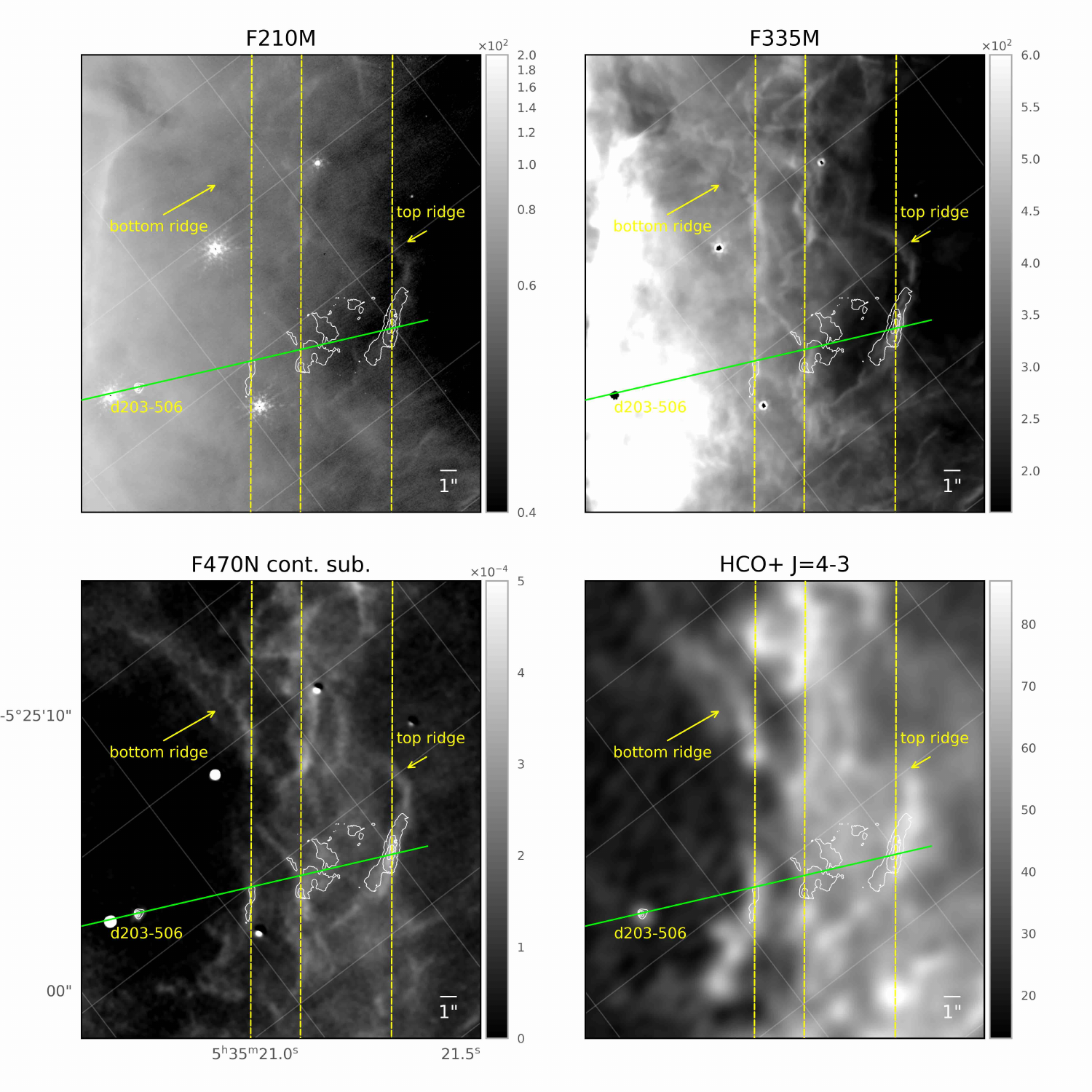}
\caption{Zoom into the   Bar covered in part by the NIRSpec observations. From upper left to bottom right: NIRCam maps in the F210M filter (tracing dust scattered light), F335M (emission from the 3.3-3.4$\mu$m aromatic and aliphatic CH stretching modes), continuum 
subtracted F470N (tracing H$_2$ 0-0 S(9)) and the ALMA HCO$^+$ J=4-3 map. Contours of the NIRSpec map in the 0-0 S(9) line with levels equal to 1.5, 2.3, 3.1, 3.9, 4.6 10$^{-4}$ erg ${\rm s^{-1}}$ ${\rm cm^{-2}}$ ${\rm sr^{-1}}$ are shown in white. The vertical dotted yellow lines indicate the average position of the main DFs. The inclined line in green gives the position of the cut in the NIRSpec field shown in Fig.~\ref{fig:cut_Bar_NIRSpec}.}
\label{fig:map_Bar_ZOOM_NIRSPec}
\end{center}
\end{figure*}

\begin{figure*}[h!]
\begin{center}
\vspace*{0cm}
\includegraphics[width=0.95\textwidth]{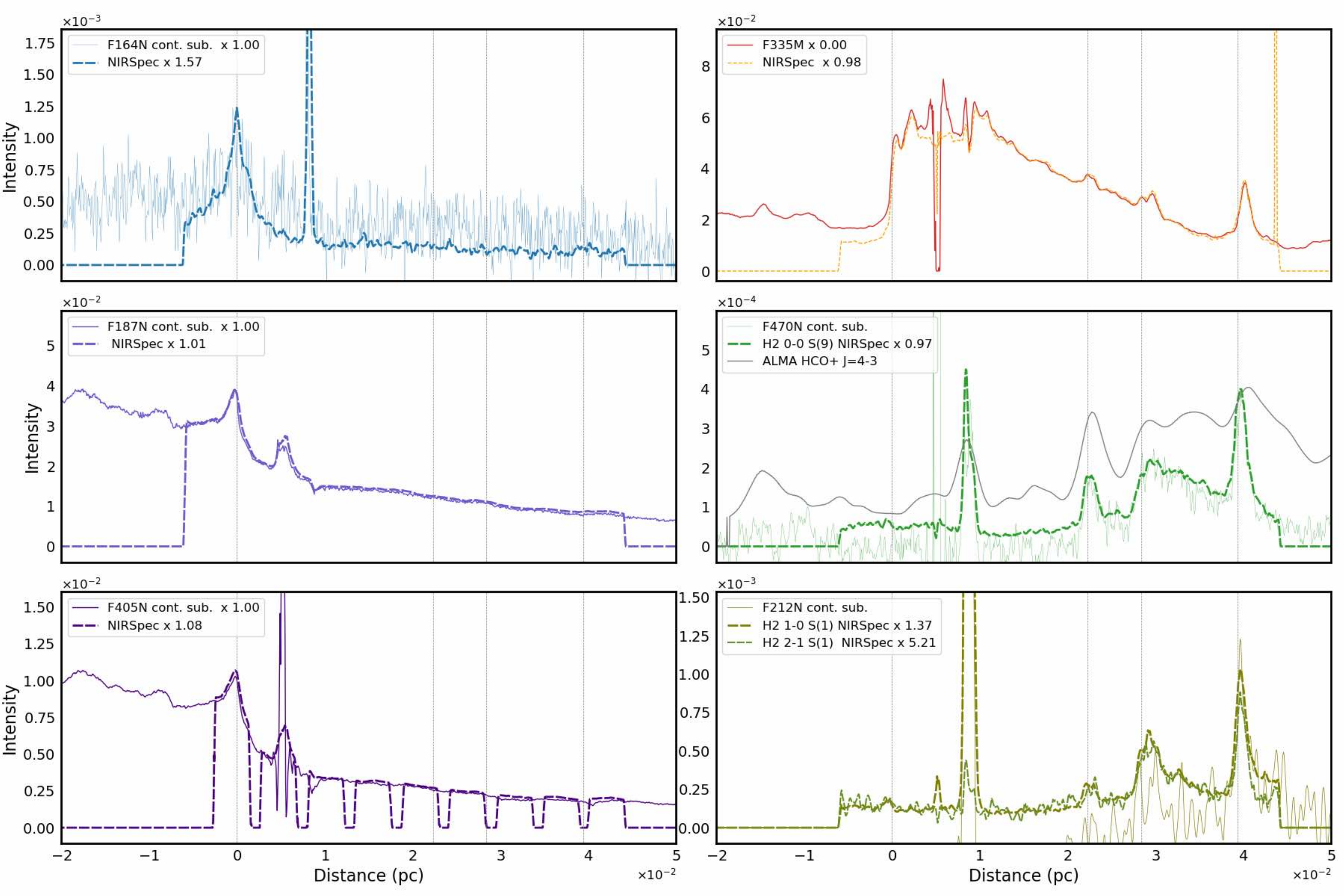}
\caption{Intensity profiles shown as a function of the distance from IF along the NIRSpec cut (Fig.~\ref{fig:map_Bar_ZOOM_NIRSPec}.) comparing NIRCam imaging data in selected filters (continuum substracted) and NIRSpec line intensities.   Units are in erg cm$^{-2}$ s$^{-1}$ sr$^{-1}$.
The  emission line intensities are continuum 
subtracted. The vertical dotted lines indicate the average position of the IF and the main DFs. In comparison to the cuts perpendicular to the Bar, it can be noted that the positions of the DFs are slightly shifted by approximately 0.0016 pc due to the inclination of the  NIRSpec cut. The intensity profiles of the lines observed with NIRSpec have been scaled by the factors indicated in the legends. The intensities have not been corrected for dust extinction.}
\label{fig:cut_Bar_NIRSpec}
\end{center}
\end{figure*}

In Figs. \ref{fig:cut_Bar_NIRCAM_H2}, \ref{fig:map_Bar_ZOOM_NIRSPec} and \ref{fig:cut_Bar_NIRSpec}   the NIR H$_2$ line emission (delineating the H$^0$/H$_2$ transition) show several bright ridges which are spatially resolved and small scale structures. 
The H$^0$/H$_2$ fronts appear highly structured with several ridges and the emission rise is extremely sharp with a width of $0.5$ to $1\arcsec$ (0.001-0.002~pc or 200-400~AU). The ridges run parallel to the Bar but a succession of bright substructures is also observed from the edge towards the molecular region. This is particularly clear in the southwest part of the Bar which corresponds to the upper part of the map displayed in Fig. \ref{fig:cut_Bar_NIRCAM_H2}. In this area, the structure of the Bar is very complex and irregular. 
The H$_2$ emission ridges appear in an area that starts at about $10\arcsec$ from the IF ($d_{PDR}$=0.02 pc) and up to $20\arcsec$ ($d_{PDR}$=0.04 pc) as shown in Fig. \ref{fig:cut_Bar_NIRCAM_H2}. 
We interpret the three main ridges that appear as three edge-on portions of the DF surface which are successively more and more distant from the IF.
These edge-on portions of the DF, denoted DF1, 2, 3 thereafter,  are located at a projected distance from the IF of $d_{PDR}\sim$0.02 pc, 0.027 pc, and 0.038 pc respectively, as indicated by vertical dashed lines in Figs. \ref{fig:cut_Bar_NIRCAM_H2} and \ref{fig:map_Bar_ZOOM_NIRSPec}. The DF2 is the one which coincides best with the average position of H$_2$ emission ridges all along the Bar.

{\bf A terraced-field-like structure.} 
A terraced-field-like structure with several steps seen from above as shown in Fig. \ref{fig:Geometry} can explain the succession of H$_2$ ridges across the Bar. In that geometry, each H$_2$ emission ridge corresponds to a portion of the DF seen edge-on, i.e., a step.  
Since highly rotationally and ro-vibrationally excited H$_2$ emission profiles are sensitive to the gas density, the very narrow and bright ridges must be due to irradiated dense material. For low density interface, the H$_2$ emission is spatially more extended and weaker. 
In the isobaric hypothesis, as discussed previously, the gas density rises as the gas cools at the DF \citep[e.g., ][]{Allers05,Joblin2018}.

Additional evidence for the terraced-field-like structure comes from the difference in visual extinction $A_{V\text{Bar}}^{\text{los}}$ along the line of sight \citep[see Fig. \ref{fig:Geometry} and Sect. \ref{sec:H2_extinction};][]{Peeters-spectro}, as well as the comparison between NIR and millimeter data (Sect. \ref{sec:ALMA}), showing that  $A_{V\text{Bar}}^{\text{los}}$ is higher for the DF1 than for the DF2 and the DF3 dissociation fronts. 
Furthermore, DF1 remains visible in the NIRCam filter F335M but is no longer discernible in the F210M filter (see Fig. \ref{fig:map_Bar_ZOOM_NIRSPec}). This last filter is most likely due to the dust scattered light. The fact of not seeing DF1 in F210M confirms that there is more material along the line of sight at the DF1 position. 

An additional morphological point to highlight is the contrast between a relatively smooth and unstructured IF (see the Pa $\alpha$ line map in Fig. \ref{fig:cut_Bar_NIRCAM_H2}) and a complex, structured,  folded DF surface as traced by the H$_2$ 0-0 S(9) line in Fig. \ref{fig:cut_Bar_NIRCAM_H2}). Moreover, the southwest part of the Bar which corresponds to the upper part of the map displayed in Fig. \ref{fig:cut_Bar_NIRCAM_H2} is much more structured than the other regions of the Bar. In the northeast, a single main DF is observed. This could be related to previous ground-based observations of the molecular condensations deeper inside the PDR which showed that the northeast part of the Bar has a main condensation while the southwest part is fragmented into several components \citep[e.g.,][]{Lis03,Lee2013}.

{\bf Physical origin of the terraced-field-like structure.} 
These structures may result from pre-existing high density structures shaped by the high FUV field inducing a compression. 
The density contrasts increase due to compression. 
Another potential explanation is that the region exposed to stellar winds and protostellar outflows make it especially turbulent.
On larger spatial scales, SOFIA observations of C$^+$ reveal that stellar winds and protostellar outflows shape the molecular cloud and also inject mechanical energy \citep{Pabst20}.
Regularly spaced ridges that run parallel to the photo-dissociation front could also suggest that large-scale magnetic fields are dynamically-important 
\citep{Mackey2011} and raise the question if they could be associated with magnetic-driven density peaks.
If the gas thermal pressure is very large then one needs strong magnetic fields for such an hypothesis to be dynamically relevant \citep[e.g., for $P_{th}{\sim} 2~10^8$~K~cm$^{-3}$, one needs 800 $\mu$G;][]{Pellegrini09,goicoechea_compression_2016}. SOFIA HAWC+ observations of the dust polarization toward the Bar reveal a magnetic field strength of $\sim 300 ~\mu$G \citep{Chuss19,Guerra21}. However, high angular observations of the dust polarization are needed to confirm its relevance.
\citet{berne2010waves} observed similar ridge-like structures, coined the "Ripples," in the western part of the Orion Nebula, also at the interface between the molecular cloud and the HII region. They interpreted the formation of these structures as the result of the Kelvin-Helmholtz (KH) instability occurring due to gas shearing at the interface. Interestingly, the spatial extension of these structures is about 10 times larger ($\sim 0.1 pc$) than that of the ridges observed in the Bar ($\sim 0.01$ pc, Fig.~\ref{fig:sketck-profile}). This can be explained by the larger density of the molecular gas in the   Bar (close to $10^5$ cm$^{-3}$ at the DFs, Table~\ref{tab:param_OB})  as compared to the density in the Ripples (closer to $10^4$ cm$^{-3}$). The fact that the ridges in the Bar appear less well-aligned than in the Ripples also suggests that the gas is in a more turbulent phase, which could correspond to the decay of the KH instability \citep{berne2012kelvin}.

\subsubsection{Extinction attenuation of the H$_2$ NIR lines along the line of sight}
\label{sec:H2_extinction}

The intensity variations in the different H$_2$ 1-0 S(1) at 2.12~$\mu$m emission peaks ranging from $\sim 2$ to $\sim 10 \times 10^{-4}$ erg s$^{-1}$ cm$^{-2}$ sr$^{-1}$ (Fig.~\ref{fig:cut_Bar_NIRSpec}) may result from a combination of effects due to the local gas densities, geometry (length of the edge-on portion along the line of sight) and dust extinction  along the line of sight. 
For the H$_2$ emission in the PDR, extinction along the line of sight due to the dust in the Bar itself between the ionized gas and the region of 
excited H$_2$ may significantly attenuate the NIR emission. 
The extinction might be variable depending on the sightline and the density of the region crossed. From the radio and NIR H$_2$ line maps, \citet{walmsley2000structure} suggests, in fact, that extinction can vary rapidly as a function of position in the Bar. 
Precise spatial estimates of the internal PDR extinction are thus required. This is possible with the high angular resolution near-IR line maps we obtained with \textit{JWST} 
that constraint in detail how dust extinction affect the apparent morphology of the NIR H$_2$ line emission and how the matter is distributed along the line of sight.

{\bf Extinctions towards the edge-on DFs.} 
An effective way to measure extinction is comparing the
observed-to-theoretical H$_2$ ro-vibrational line flux ratios from
pairs of lines arising from the same upper level that are 
separated in wavelength. The dust absorption cross-section rapidly drops with increasing wavelength.
Although NIRCam maps can give an overall view of the extinction across the entire front, we cannot use the maps in the F212N filter (centered on H$_2$ 1-0 S(1) line at 2.12~$\mu$m) and the F323N filter (centered on 1-0 O(5) line at 3.23~$\mu$m with same upper level) since this last filter is dominated by aromatic band emission (see Sect. \ref{sec:line_AIB_continuum_Filter}). NIRSpec observations from \citet{Peeters-spectro} where line intensity can be measured without being contaminated by bands were used.
The extinction map and profile derived along the line of sight, $A_{V\text{Bar}}^{\text{los}}$, in the NIRSpec field is shown in \citet{Peeters-spectro}.   
$A_{V\text{Bar}}^{\text{los}}$ is found equal to 10-12 on DF1 and decreases to 5-3 on DF2 and DF3. This shows that  DF1 is farther along the line of sight and  is in agreement with 
 the stepwise structure (see Fig. \ref{fig:Geometry}) with the column density along the line of sight increasing for  
 the first DFs which are more distant from the observer (but closer in projected distance from the IF).

{\bf NIR H$_2$ line intensity variations due to extinction.} Due to extinction effects along the line of sight, the H$_2$ 1-0 S(1) line at 2.12~$\mu$m is significantly attenuated compared to the 0-0 S(9) line at 4.69~$\mu$m at the DF1 (by about 50\%, see the
line profiles shown in Fig. \ref{fig:cut_Bar_NIRSpec} towards the NIRSpec cut). 
Comparing the NIRCam H$_2$ 0-0 S(9) 4.69~$\mu$m and Keck 1-0 S(1) 2.12~$\mu$m line maps, one can see that the latter is  systematically weaker all along the DF1 on the entire southern part of the   Bar (Fig. \ref{fig:cut_Bar_H2_HCO_CO}). 
However, the extinction alone cannot account for the intensity variations observed between the DFs.
The H$_2$ at 4.69~$\mu$m, which is little attenuated by the dust extinction  effects (decrease less than 10-20\% for $A_V=10$), is in fact twice as high at DF3 than at DF1.
This intensity variation must result from geometrical or density effects. 

\subsubsection{Highly rotationally and ro-vibrationally excited H$_2$ lines profiles}
\label{sec:H2_density}

Here, we compare the surface brightness profiles of several rotationally and ro-vibrationally excited H$_2$ lines,  [FeII]~1.664~$\mu$m, Pa and Br $\alpha$ lines and AIBs  measured with NIRCam and NIRSpec (see Figure \ref{fig:cut_Bar_NIRSpec}). A very good agreement between NIRCam and NIRSpec is found in terms of line distribution and intensity. 

{\bf The spatial emission profiles of H$_2$ lines 
agree in remarkable detail.} 
The H$_2$ 1-0 S(1) (v=1, J=3, $E_u$=6951~K), 0-0 S(9) (v=0, J=11, $E_u$=10261~K), 2-1 S(1) (v=2, J=3, $E_u$=12550~K) line emission show the same  spatial behavior with a strong increase at the edge-on DF1, DF2 and DF3. 
The emission peaks of the different H$_2$ lines at the edge-on DFs spatially coincide at the spatial resolution of our observations (Figure \ref{fig:cut_Bar_NIRSpec}). 
The H$_2$ line profiles follow each other very well for $d_{PDR}>$0.01 pc, with small line ratio variations. 
Along the NIRSpec cut, the most significant line ratio variation with a strong excess of the 1-0 S(1) line is observed on the irradiated disk d203-506. This is due to a density increase \citep{Berne:proplyd}. 
For dense highly irradiated conditions, collisional population of the v=1 J=3 level becomes competitive, and the 1-0 S(1) / 2-1 S(1) lines ratio is thus expected to increase from a pure radiative cascade value ($\sim$2) to a collisional excitation value (of the order of 10). In the Bar, the 1-0 S(1) / 2-1 S(1) line ratio is on the order of ${\sim}5$. This ratio varies little between the different edge-on DFs. Gas density of the H$_2$ emission zone must remain  comparable along the folded DFs surface.

{\bf Background H$_2$ emission toward the ionized and atomic region.} Along the profiles, the H$_2$ emission seen in projection in front of the Bar and in the atomic region mostly comes from the surface of the OMC-1 in the background, not from the Bar itself. This is demonstrated by several points as explained below. First, NIRCam and NIRSpec emission line profiles are flat and at the same level of intensity in the ionized and atomic regions.
Second, the NIRSpec H$_2$ excitation diagrams are very similar in the ionized and atomic region \citep{Peeters-spectro}.
The emission in the atomic region itself is predicted by PDR models to be very weak. For example, for an isobaric model (with the Meudon PDR code, \citealt{le_petit_model_2006})  with $P=5 \times 10^7 - 10^8$~K~cm$^{-3}$ (corresponding to $n\sim$5~10$^4$\,cm$^{-3}$ in the atomic region and $n\sim 10^5$ cm$^{-3}$ in the zone where the  H$_2$ abundance increases sharply and high-$J$ and $v$ H$_2$ lines emit), the predicted H$_2$ 0-0 S(9) line emissivity is on average 100-50 times lower in the atomic region than at the H$_2$ peak.

{\bf Density in the H$_2$ emission zone compared with that estimated in the atomic region.}  By fitting the intensities of a hundred H$_2$ lines measured with NIRSpec 
to the grid of Meudon PDR models
\citet{Peeters-spectro} found that densities are about 
$n_H=(3.5-10) \times 10^4$ cm$^{-3}$ 
 in the H$_2$ emission zone in the Bar, and similar towards the face-on background OMC-1 PDR. 
This density is of similar magnitude with density estimates from AIB emission profiles in Sect. \ref{sec:spatial-distribution-Bar-AIBs-FUV-extinction}.
Density must be roughly constant from the PDR edge (where AIB emission peak) to the beginning of the ro-vibrationally excited H$_2$ emission layer  
where the density and H$_2$ abundance starts to increase sharply. This is consistent with the average atomic PDR density derived from the observed location of the H$^0$/H$_2$ transition.  The H$^0$/H$_2$ is predicted by PDR models to be displaced inward from the IF by $A_V \sim 1$ for the   Bar physical conditions (i.e., high $G_0/n_H$ regime). Using $A_V /N_H = 3.5 \times 10^{-22}$~mag/cm$^{-2}$, this translates into an average atomic PDR density of  1 mag /($3.5\times 10^{-22}$~mag/cm$^{-2}$)/($d_{IF-DF}$)=$4.6\times 10^4 \times$ (0.02~pc/$d_{IF-DF}$)  cm$^{-3}$ with $d_{IF-DF}$ the distance between IF and DF. By taking the values of $d_{IF-DF}$ given in Sect. \ref{sec:H2_DF}, we obtain densities 
comparable to those derived at the PDR edge from AIB emission.

\subsubsection{Spatial distribution between the H$_2$ and HCO$^+$ J=4-3 emission}
\label{sec:ALMA}
 
{\bf Common substructures in the H$_2$ and HCO$^+$ J=4-3 lines.} Figs.~\ref{fig:map_Bar_ZOOM_NIRSPec} and ~\ref{fig:cut_Bar_H2_HCO_CO} compare the H$_2$ maps to HCO$^+$ J=4–3 line map across the same field of view. Most of the substructures are common to both maps and show a very similar distribution.
The overall  spatial coincidence between the H$_2$ and HCO$^+$ line emission shows that they both come from the edge of dense structures and that they are chemically linked.
Because of its high dipole moment, and thus critical density for collisional excitation (a few 10$^6$\,cm$^{-3}$ for optically thin emission), the HCO$^+$ J=4–3 rotational line is a good indicator of dense molecular gas. 
Thus, some of the densest portions of the Bar lie very near the DFs.
Detection of both bright HCO$^+$ and CO emission by ALMA towards the H$_2$ vibrational emission layers suggests that the C$^+$/CO transition nearly coincides with the H$^0$/H$_2$ transition \cite[in agreement with][]{goicoechea_compression_2016}.
In dense PDRs, the reaction between vibrationally excited H$_2$ molecules and C$^+$ ions becomes exothermic
and leads to the formation of CH$^+$. Fast exothermic reactions with H$_2$ subsequently lead to the formation of  CH$_{3}^{+}$. This key hydrocarbon ion  reacts with abundant oxygen atoms
and enhances the HCO$^+$ abundance in the H$_2$ emitting PDR layers 
\citep{goicoechea_compression_2016}.
We estimated that the average offset between H$_2$ and HCO$^+$ is less than $1\arcsec$, about $\sim 0.6\arcsec$ (or 0.0012~pc). This is close to the distance between the H$^0$/H$_2$ and C$^+$/C/CO transition as predicted by high-pressure isobaric stationary PDR models \citep{Joblin2018}.

{\bf Bright emission from the surface of the molecular condensations.} The small H$_2$ and HCO$^+$ J=4-3 structures localized at the DF are in general shifted by a about $\sim$10-20$\arcsec$ relative to the center of the bigger ($5\arcsec-10\arcsec$) molecular condensations seen more inside the molecular cloud \citep{YoungOwl00,Lis03}.
However, some bright H$_2$ fronts detected in the northwest end of the Bar (sixth cut in Fig. \ref{fig:cut_Bar_NIRCAM_H2}) and between the center and the southwest end of the Bar (zone C in Figs. \ref{fig:composite-image-inside-Orion-nebula} and \ref{fig:zoom-image-inside-Orion-nebula}) correspond to the irradiated superficial layers of the bright cold cores detected in CS J=2-1 \citep[cores denoted 3 in the north and 1 in the center-south respectively in][]{Lee2013}. These starless cores are fragmented into 3-5 components, and their fragments are embedded in larger filamentary structures. Some of the clumps are likely collapsing to form a low-mass star \cite[][]{Lis03}.
JWST observations could therefore provide very strong constraints on the external boundary conditions ($n_H, T_g$) of these molecular condensations.

{\bf Substructures located further along the line of sight.} The HCO$^+$ J=4-3 substructures found at DF1 have emission velocity  ($v_{LSR}= 8-9$ km s$^{-1}$) more consistent with emission from the background OMC-1 than from the Bar ($v_{LSR}= 10.5$ km s$^{-1}$). These structures may thus be located further along the line of sight. This is consistent with the stepwise structure (Fig. \ref{fig:Geometry}) and estimates of extinction from H$_2$ lines (see Sect. \ref{sec:H2_extinction}). Towards the HCO$^+$ J=4-3 and H$_2$ common substructures, those found at DF1 are accordingly faintly visible in the H$_2$ ro-vibrational emission at 2.12 $\mu$m (Figs. \ref{fig:cut_Bar_NIRSpec} and \ref{fig:cut_Bar_H2_HCO_CO}), since they are affected by extinction along the line of sight.

\section{Photoevaporating protoplanetary disks}
\label{sec:proplyds}
%
In this section, we describe the photoevaporating proto-planetary disks observed in the whole NIRCam fields. 
The NIRCam images of the Orion Nebula show, in several passbands, numerous spatially resolved externally illuminated protoplanetary disks surrounding young stars, also known as proplyds \citep{O'Dell1993}. They are mostly found in the M42 region, clustering around the Trapezium stars and $\theta ^2$ Ori\,A, south of the Bar. A couple of proplyds were also identified in M43, located nearby NU\,Ori (HD\,37061), the B0.5V star creating the almost spherical HII region of M43 (see Appendix~\ref{Appendix:north-nebula-morphology}). Proplyds inside the HII region typically show bright heads of ionized gas (ionization front) and tails pointing directly away from the brightest OB stars. The proplyd family (as defined in \citealt{O'Dell1993}) also include YSO's seen in silhouette against the background HII region. The disks appear as dark ellipses on top of a bright nebula background and are called pure silhouettes.

The Hubble Space Telescope was the first observatory to spatially resolve proplyd anatomy and show detailed structure of each one of its components: ionized cocoon, embedded photoevaporating disk and jets/outflows. Nearly 200 proplyds in the Orion Nebula were discovered in HST images, mostly using narrow-band filters centered on the H$\alpha$ $\lambda$6563 and forbidden lines of  [NII] $\lambda$6583, [O I] $\lambda$6300, [O III] $\lambda$5007, and [S II] $\lambda$6717 + 6731 lines \citep{Prosser1994, O'Dell1994, O'Dell1996, McCaughrean1996, Bally1998, bally2000, O'Dell2001, Smith2005}. The most complete catalog of proplyds found in M42 and M43 is presented in \citet{Ricci2008}. The atlas is based on HST/ACS/WFC observations obtained under the Treasury Program on the Orion Nebula (PI: M. Robberto, GO 10249) using B, V, I, $z$ passbands and the H$\alpha$ narrow-band filter. They compile proplyds from previous studies and find new ones, including disks identified by their bipolar nebulae and jets, if closer than 1$\arcsec$ to the stellar source. They exclude HH objects, bow-shocks and elongated jets, but include candidate background galaxies or filaments. 
The catalog lists 178 bright proplyds with tails, 28 silhouettes, 8 bipolar nebulae and 5 jets, on a total of 219 sources. 

In order to identify known proplyds, and possibly find new ones, we cross-matched the sources in the NIRCam images with the \citet{Ricci2008} catalog.  We used the F187N narrow-band filter, centered on the Pa$\alpha$ emission line, because proplyd ionization fronts and tails are better traced in hydrogen recombination lines. In addition, the JWST diffraction limit at this wavelength (1.87 $\mu$m) is nearly the same as HST  at 0.656 $\mu$m or H$\alpha$, that is $\sim$ 0.07$\arcsec$, allowing for a direct comparison of proplyd morphology in the optical and in the near-IR. Narrow-band filters centered on emission lines are always preferred to observe proplyd structure because they cancel most of the continuum from the central star. 
Nevertheless, and due to the extreme sensitivity of JWST/NIRCam instrument, most proplyds remain unseen, obscured by the bright ``snow-flake" shape PSF of the JWST, even in the narrow-band filters.
The best targets are proplyds with nearly edge-on disks that still remain optically thick at near-IR wavelengths and hence are able to cover the young star. A few proplyds show wind-wind arcs and a couple show collimated jets. A detailed analysis of a few of these objects will be the subject of another article.

\begin{figure*}[h!]
\begin{center}
\vspace*{0cm}
\includegraphics[width=1\textwidth]{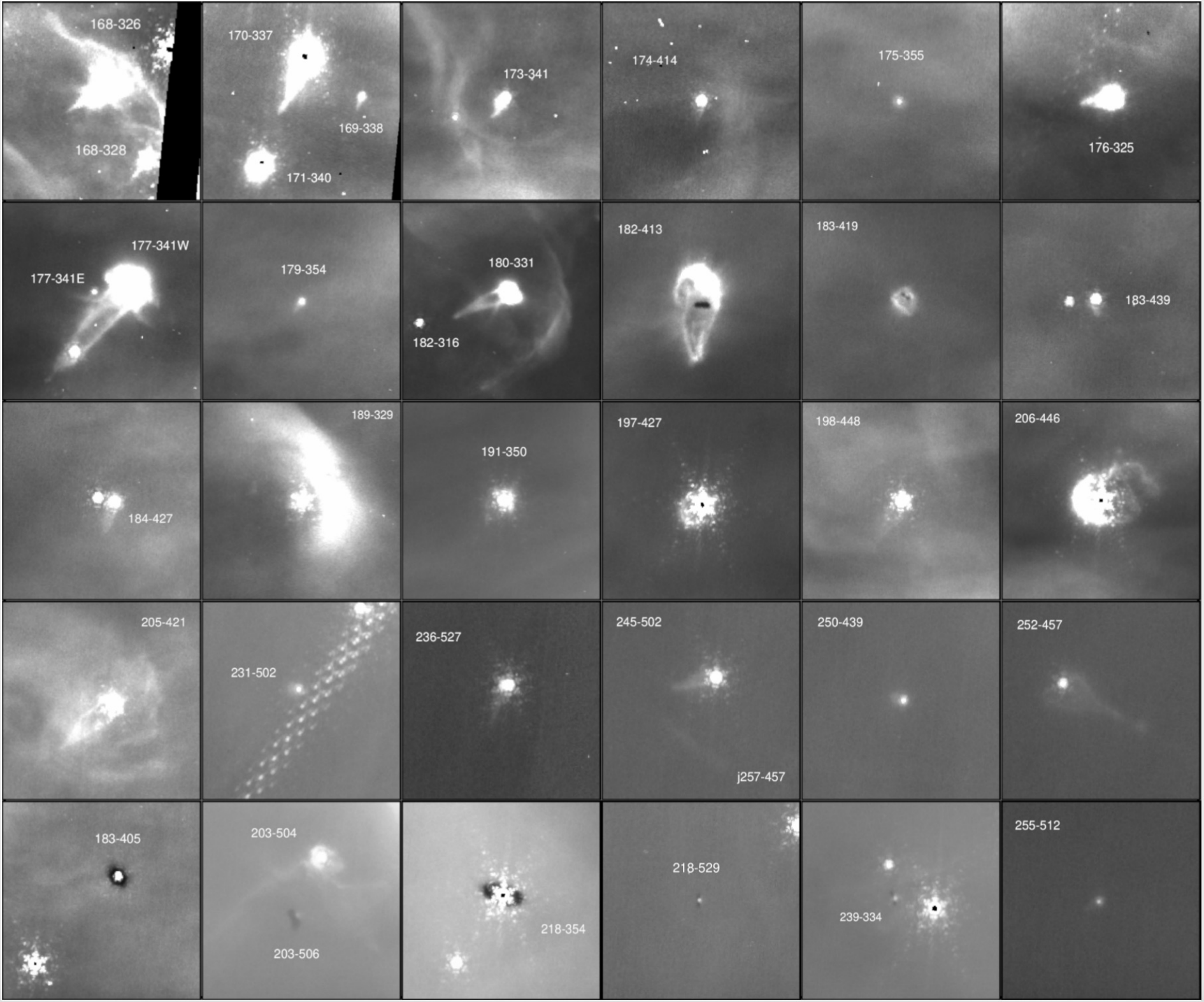}
\caption{Zoom on the 29 bright proplyds and 5 silhouettes showing extended structure in the F187N image of M42 covering the southeast region of the Trapezium and the Bar (detector B). The tiles are $5\arcsec \times 5\arcsec$ with north up and east to the left. Some images suffer from instrumental effects such as the diffraction pattern of bright stars and uncorrected cosmic ray events, affecting particularly the edge of the images because of lack of redundancy. }
\label{fig:proplyds_M42B}
\end{center}
\end{figure*}


\begin{figure}[h!]
\begin{center}
\vspace*{0cm}
\includegraphics[width=0.48\textwidth]{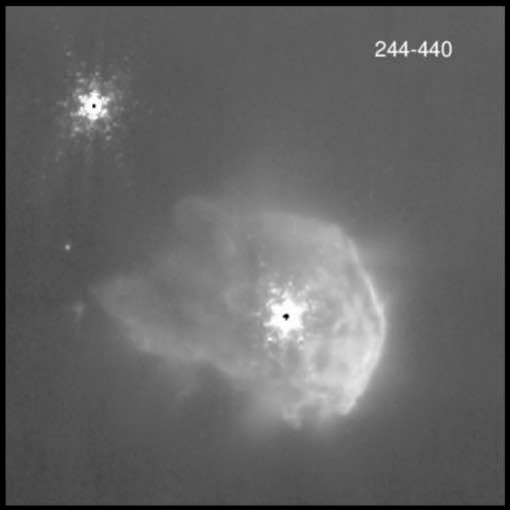}
\caption{Giant proplyd 244-440 located southeast of the Bar and illuminated by both stars, $\theta ^1$\,Ori\,C and $\theta ^2$\,Ori\,A. The image is $10\arcsec \times 10\arcsec$ in the F187N NIRCam filter, with north up and east to the left.}
\label{fig:244-440}
\end{center}
\end{figure}


\subsection{M42 region}

The NIRCam images were divided into module A, covering the North of the Dark Bay and corresponding to the northeast region of the Trapezium stars, and module B, covering the southeast region of the Trapezium stars, the Bar and $\theta^2$\,Ori\,A.
The M42 south field of view includes 62 known proplyds but only 34 of these were identified as having extended proplyd structure in the F187N image: 29 bright proplyds and 5 pure silhouettes.  They are shown in Fig.~\ref{fig:proplyds_M42B}. The giant proplyd 244-440 is shown in Fig.~\ref{fig:244-440} due to its much larger size. No new proplyds were found. 

The field of view north of the Dark Bay 
includes 16 sources from the \citet{Ricci2008} catalog, of which 12 are bright proplyds, 3 are pure silhouettes, and one is a bipolar reflection nebula. Only 7 of the 12 proplyds show extended structure, and only one silhouette of the 3 is visible in the F187N image because it is an edge-on disk. The other proplyds are simply point-like sources. The proplyd \mbox{215-106} shows a bright spot in the PSF that can be associated to a jet or a fainter smaller companion, requiring further analysis. The reflection bipolar nebula 208-122 does not appear in the F187N image but it reveals instead a close binary. The reflection polar nebulae and a dark disk are clearly seen in the F164N image centered at the [FeII] line at 1.64~$\mu$m. These objects are shown in Fig. \ref{fig:proplyds_M42A}. We also find 3 new proplyd candidates (shown in Fig.~\ref{fig:new_proplyds_M42A}), very faint and small, and named them 171-212, 180-218 and 234-104, following \citet{O'Dell1994} coordinate-based naming convention. 171-212 is a small proplyd in the north with a faint tail. 234-104 shows a faint ionized cusp in Pa$\alpha$ but not a tail. It lies in the northeast, at a distance greater than 140$\arcsec$ from the Trapezium. Both objects face the Trapezium stars. 180-218 lies along what seems to be a faint ionized filament or shock front. A round cusp surrounding a star is seen in Pa$\alpha$ and Br$\alpha$ hydrogen recombination lines, but not in other line tracers. It also shows no tail. The cusp can be the ionization front or a bow-shock caused by wind-wind interaction, requiring further analysis. The 3 objects are proplyds in nature because they fill in the proplyd criteria which are a disk and/or envelope that is being photoevaporated by external UV radiation. They show an ionization front and sometimes a tail and a bow-shock and have already a YSO forming inside. They differ from Evaporating Gaseous Globules (EGGs) which do not have yet a star forming and are just condensations of dense gas experiencing external photoevaporation. The 3 objects are not part of the 2MASS catalog of point sources and hence they must be very-low mass YSOs.

\subsection{M43 region}

The NIRCam images obtained in the parallel mode were divided into module A, covering M43, and module B, covering a region in the north of M42.
The NIRCam images of M43 include 4 proplyds from \citet{Ricci2008} catalog but only one of them shows extended structure. That is proplyd 332-1605 pointing directly to the ionizing star NU\,Ori, at $27\arcsec$ to the west, and showing a long tail, with a head-to-tail extension of 11.6$\arcsec$ or 4\,600 au at 414~pc, measured in the F187N NIRCam image. That is nearly 10 times larger than the proplyd HST10 which makes it  a giant proplyd candidate. This proplyd was first discovered in HST/WFPC2 parallel images (PI: Rubin, GO~6065) analyzed by \citet{O'Dell2001} as having a faint ionization front visible in H$\alpha$ and [SII] images but not in [OIII], and no tail. This object was also imaged with HST/ACS in H$\alpha$ by \citet{Ricci2008} who confirmed its tailess structure. The fact that we see a long tail in Pa$\alpha$ and not in H$\alpha$ is consistent with the low ionizing power of the star NU\,Ori (B0.5V) when compared to the Trapezium stars (O7V for $\theta ^1$\,Ori\,C) or $\theta ^2$\,Ori\,A (O9.5V). A new proplyd candidate is found in the F187N image with a prominent jet rendered visible by a chain of knots or HH objects. The ionization front and the knots are also visible in the HST/ACS image in H$\alpha$, but not the central star. The powerful jet and the fact that it is not visible in the optical means this object must be very young, still embedded in its circumstellar envelope of gas and dust and experiencing high accretion, that is, it is still a protostar. This object is part of the 2MASS catalog of point sources. 
We named it 269-1713 following \citet{O'Dell1994} coordinate-based naming convention. The new proplyd is located at $97\arcsec$ to the southwest of the ionizing star, NU\,Ori. Fig.~\ref{fig:giant_proplyd_M43} and Fig.~\ref{fig:new_proplyd_M43} show respectively the giant proplyd 332-1605 and the new proplyd candidate 269-1713 in NIRCam Pa$\alpha$ vs. HST/ACS H$\alpha$ (PI: J.Bally, GO~9825) images.
In the NIRCam images of M43 and M42 north we find numerous extended, elliptical and diffuse objects, sometimes with spiral arms. These are background galaxies.

\section{Conclusions}
\label{sec:conclusions}

The JWST/NIRCam and MIRI imaging observations of the Orion Nebula allow us to probe the global fundamental structure and small-scale structures of an interstellar cloud strongly illuminated by UV radiation.
We have access to the multiple scales  of the nebula with resolution of 0.1 to 1$\arcsec$ from 2 to 25 $\mu$m, equivalent to $\sim \rm 2\times10^{-4}$ to $\rm 2\times10^{-3}$  pc  or 40 to 400 au at 414~pc, over a field of view of 150$\arcsec$ and 42$\arcsec$, equivalent to $\sim$0.3 and 0.08~pc (at the Orion distance of 414~pc) for NIRCAM and MIRI images centered on the Bar.
Our main results can be summarized as follows.

\begin{itemize}

\item One of the most striking features observed in all our NIRCAM and MIRI images is that the molecular cloud borders appear structured at small scales. 
Numerous patterns are observed, such as ridges, waves and globules. This highlights a very intricate irradiated cloud surface (most likely turbulent) and sub-dense structures at such small scales that they were inaccessible to previous IR observations. Several bright emission features associated with the highly irradiated surroundings of dense molecular condensations, embedded young star and photoevaporated protoplanetary disks are  detected in the extended PDR layers.

\item 
The observations spatially resolve  the transition 
from the ionization front, the dissociation front to the molecular cloud of the prototypical highly irradiated extended dense Orion Bar PDR. This allows us to study the PDR along all its fronts and to spatially resolve the FUV radiation penetration scales inside the molecular cloud.
A progressive structure is evident in agreement with previous studies. However, instead of a smooth PDR transition, JWST unambiguously reveals a highly sculpted interface with very sharp edges and multiple ridges.

\item The spatial distribution of the AIB emission reveal 
a very sharp illuminated edge at the IF (on scales of 1$\arcsec$ or 0.002 pc) with a strong density rise in the neutral zone. 
This is expected due to the sharp decrease in gas temperature at the ionization front if the thermal pressures in the ionized and neutral region are of similar magnitude. The density we derived 
in the atomic region ($n_H \sim (5-10)\times 10^4$~cm$^{-3}$) is much higher (factor 10-20) than the electron density previously derived at the IF. 
Behind the sharp PDR edge, an extensive warm layer 
of neutral material, essentially atomic with strong emission from the AIBs, is observed up to the H$^0$/H$_2$ dissociation front at 10-20$\arcsec$ or 0.02-0.04 pc from the IF.

\item In contrast to the IF, a very complex, structured and folded H$^0$/H$_2$ dissociation front surface is traced by the H$_2$ lines. This is particularly apparent in the southwestern part of the Bar. A terraced-field-like structure with several steps seen from above can explain the succession of H$_2$ ridges across the Bar. In that geometry, each observed H$_2$ emission ridge corresponds to a portion of the DF seen edge-on.

\item The line spatial profiles of the highly rotationally and ro-vibrationally excited H$_2$ agree in remarkable detail. Physical conditions must be comparable along the folded DFs surface. 
Very thin and bright H$_2$ emission layers ($\sim 10^{-3}$ pc) are spatially resolved at the irradiated surface of the dense molecular regions.  
The highly excited H$_2$ emission arises from the very thin zone where the gas density and H$_2$ abundance starts to increase sharply. 

\item A remarkable agreement in the spatial distribution between the rotationaly and rovibrationaly excited H$_2$ and ALMA HCO$^+$ J=4-3 emission maps is observed. This indicates that they both come from the edge of dense structures and that they are chemically linked. Some of the densest portions of the Bar lie very near the DFs. 
This is in agreement with previous analysis of ALMA  and \textit{Herschel} observations. However, the small structures were unresolved in most tracers until now.   
JWST observations provide very strong constraints on the external boundary conditions of the dense molecular condensations.

\item  In M42, several outflows interacting with the ambient ionized gas of the nebula or the molecular gas are detected. 
Crenellated structures and various arches which are most likely bow-shocks are observed. 
Regions exposed to stellar winds and protostellar outflows might be especially turbulent.

\item Numerous proplyds are identified in the NIRCam images of M42 and M43. Nevertheless, many remain unseen, obscured by the bright snow-flake PSF of the very sensitive JWST. The best observed targets are proplyds with nearly edge-on disks that remain optically thick at near-IR wavelengths and thus are able to cover the young star. For these proplyds, the NIRCam instrument offers a unique opportunity to study proplyd morphology in the near-IR with a spatial resolution comparable to the HST in the optical. We find 4 new proplyds identified in the F187N images, 3 located at the northeast from the Trapezium stars and one in the M43 region. They were named 171-212, 180-218, 234-104 and 269-1713, following \citet{O'Dell1994} coordinate-based naming convention.

\end{itemize}

The JWST ERS program on the Orion Bar PDR \citep{pdrs4all} gives access to IFU spectroscopy with NIRSpec and MIRI which will be published in other articles.
IFU spectroscopy provides insight into the local gas physical conditions (temperature, density, and pressure), the dust properties and the chemical composition of the warm, very structured irradiated medium.  
It will be possible to probe the dust properties and physical conditions in the dense substructures detected with NIRCam and MIRI images  described in this article. To determine the pressure and density variations at the PDR edge, 
future detailed spatial studies of both the H$_2$  pure rotational and rovibrational lines will be carried out \cite[]{Peeters-spectro,lettergas}.
These constraints on the physical conditions may allow us to better understand the dynamical effects in PDRs, such as compression waves and photo-evaporative flows.

\begin{acknowledgements}
We are grateful to the referee O'Dell C. Robert for relevant and constructive comments.
MIRI data reduction is performed at the French MIRI centre of expertise with the support of CNES and the ANR-labcom INCLASS between IAS and the company ACRI-ST. NIRCam data reduction is performed by both IRAP and IAS. EP and JC acknowledge support from the University of Western Ontario, the Institute for Earth and Space Exploration, the Canadian Space Agency, and the Natural Sciences and Engineering Research Council of Canada. OB is funded by a CNES APR program. Part of this work was supported by the Programme National ``Physique et Chimie du Milieu Interstellaire'' (PCMI) of CNRS/INSU with INC/INP co-funded by CEA and CNES.
JRG and SC thank the Spanish MCINN for funding support under grant \mbox{PID2019-106110GB-I00}. Studies of interstellar PAHs at Leiden Observatory (AT) are supported by a Spinoza premie from the Dutch Science Agency, NWO. CB is grateful for an appointment at NASA Ames Research Center through the San Jos\'e State University Research Foundation (80NSSC22M0107). Work by YO and MR is carried out within the Collaborative Research Centre 956, sub-project C1, funded by the Deutsche Forschungsgemeinschaft (DFG) – project ID 184018867.  TO is supported by JSPS Bilateral Program, Grant Number 120219939.
AP would like to acknowledge financial support from Department of Science and Technology - SERB via Core Research Grant (DST-CRG) grant (SERB-CRG/2021/000907), Institutes of Eminence (IoE) incentive grant, BHU (incentive/2021- 22/32439), Banaras Hindu University, Varanasi and thanks the Inter-University Centre for Astronomy and Astrophysics, Pune for associateship.
This work is based on observations made with the NASA/ESA/CSA James Webb Space Telescope. The data were obtained from the Mikulski Archive for Space Telescopes at the Space Telescope Science Institute, which is operated by the Association of Universities for Research in Astronomy, Inc., under NASA contract NAS 5-03127 for JWST. These observations are associated with program \#1288.
Support for program \#1288 was provided by NASA through a grant from the Space Telescope Science Institute, which is operated by the Association of Universities for Research in Astronomy, Inc., under NASA contract NAS 5-03127. MB acknowledges DST for the DST INSPIRE Faculty fellowship. This work is sponsored in part by the CAS, through a grant to the CAS South America Center for Astronomy (CASSACA) in Santiago, Chile.
HZ acknowledges support from the Swedish Research Council (contract No 2020-03437). 
AR gratefully acknowledges support from the directed Work Package at NASA Ames titled: ‘Laboratory Astrophysics –The NASA Ames PAH IR Spectroscopic Database’.

\end{acknowledgements}


\bibliographystyle{aa}
\bibliography{mainbib}

\begin{appendix} 

\section{Contribution of lines and AIBs in imaging bands}
\label{Appendix:contribution_lines}

\begin{figure*}[h]
\begin{center}
\includegraphics[width=0.8\textwidth]{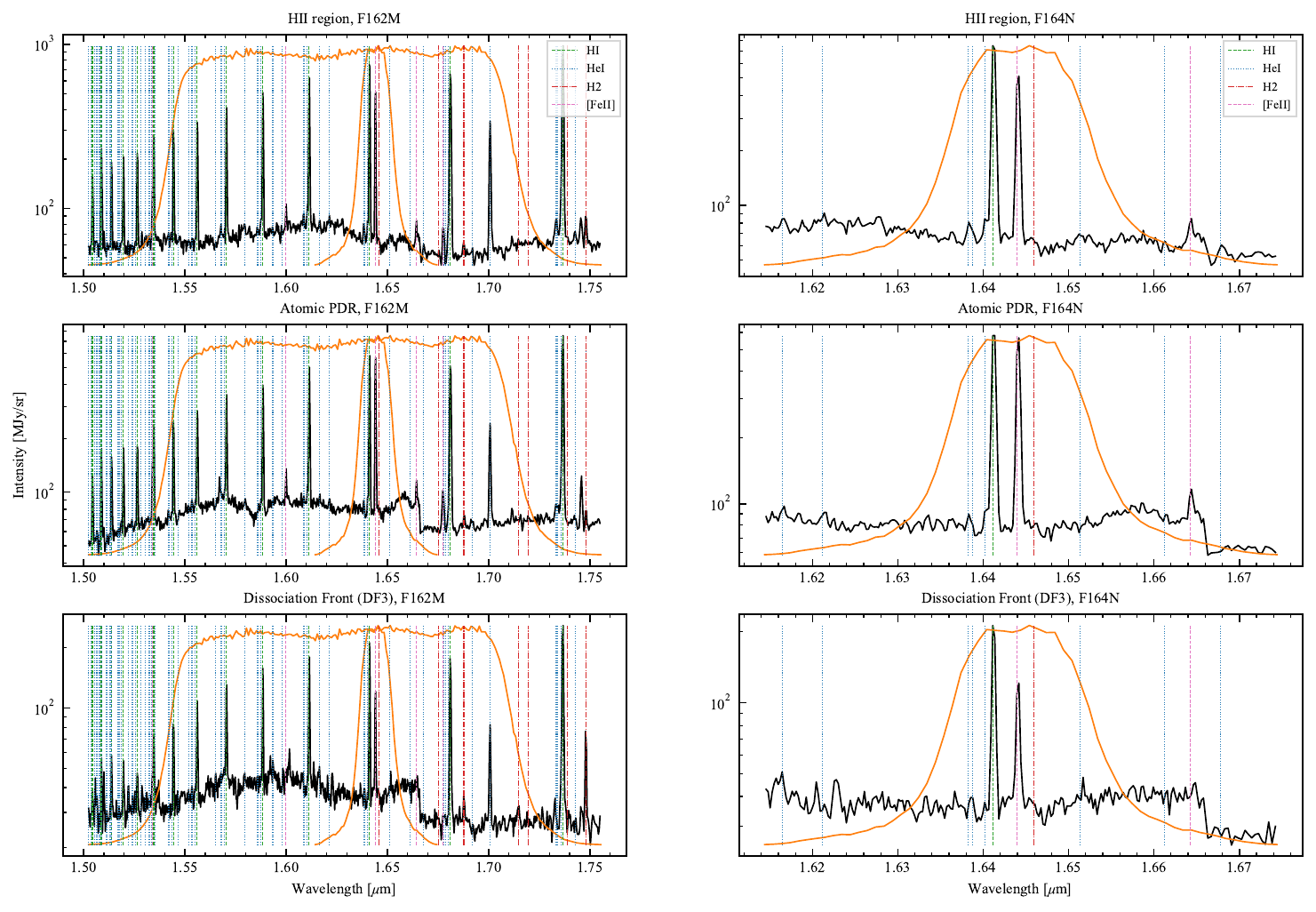}
\caption{Three template spectra (from top to bottom) shown in  \citet{Peeters-spectro} in the range covered by F162M (left) and F164N (right). Response functions of the filters are shown with thin orange lines. Expected positions of HI, HeI, H$_2$, and [FeII] lines are marked.}
\label{fig:template_FeII}
\end{center}
\end{figure*}

\begin{figure*}[h!]
\begin{center}
\includegraphics[width=0.8\textwidth]{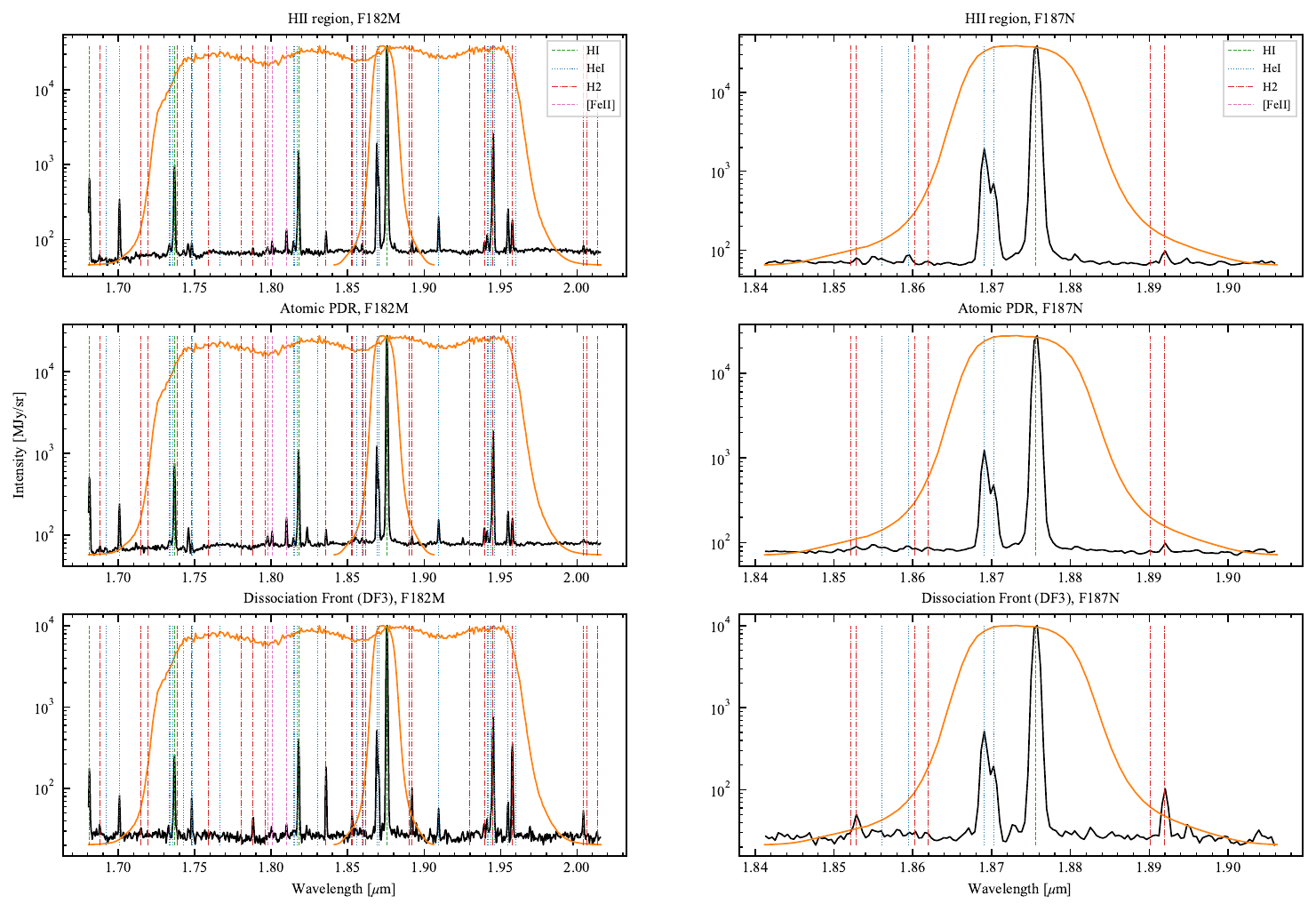}
\caption{Same as Fig.~\ref{fig:template_FeII}, but for F182M (left) and F187N (right).}
\label{fig:template_Pa_a}
\end{center}
\end{figure*}

\begin{figure*}[h!]
\begin{center}
\includegraphics[width=0.8\textwidth]{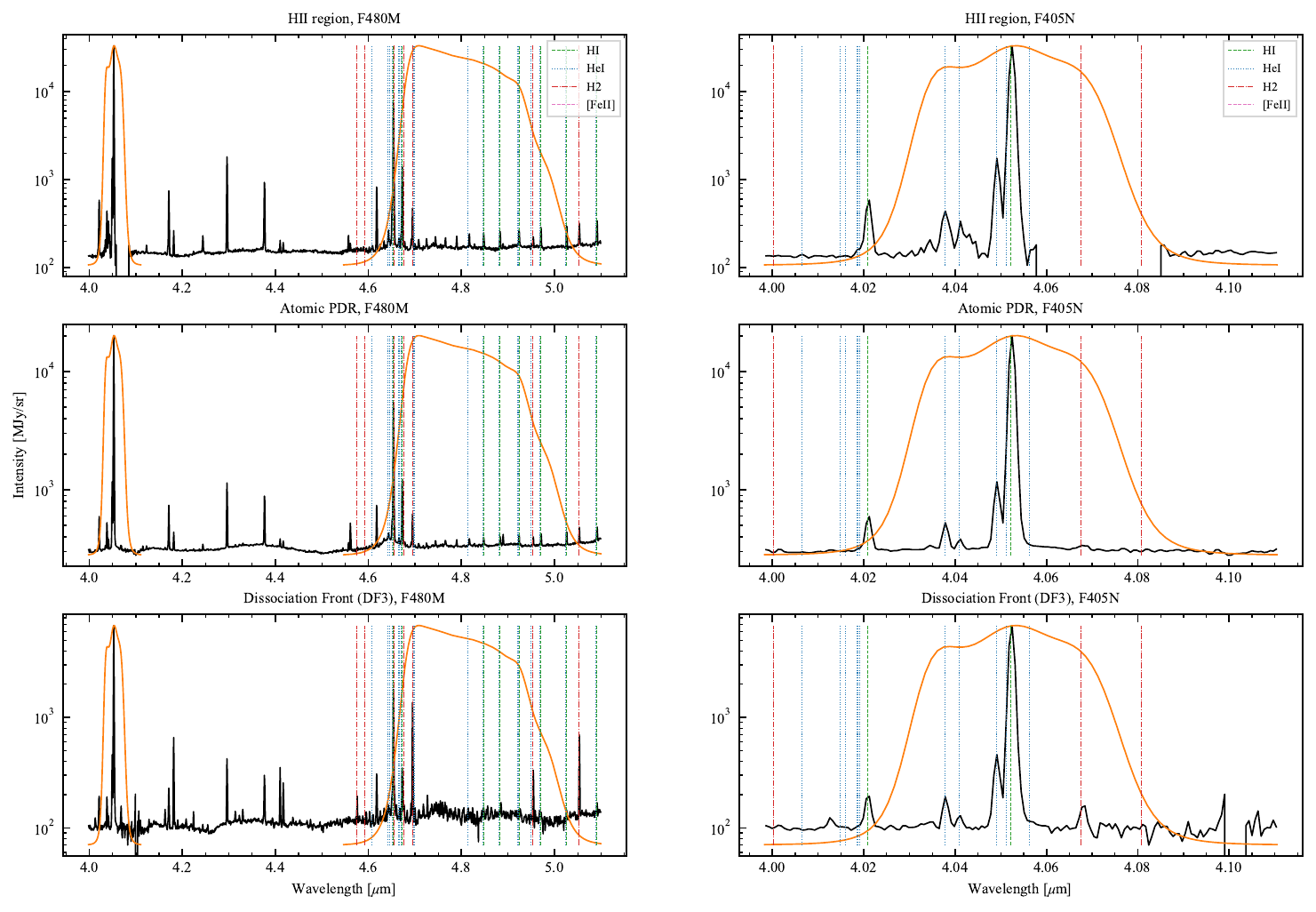}
\caption{Same as Fig.~\ref{fig:template_FeII}, but for F480M (left) and F405N (right).}
\label{fig:template_Br_a}
\end{center}
\end{figure*}

\begin{figure*}[h!]
\begin{center}
\includegraphics[width=0.8\textwidth]{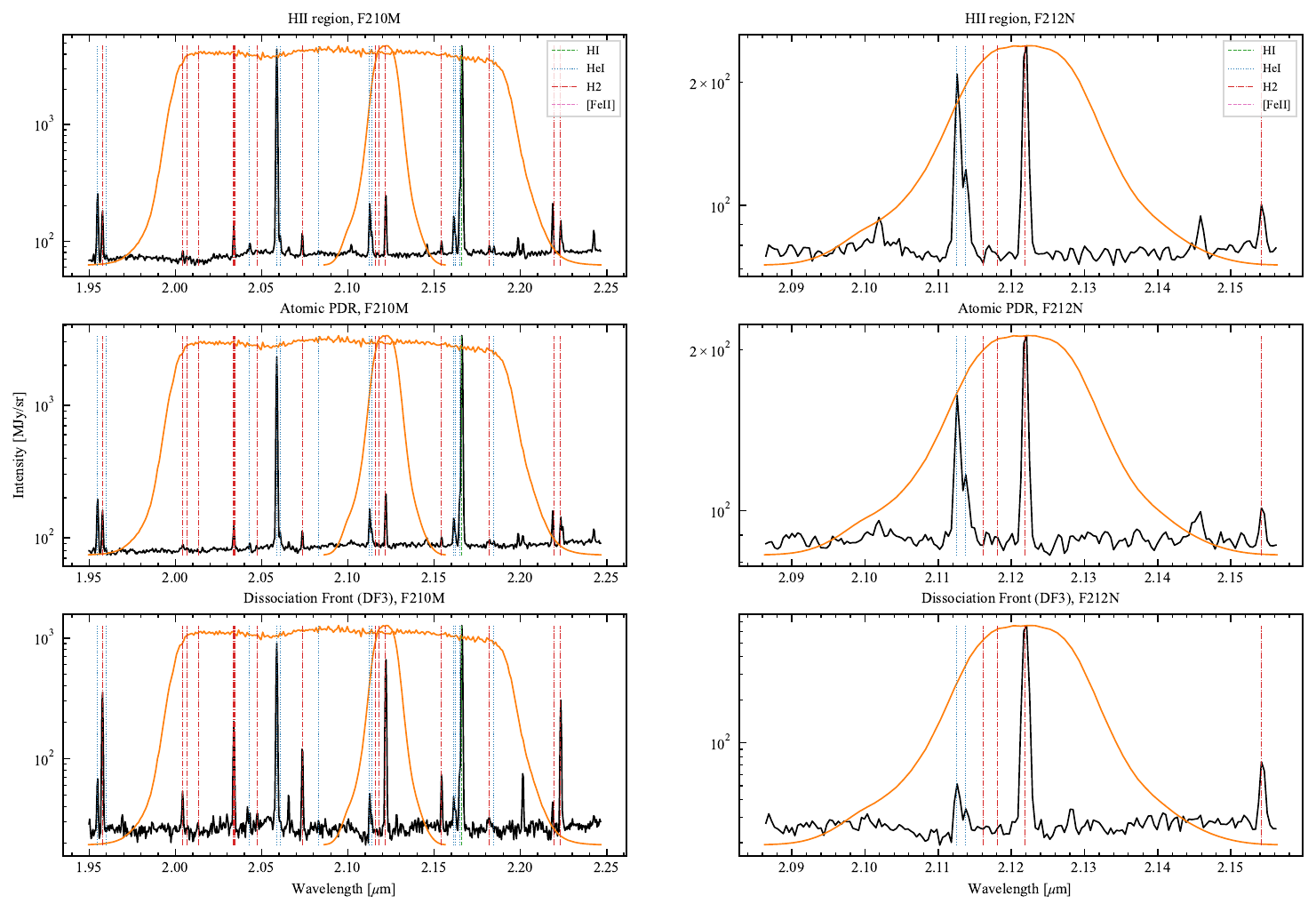}
\caption{Same as Fig.~\ref{fig:template_FeII}, but for F210M (left) and F212N (right).}
\label{fig:template_H2_212}
\end{center}
\end{figure*}

\begin{figure*}[h!]
\begin{center}
\includegraphics[width=0.8\textwidth]{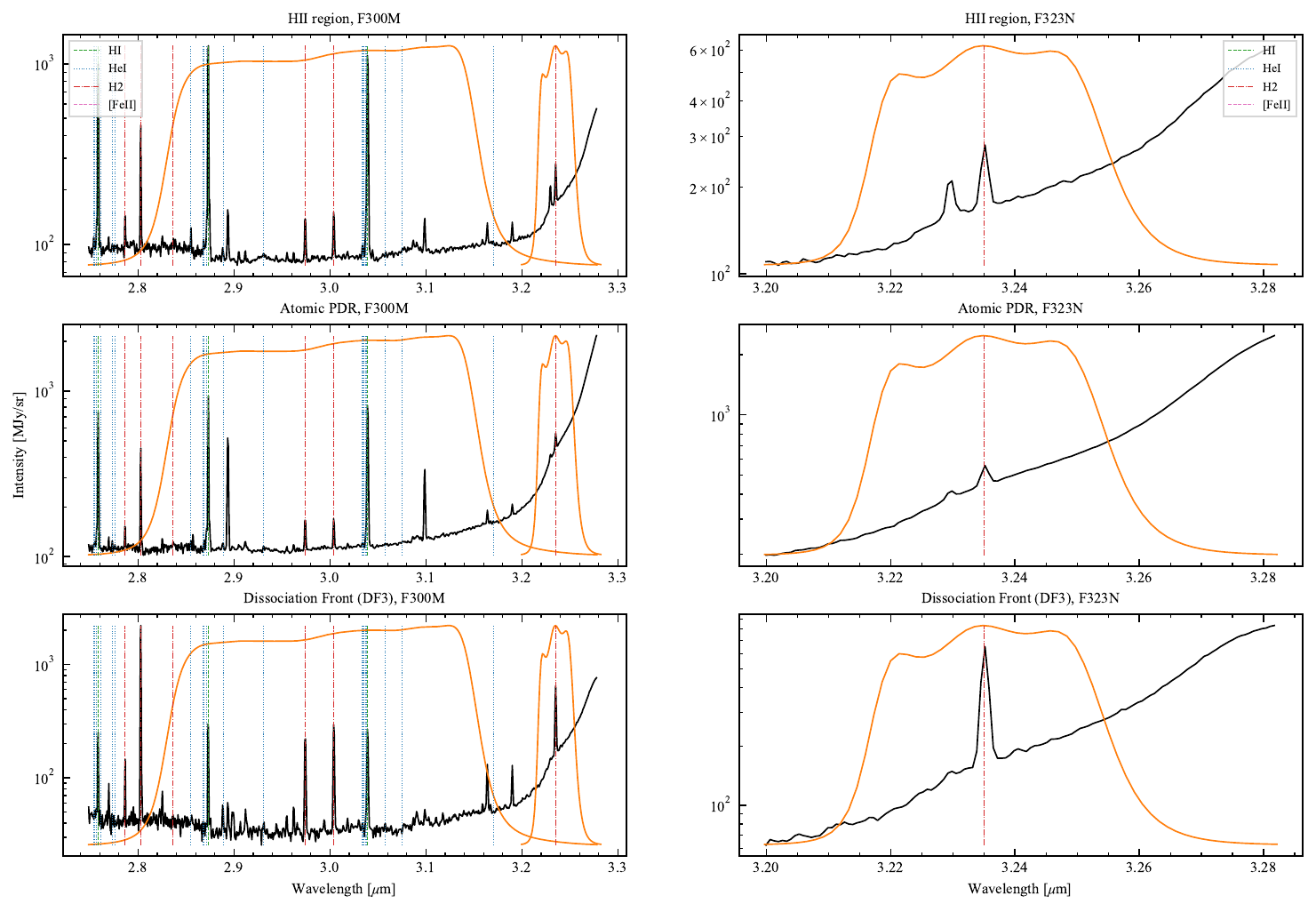}
\caption{Same as Fig.~\ref{fig:template_FeII}, but for F300M (left) and F323N (right).}
\label{fig:template_H2_324}
\end{center}
\end{figure*}

\begin{figure*}[h!]
\begin{center}
\includegraphics[width=0.8\textwidth]{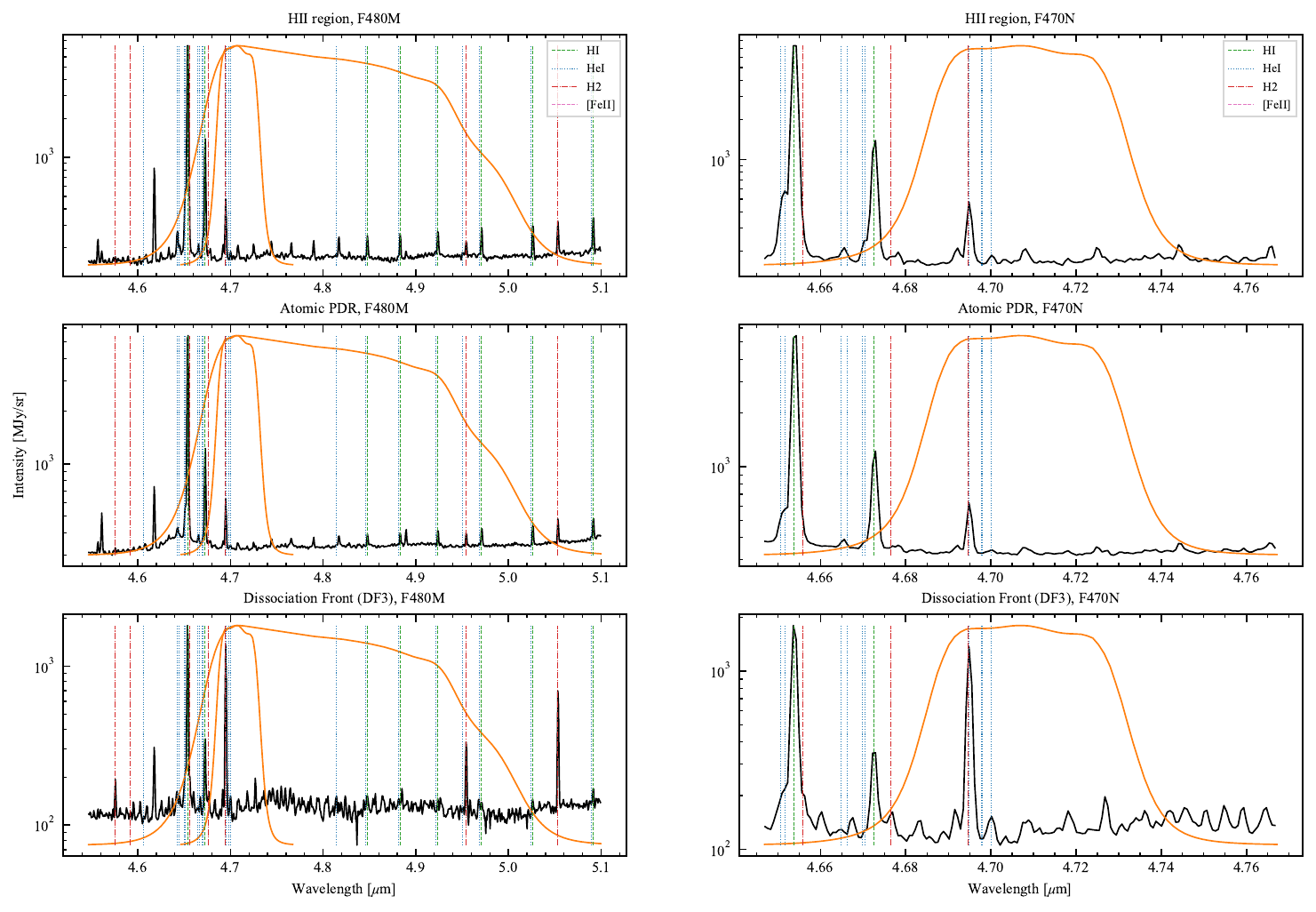}
\caption{Same as Fig.~\ref{fig:template_FeII}, but for F480 (left) and F470N (right).}
\label{fig:template_H2_469}
\end{center}
\end{figure*}

\begin{figure*}[h!]
\begin{center}
\includegraphics[width=0.8\textwidth]{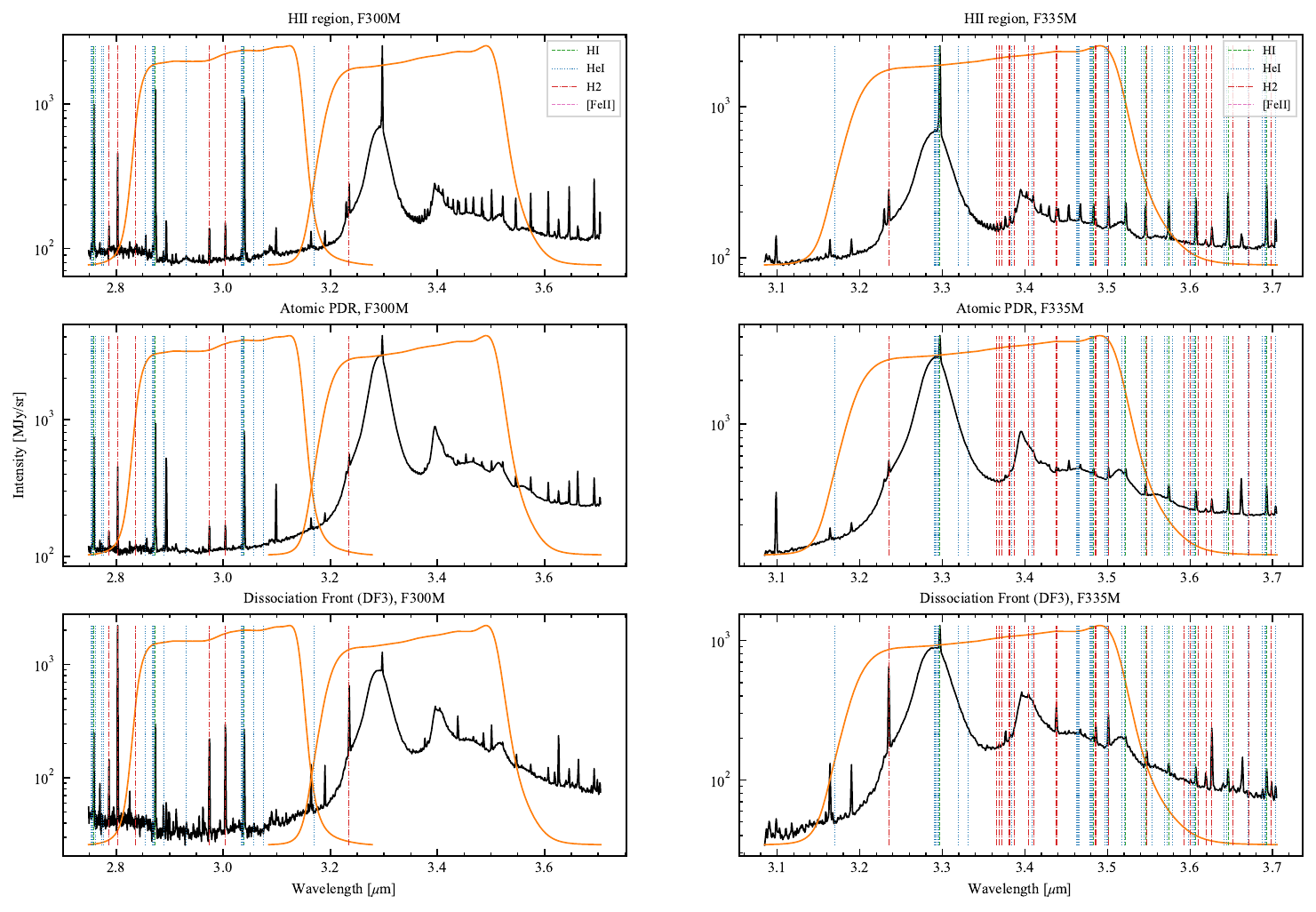}
\caption{Same as Fig.~\ref{fig:template_AIB}, but for F300M (left) and F335M (right).}
\label{fig:template_AIB}
\end{center}
\end{figure*}

Figs.~\ref{fig:template_FeII} to \ref{fig:template_AIB} show the three template spectra in \citet{Peeters-spectro}, illustrating the variation of the contribution of different lines into each imaging band. 
More detailed analysis with a full set of lines will be presented in the science enable product and the associated article \citep{chown23}.

\clearpage

\clearpage

\section{JWST and ground-based telescopes images of the Bar}
\label{Appendix:Images_all_filters_Orion_Bar}

\begin{figure*}[h!]
\begin{center}
\vspace*{0cm}
\includegraphics[width=0.8\textwidth]{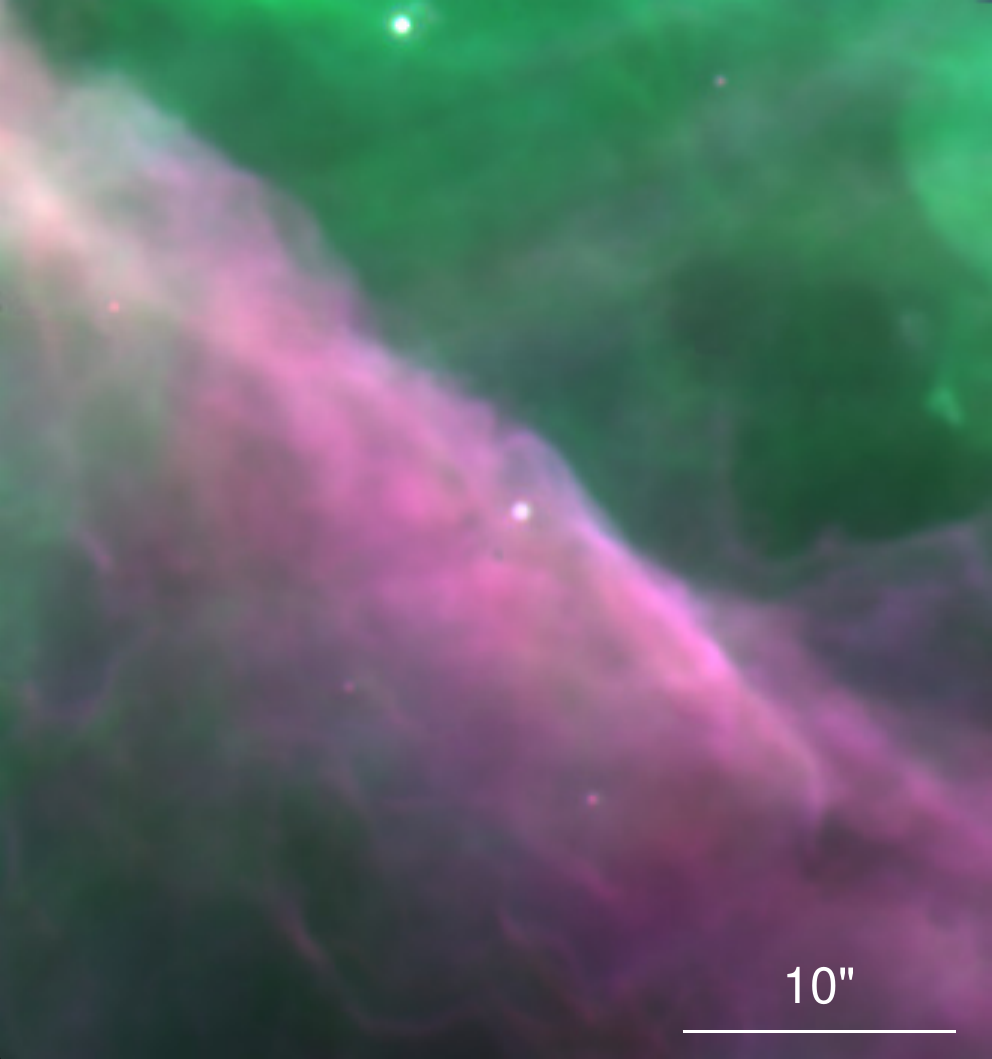}
\includegraphics[width=0.3\textwidth]{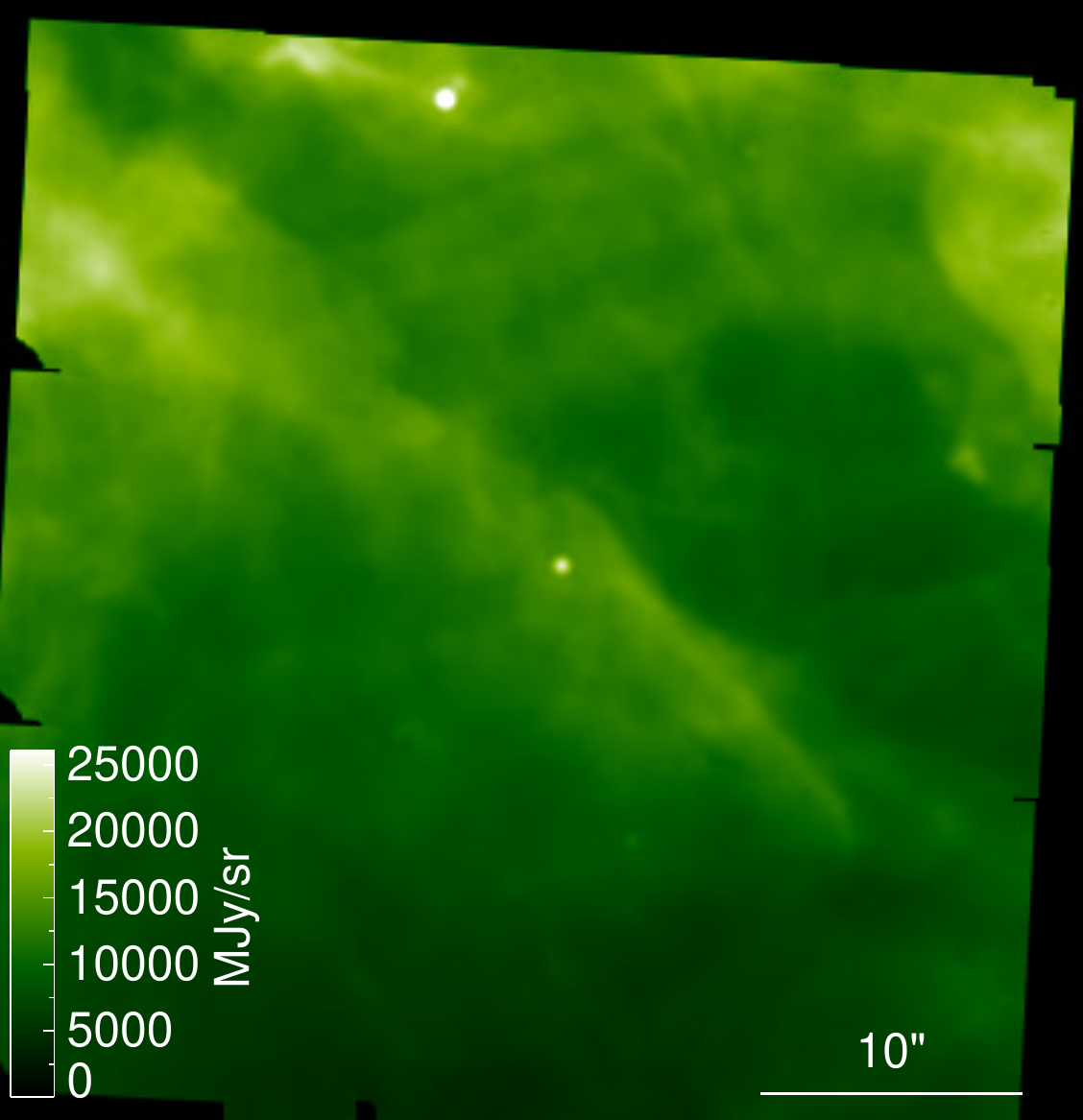}
\includegraphics[width=0.3\textwidth]{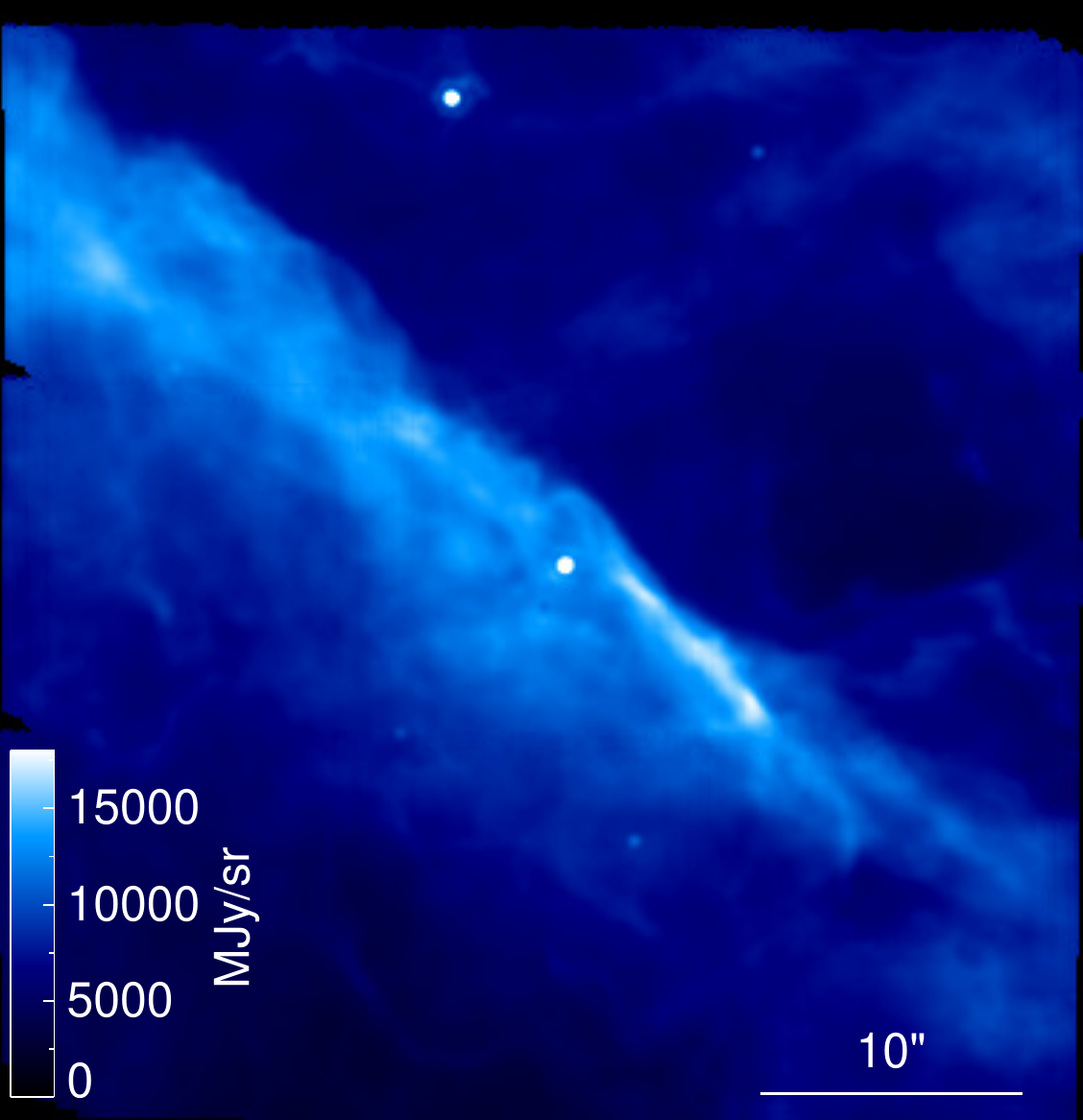}
\includegraphics[width=0.3\textwidth]{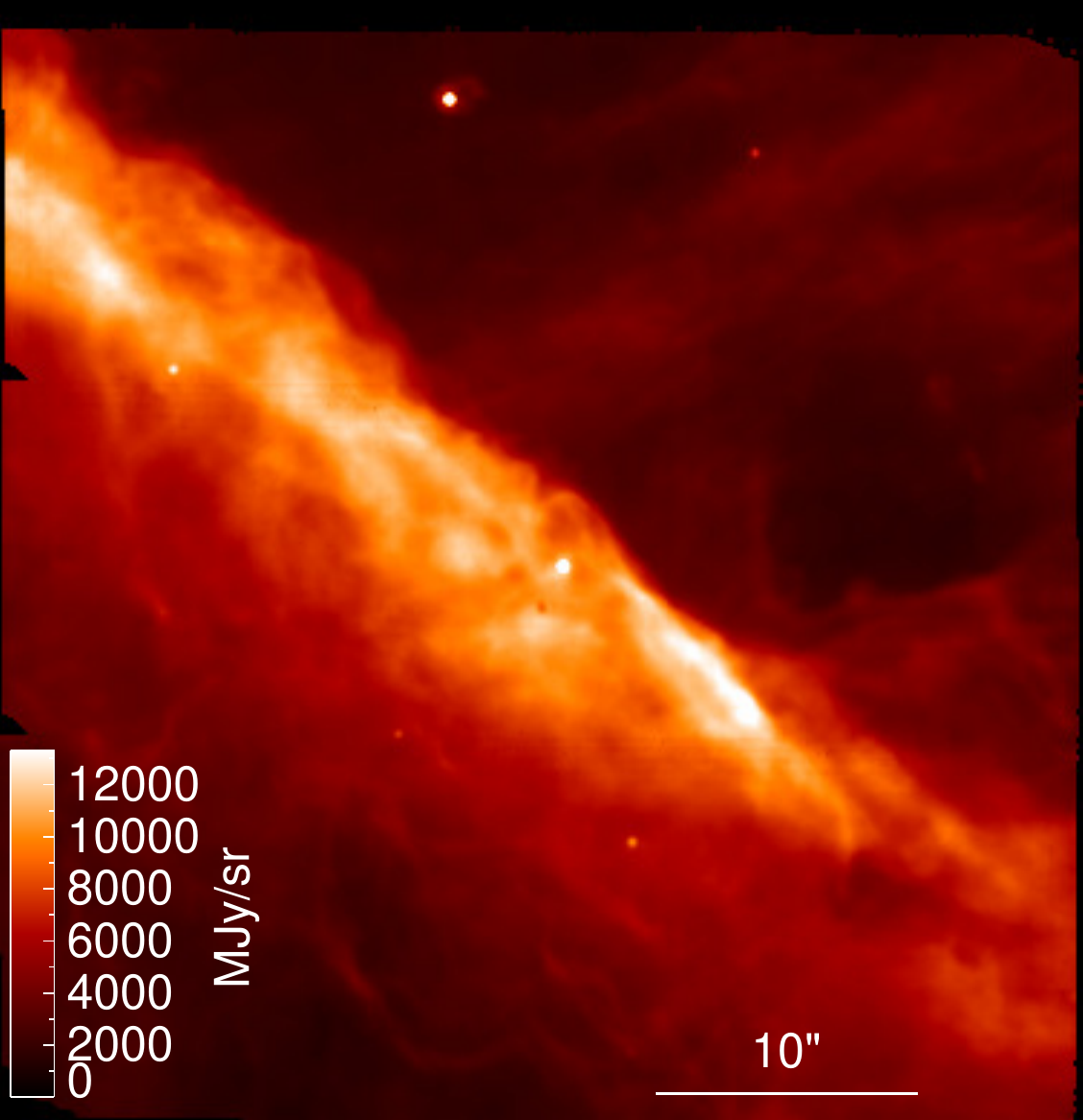}
\caption{Southwestern part of Bar as seen by the JWST’s MIRI instrument with north up and east left. 
Several images in different filters were combined to produce an RGB composite image: F1500W (green), F1130W (blue) and F770W (red)
that traces respectively continuum emission from hot/warm dust and hydrocarbons (AIBs). The individual images used to make the RGB composite one are shown below. The size of the images is $\sim$42$\arcsec \times 42 \arcsec$ and it is centered on RA=05$^{\rm h}$35$^{\rm m}$20$\fs$33, DEC=-05$\degr$25$\arcmin$03.77$\arcsec$
very close the irradiated proto-planetary disk 203-504. }
\label{fig:composite-image-MIRI}
\end{center}
\end{figure*}

Fig.~\ref{fig:composite-image-MIRI} shows the composite MIRI images of the southwest part of Bar. This region designated the "SW Bright Bar" in \cite{Odell2017b} presents the sharpest ionization front of the Bar. Its structure and physical conditions near the MIF are studied in detail from HST, MUSE and velocity data
in their Section 4.3.

\begin{figure*}[h!]
\begin{center}
\vspace*{0cm}
\includegraphics[width=1\textwidth]{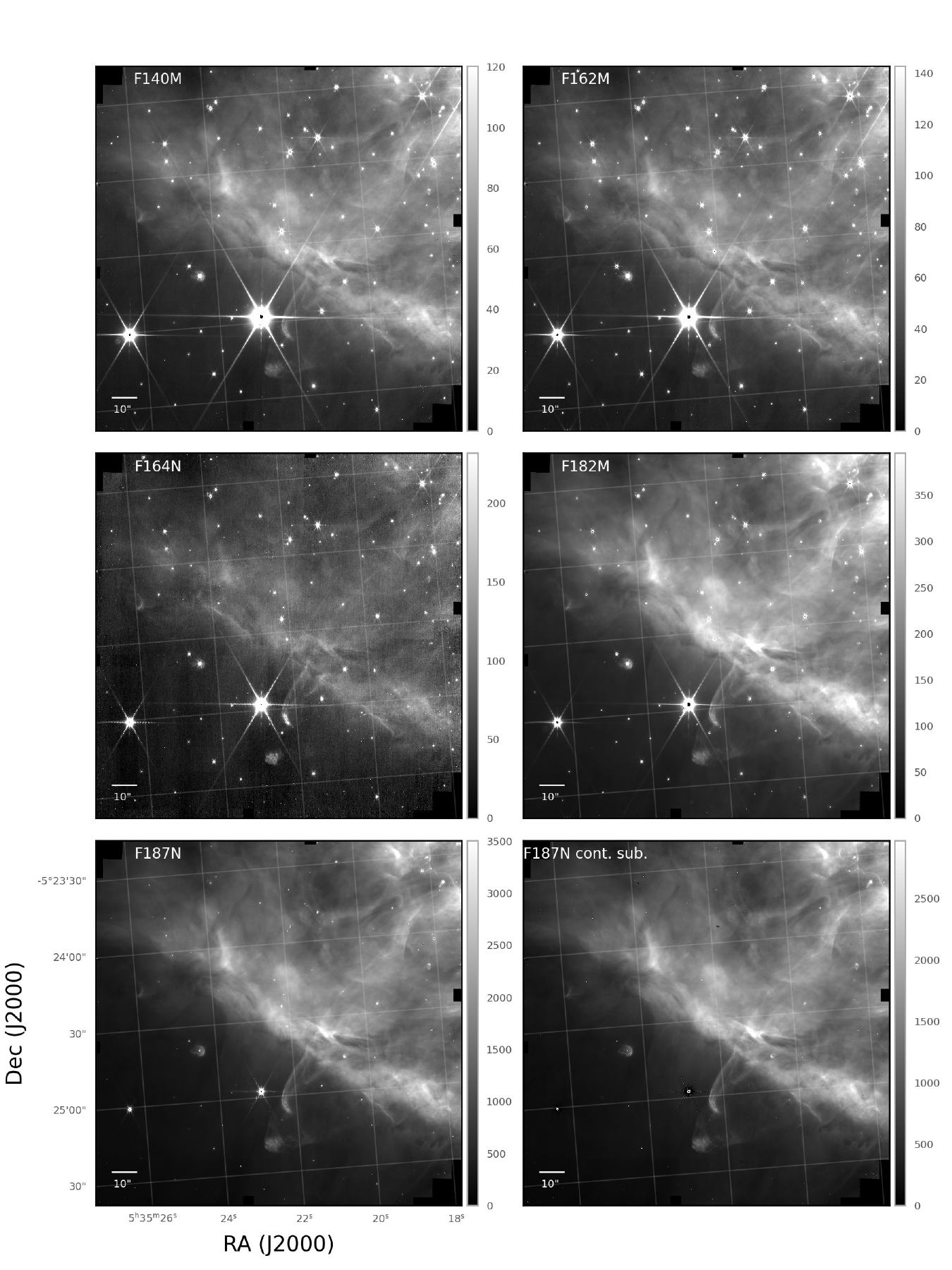}
\caption{NIRCam maps of the Bar in different filters. Units are in MJy/sr.}
\label{fig:Annexe_Orion_bar_all_filters_1}
\end{center}
\end{figure*}

\begin{figure*}[h!]
\begin{center}
\vspace*{0cm}
\includegraphics[width=1\textwidth]{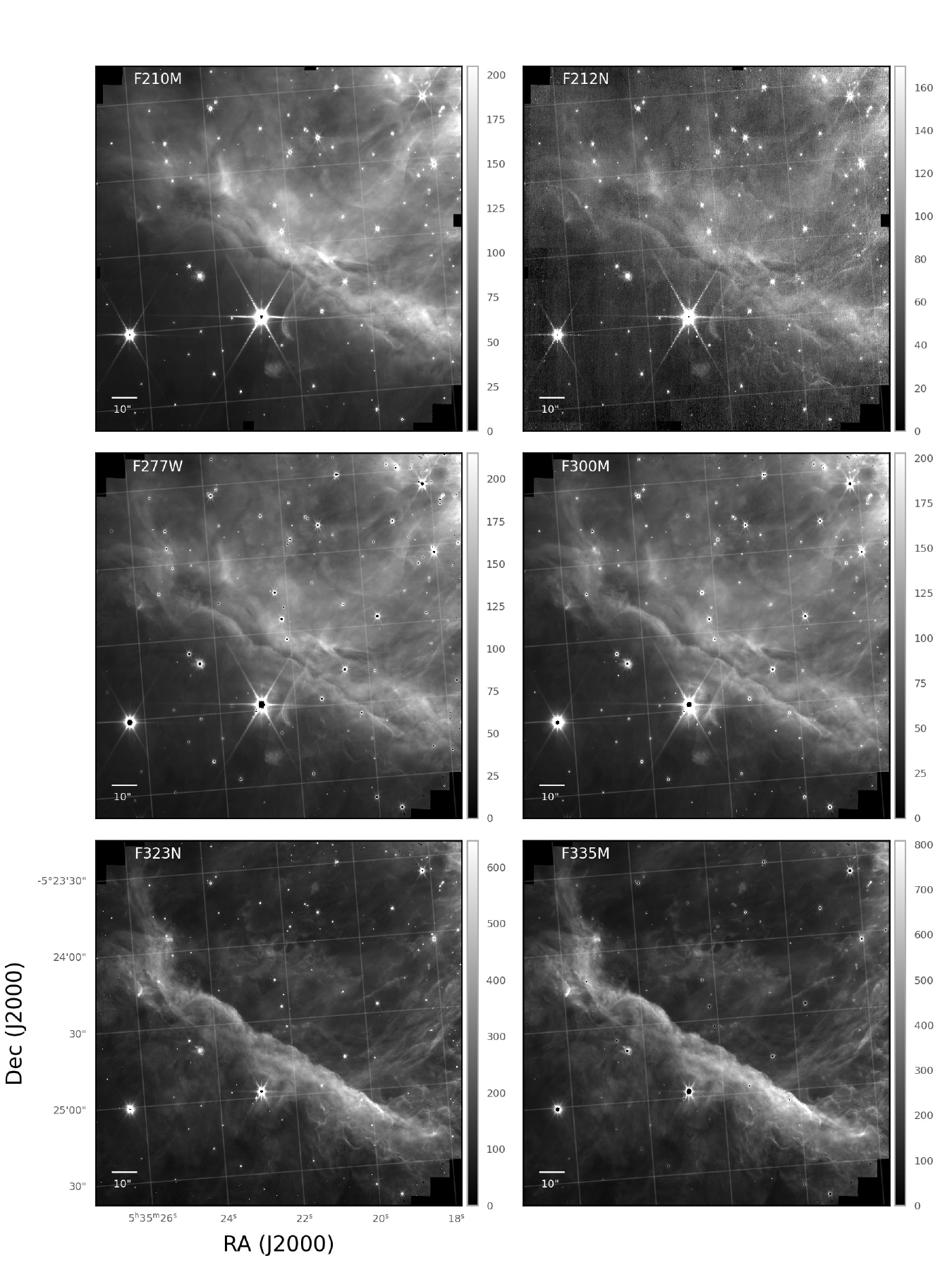}
\caption{Same as Fig.~\ref{fig:Annexe_Orion_bar_all_filters_1}, but for other filters.}
\label{fig:Annexe_Orion_bar_all_filters_2}
\end{center}
\end{figure*}

\begin{figure*}[h!]
\begin{center}
\vspace*{0cm}
\includegraphics[width=1\textwidth]{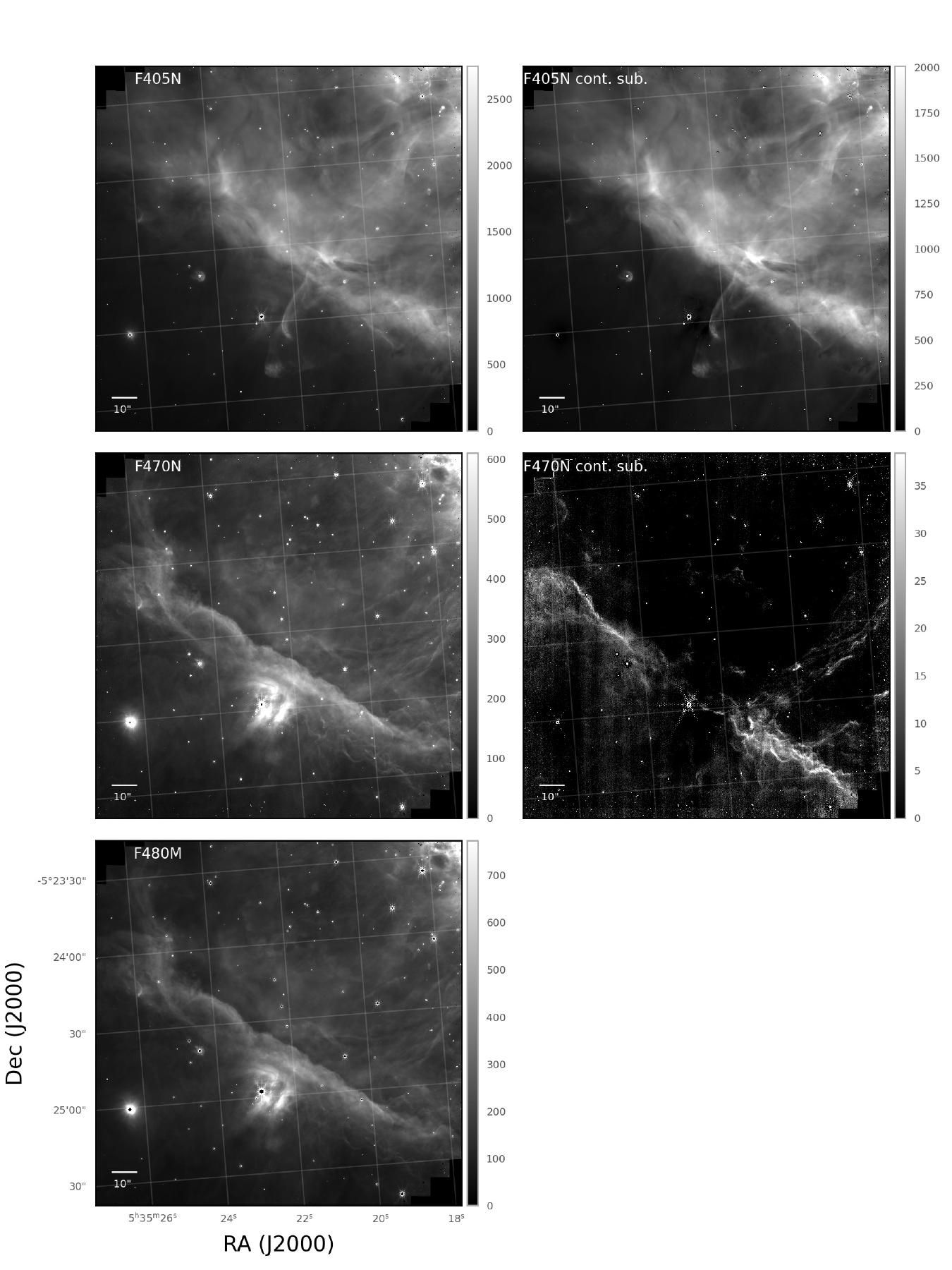}
\caption{Same as Fig.~\ref{fig:Annexe_Orion_bar_all_filters_1}, but for other filters. 
}
\label{fig:Annexe_Orion_bar_all_filters_3}
\end{center}
\end{figure*}

\begin{figure*}[h!]
\begin{center}
\vspace*{0cm}
\includegraphics[width=0.9\textwidth]{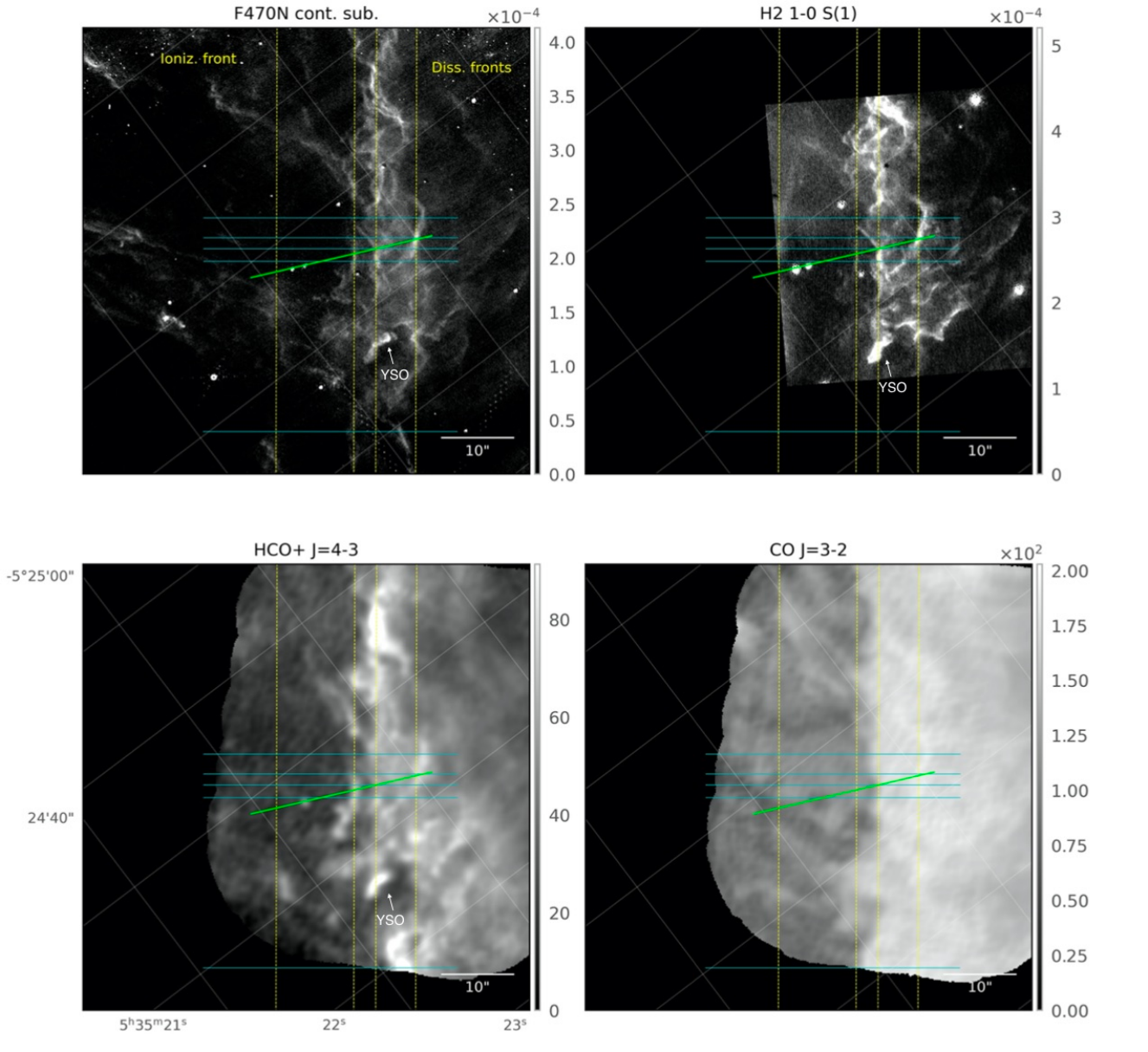}
\caption{NIRCam, Keck and ALMA maps of the southwestern part of Bar. The maps have been rotated 
so that the ionizing radiation strikes the Bar from the left. 
{\bf Upper left}: Map in the F470N filter continuum subtracted (${\rm H_2}$ 0-0 S(9) line at 4.7 $\mu$m). Units are in erg s$^{-1}$ cm$^{-2}$ sr$^{-1}$. {\bf Upper right}: Keck/NIRC2 map of the ${\rm H_2}$ 
1-0 S(1) line at 2.12 $\mu m$ with a resolution of 0.085$\arcsec$ \citep{Habart2023}.  
Our NIRCam observations show a very good agreement with Keck/NIRC2 observations in terms of line distribution and intensity. {\bf Lower left and right}: ALMA observation at 1$\arcsec$ resolution of the HCO$^+$ J=4–3 line integrated intensity (in K km s$^{-1}$) and the CO J=3-2 line peak temperatures ($T_{peak}$) \citep{goicoechea_compression_2016}. There exists a remarkably similar spatial distribution between the 
${\rm H_2}$ and HCO$^+$ emission from ALMA.
}
\label{fig:cut_Bar_H2_HCO_CO}
\end{center}
\end{figure*}

Figs.~\ref{fig:Annexe_Orion_bar_all_filters_1}-\ref{fig:Annexe_Orion_bar_all_filters_3} present all NIRCam images of the Bar obtained in the filters listed in Table \ref{tab:filters}, and some filters continuum subtracted.
Fig.~\ref{fig:cut_Bar_H2_HCO_CO} shows for the southwestern part of the Bar NIRCAM map of the 0-0 S(9) H$_2$ line at 4.69 $\mu$m, Keck/NIRC2 map of the 1-0 S(1) H$_2$ line at 2.12 $\mu$m, and ALMA maps of the HCO$^+$ J=4-3 and CO J=3-2 lines. 

\clearpage

\clearpage

\section{NIRCam images of the north of the Dark Bay, north of M42 and M43}
\label{Appendix:north-nebula-morphology}

In this Appendix, we present several prominent features arising in the  images within the fields centered on north of the Dark Bay, north of M42 and M43 obtained in the NIRCam parallel mode.
Figs. \ref{fig:composite-image-North-Orion-nebula}, \ref{fig:composite-image-Center_South_M43} and \ref{fig:composite-image-Center_M43}.
show composite images in three selected filters F187N, F335M and F470N in north of the Dark Bay and F187N, F335M and F212N in north of M42 and M43. As in the field centered on the Bar, the molecular cloud border appears quite structured. Multiple patterns emerge such as ridges, waves, evaporating globules and bow-shocks (Fig. \ref{fig:zoom-image-North-Orion-nebula-M43}). 
In Figs.~\ref{fig:Annexe_Orion_north_all_filters_1}-\ref{fig:Annexe_Center_M43_all_filters_2}, 
all the images obtained in the filters listed in Table \ref{tab:filters}, and some filters continuum subtracted, are presented. 
Figs.~\ref{fig:proplyds_M42A}, \ref{fig:new_proplyds_M42A},
\ref{fig:giant_proplyd_M43},
\ref{fig:new_proplyd_M43} show the proplyds detected in the north of the Dark Bay and M43.
 New proplyd candidates identified in the NIRCam F187N images of the north of the Dark Bay and M43
 are given in Table \ref{tab:new_proplyds}.

\subsection{North of the Dark Bay}
\label{sec:M42-North}

\begin{figure*}[h!]
\begin{center}
\vspace*{0cm}
\includegraphics[width=0.8\textwidth]{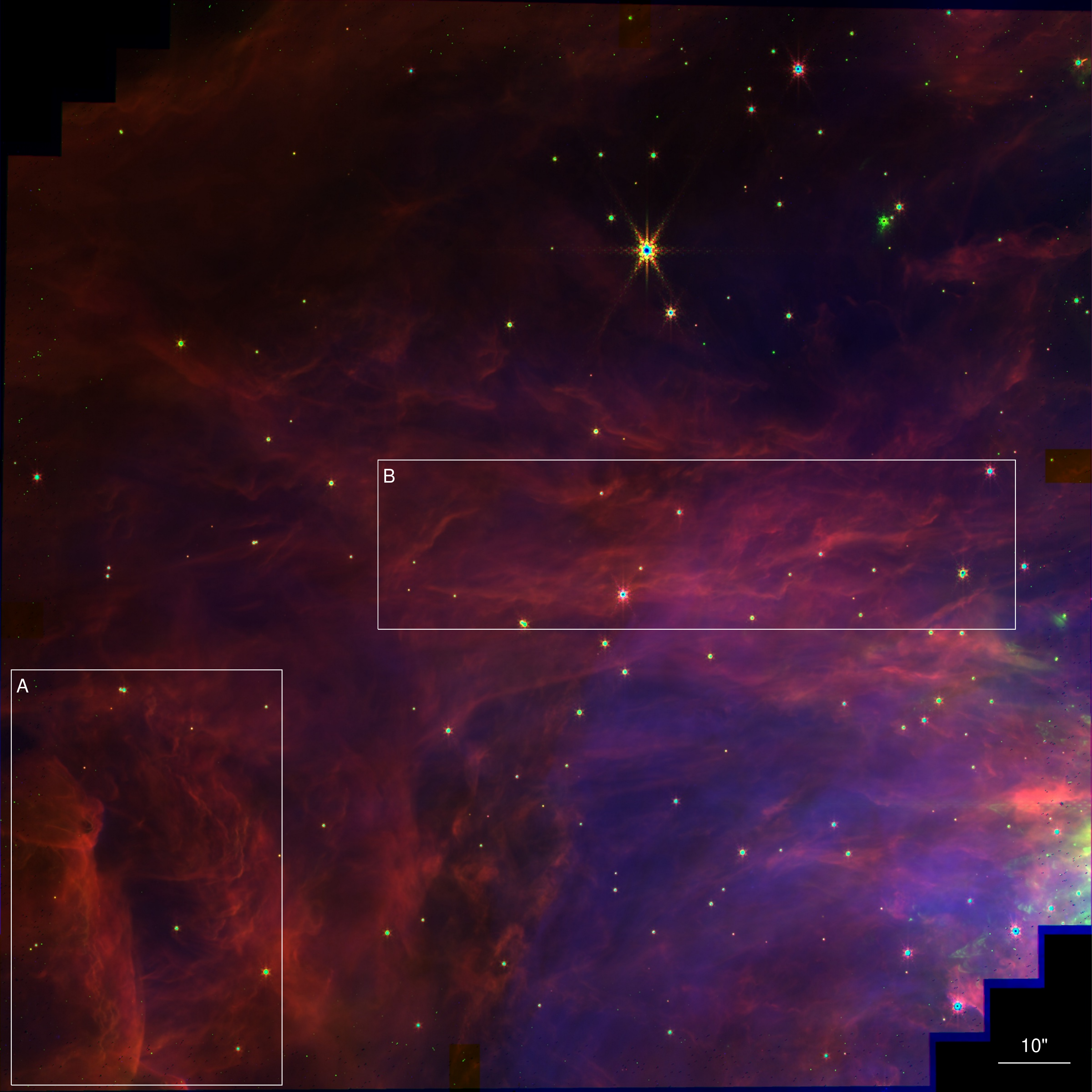}
\includegraphics[width=0.3\textwidth]{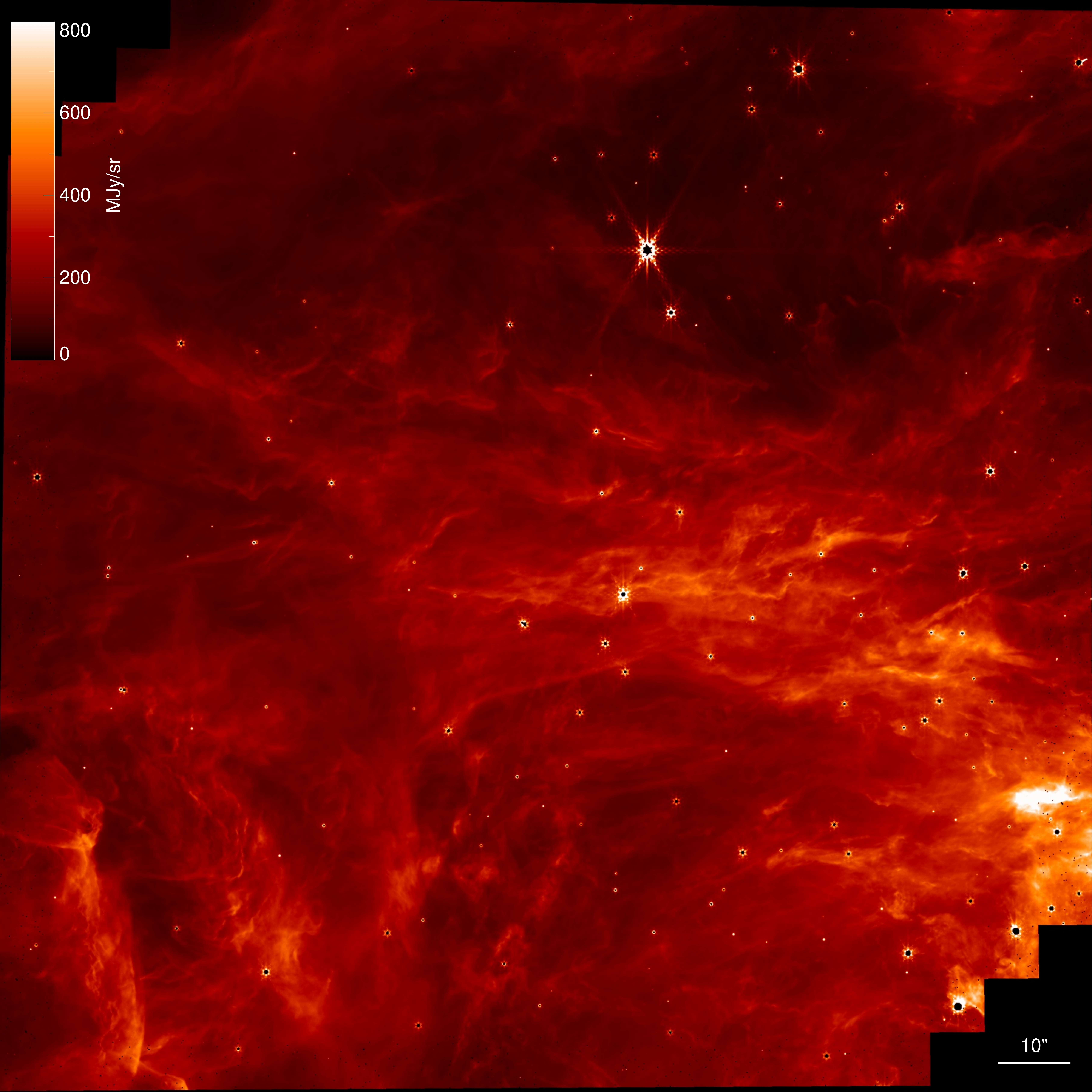}
\includegraphics[width=0.3\textwidth]{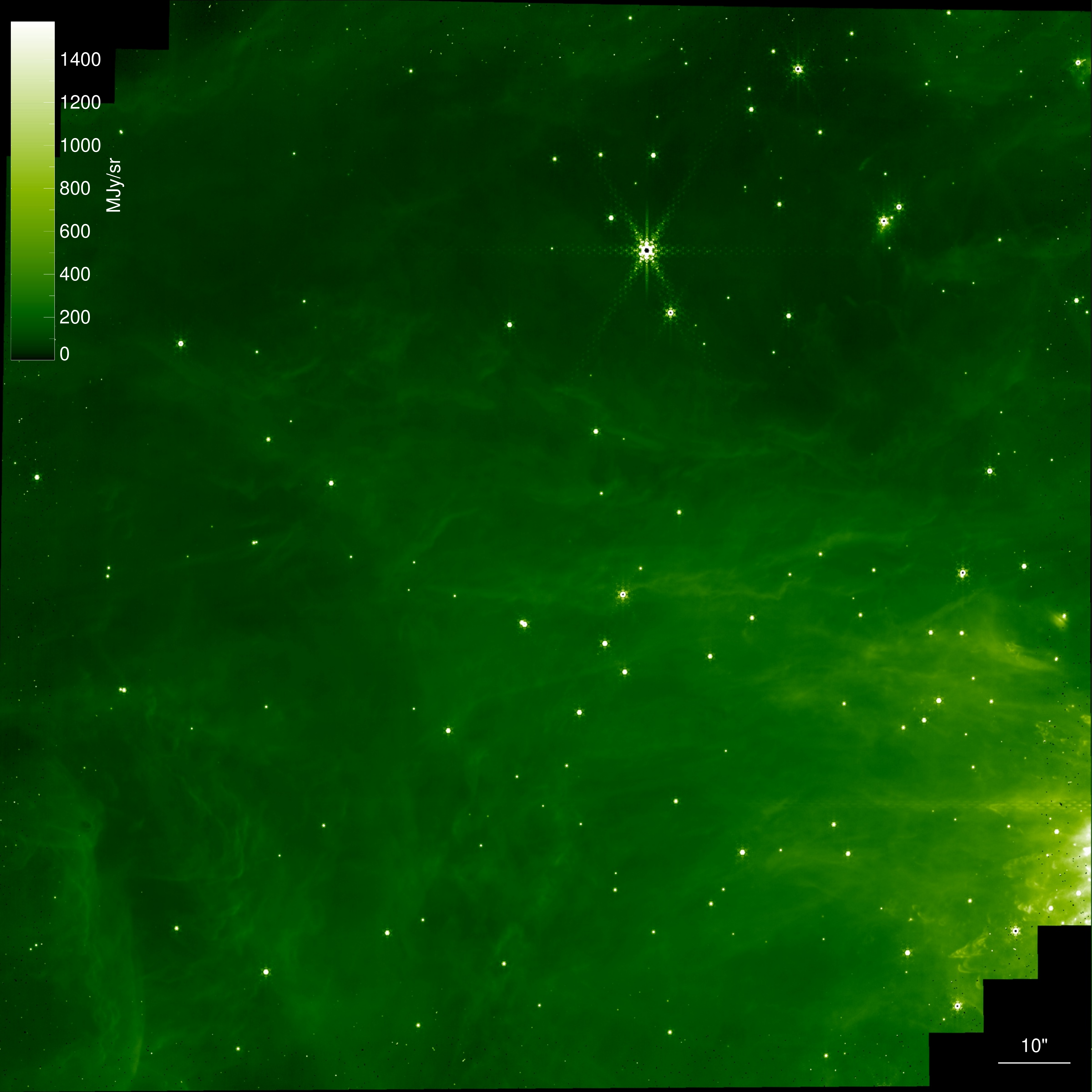}
\includegraphics[width=0.3\textwidth]{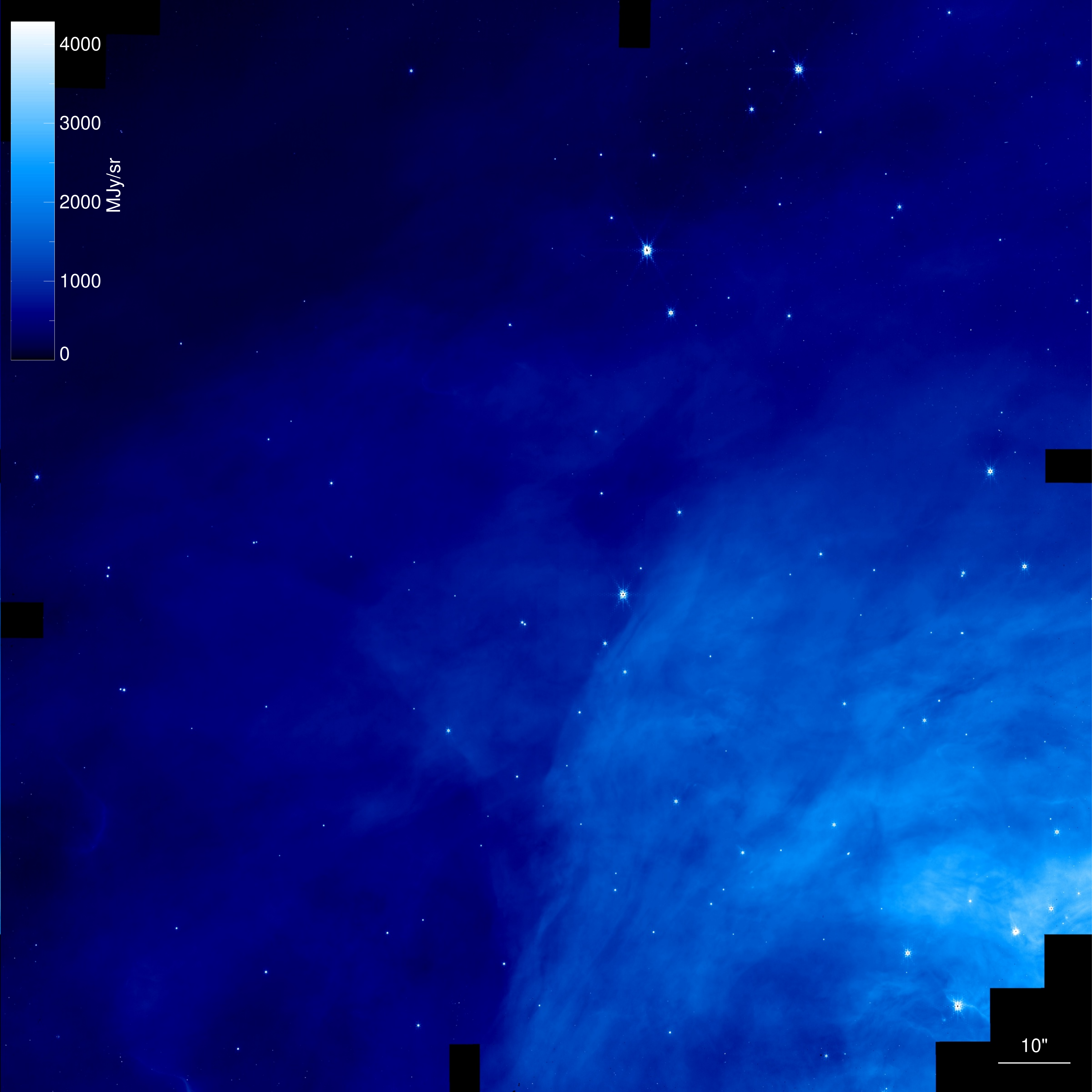}
\caption{The North of the Dark Bay as seen by the JWST’s NIRCam instrument with north up and east left. Several images in different filters were combined to produce an RGB composite image as in Fig. \ref{fig:composite-image-inside-Orion-nebula}: F335M (red), F470N (green) and F187N (blue)
that trace emission from hydrocarbons (AIBs), dust and molecular gas (H$_2$ 0-0 S(9) line), and  
ionized gas (Pa $\alpha$ line), respectively.
The individual images used to make the RGB composite one are shown below. No continuum was subtracted. The size of the images is $\sim$150$\arcsec \times 150 \arcsec$ and it is centered on RA=05$^{\rm h}$35$^{\rm m}$21$\fs$09, DEC=-05$\degr$21$\arcmin$40$\arcsec$01.  White boxes (labeled A, B)  are enlarged in Fig. \ref{fig:zoom-image-North-Orion-nebula-M43}.}
\label{fig:composite-image-North-Orion-nebula}
\end{center}
\end{figure*}

The NIRCam images of the material North of the Dark Bay reveal that there are linear bright features, such as the Bar (panels A and B in Figs.~\ref{fig:composite-image-North-Orion-nebula} and \ref{fig:zoom-image-North-Orion-nebula-M43}). Some of these features were previously seen in extinction corrected HST images \citep[e.g., eastern and northern bright bar in Fig. 6 of][]{Odell01}.  
The Dark Bay (and similar structures) are foreground extinction of visible light (Fig. \ref{fig:hst_orion_nebula}) but seen in emission in PDRs tracers, such as AIBs and H$_2$ observed with NIRCam, and [CII] cooling line emission from \textit{Herschel} and SOFIA \citep{goicoechea2015,pab21PhD}.  
The AIBs and H$_2$ line emission show the edges of these neutral warm regions with a very high level of detail. The bright PDR features (panels A and B in Figs. \ref{fig:composite-image-North-Orion-nebula} and \ref{fig:zoom-image-North-Orion-nebula-M43}), which are at a projected distance of the order of 110$\arcsec$-120$\arcsec$ (0.22-0.24~pc) from the illuminating \mbox{O7-type} star \mbox{$\theta^1$ Ori C}, are illuminated by a  strong FUV radiation field comparable to the one on the   Bar.
\cite{goicoechea2015} studied with \textit{Herschel}/HIFI the [CII] velocity emission along with the CO J=2-1 line with IRAM-30m 
and found that most of these PDRs lack detectable CO emission (CO-dark molecular gas).
In the northern-west part of the NIRCam map, there is a zone devoid of nebula emission where background stars are well visible. This is in agreement with the fact that no CO emission was previously detected and that this region does not contain much dense molecular gas. 

In the southwest corner of the continuum subtracted H$_2$ 0-0 S(9) 4.69 $\mu$m line map (Fig. \ref{fig:Annexe_Orion_north_all_filters_3}), several very bright features produced by shocks are detected. No AIB emission or Pa $\alpha$ and [FeII] line emission are detected on these features. 
The well studied outflow HH 513 lies in this region. 
Furthermore, several arches (like a bent-over cylinder) which are most likely bow-shocks are detected in the southest corner of field (see panel A in Figs. \ref{fig:composite-image-North-Orion-nebula} and \ref{fig:zoom-image-North-Orion-nebula-M43}). 
Bow-shocks can also be the result of multiple outflows that are close together in time like seen from the ground in clusters in various star-forming regions \citep{Plunkett2013,Nony2020}.

\subsection{North of M42}
\label{sec:M43-South}

\begin{figure*}[h!]
\begin{center}
\vspace*{0cm}
\includegraphics[width=0.8\textwidth]{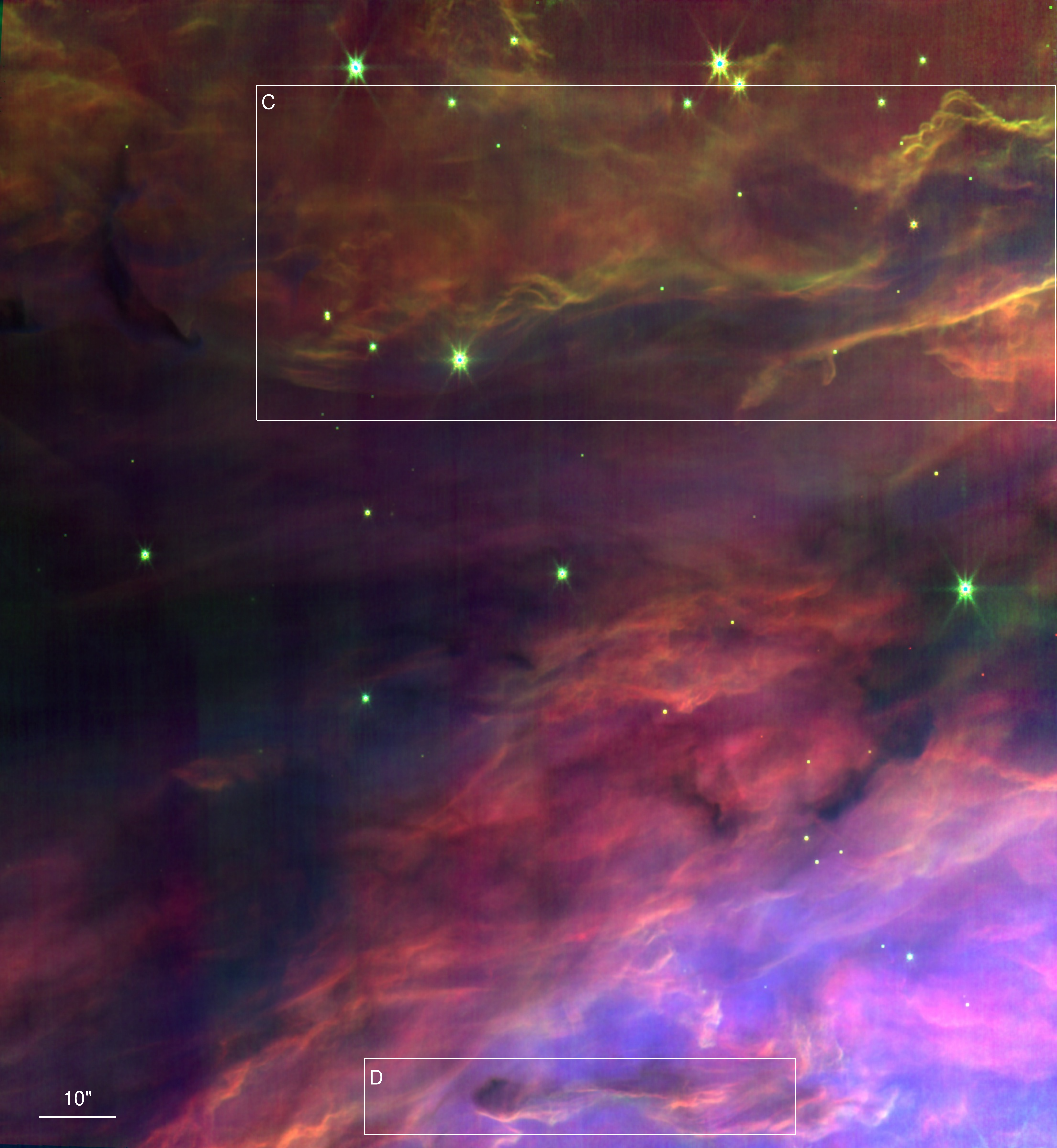}
\includegraphics[width=0.3\textwidth]{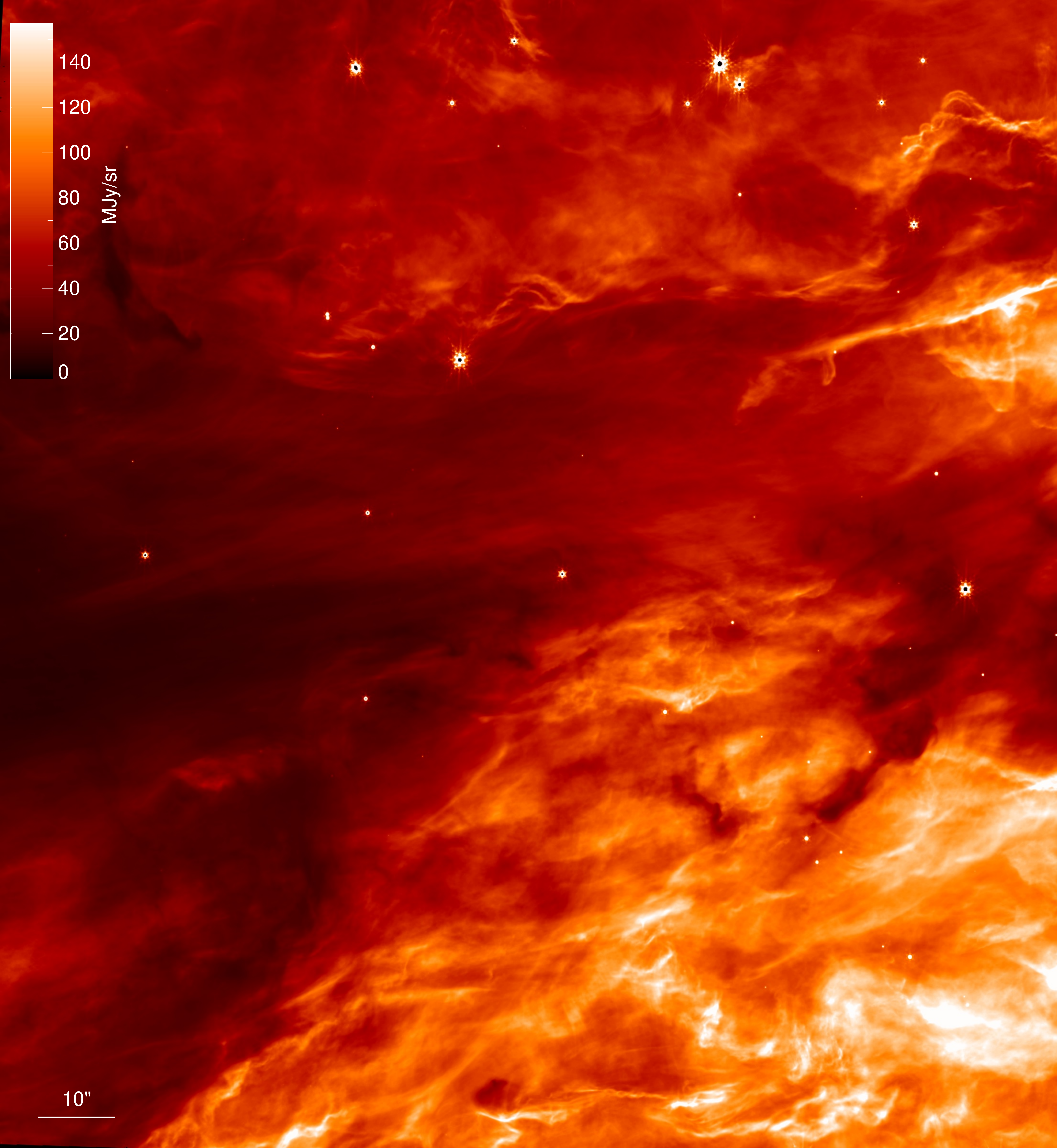}
\includegraphics[width=0.3\textwidth]{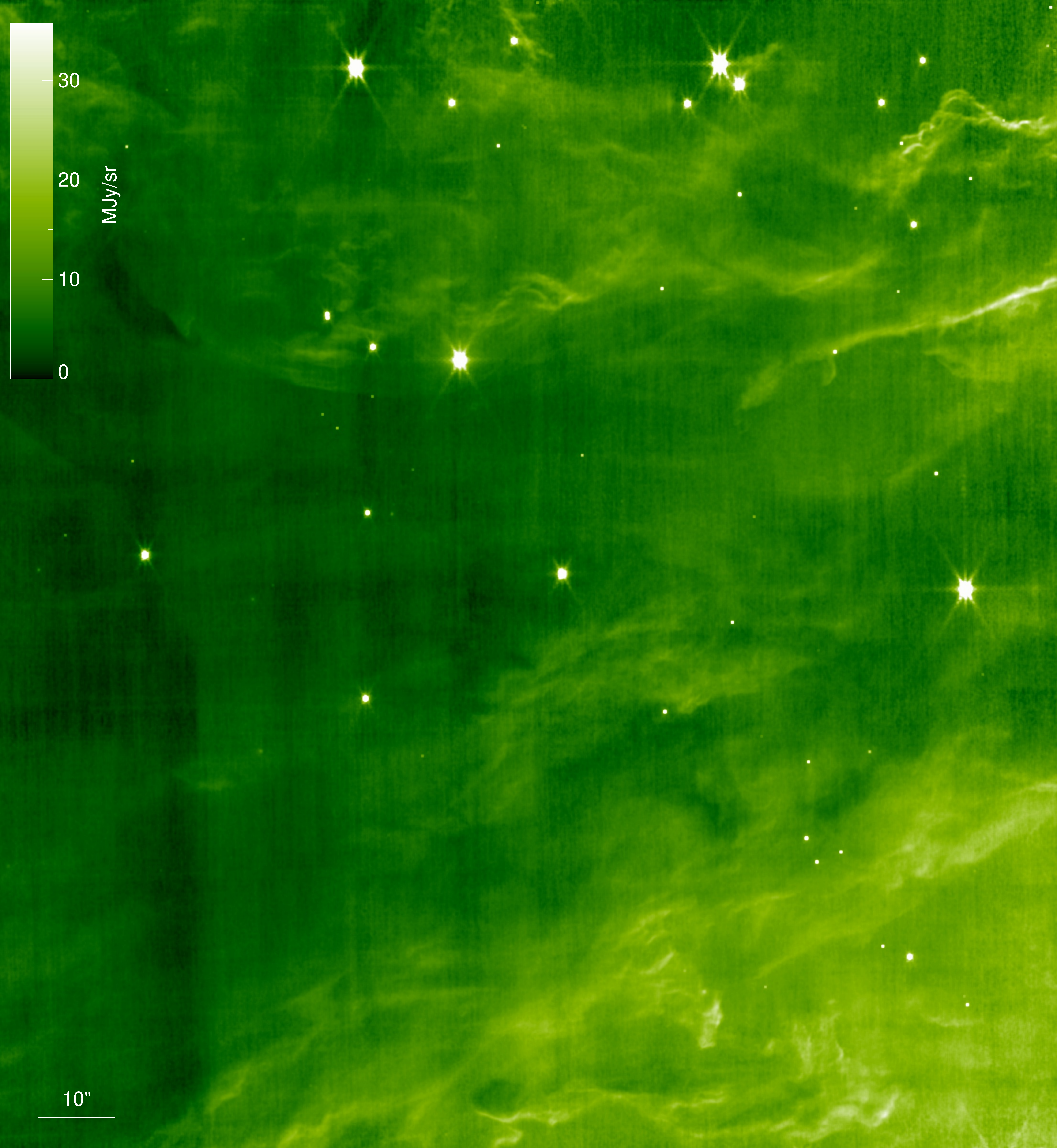}
\includegraphics[width=0.3\textwidth]{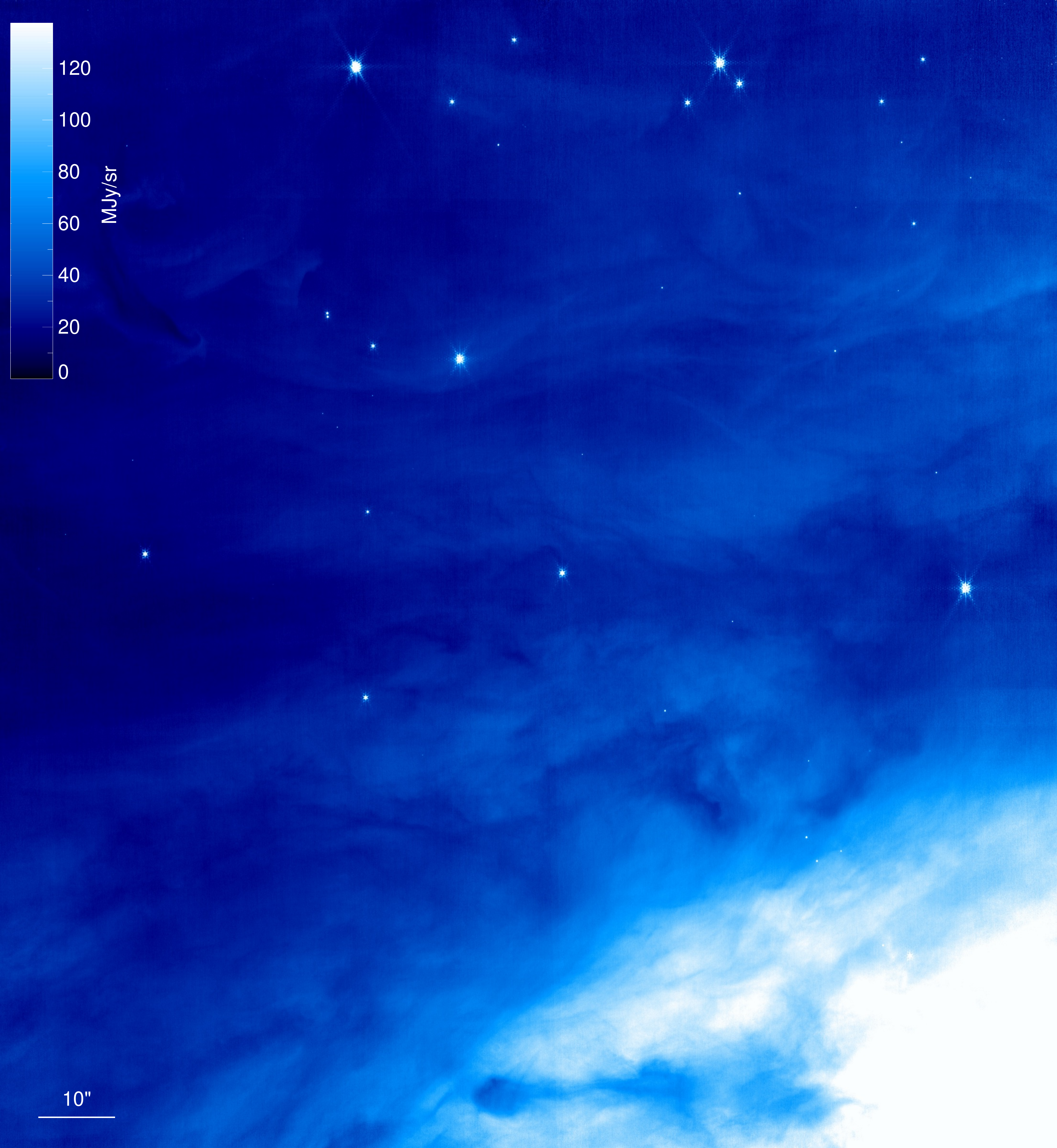}

\caption{North of M42 as seen by the JWST’s NIRCam instrument with north up and east left. Several images in different filters were combined to produce an RGB composite image : F335M (red), F212N (green) and F187N (blue) that traces respectively emission from hydrocarbons (AIBs),  molecular gas (H$_2$ 1-0 S(1) line) and ionized gas (Pa $\alpha$ line), respectively. 
The individual images used to make the RGB composite one are shown below. No continuum was subtracted. The image field was chosen to have an overlap between the different individual images. The size of the images is $\sim$135$\arcsec \times 145 \arcsec$ and it is centered on RA=05$^{\rm h}$35$^{\rm m}$29$\fs$12, DEC=-05$\degr$20$\arcmin$3.66$\arcsec$. White boxes (labeled C, D)  are enlarged in Fig. \ref{fig:zoom-image-North-Orion-nebula-M43}.
}
\label{fig:composite-image-Center_South_M43}
\end{center}
\end{figure*}

\begin{figure*}[h!]
\begin{center}
\vspace*{0cm}
\begin{minipage}{0.355\textwidth}
\includegraphics[width=\textwidth]{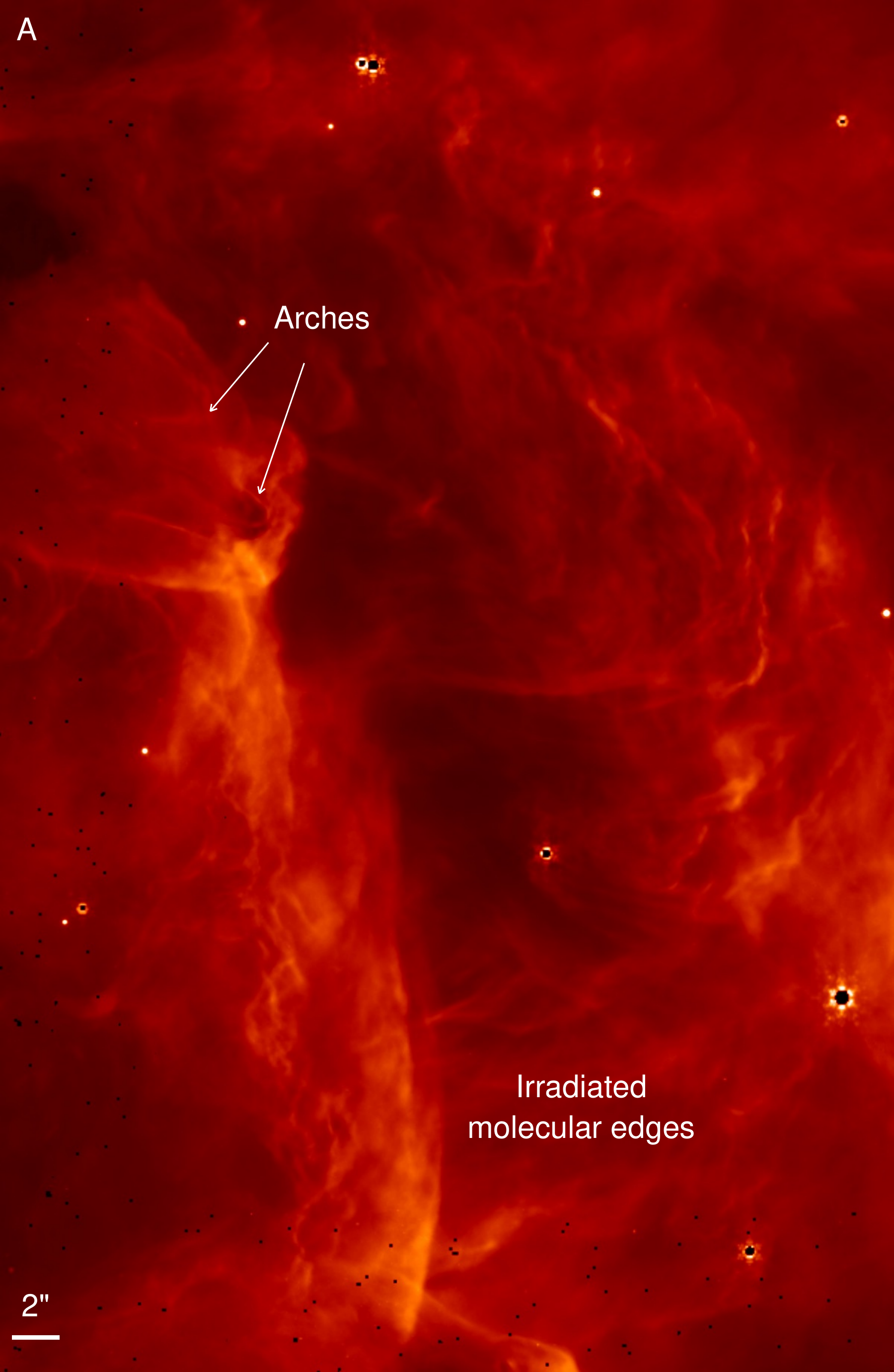}
\end{minipage}
\begin{minipage}{0.625\textwidth}
\includegraphics[width=\textwidth]{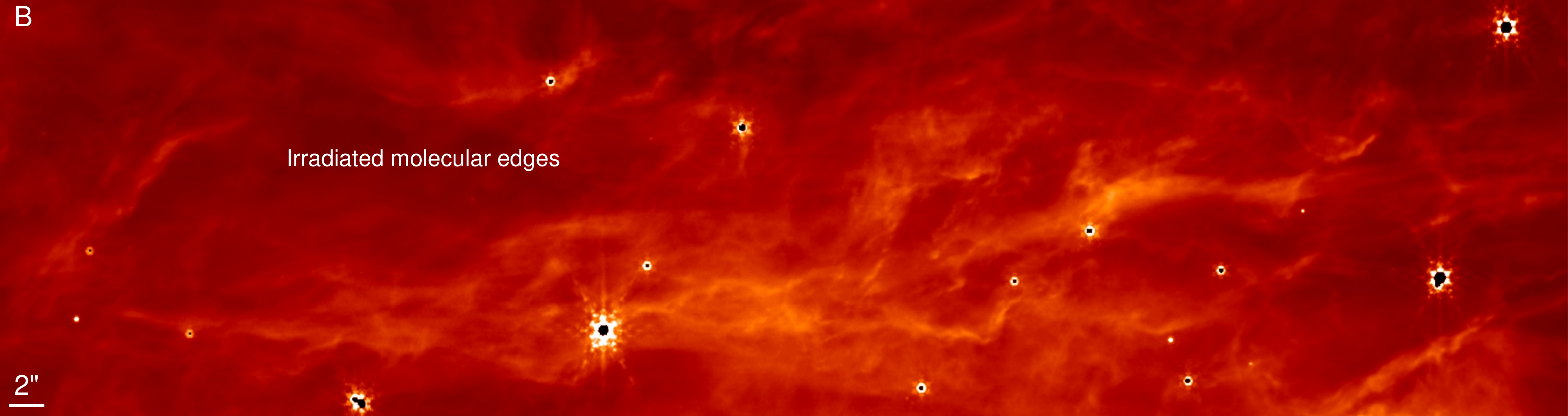}
\includegraphics[width=\textwidth]{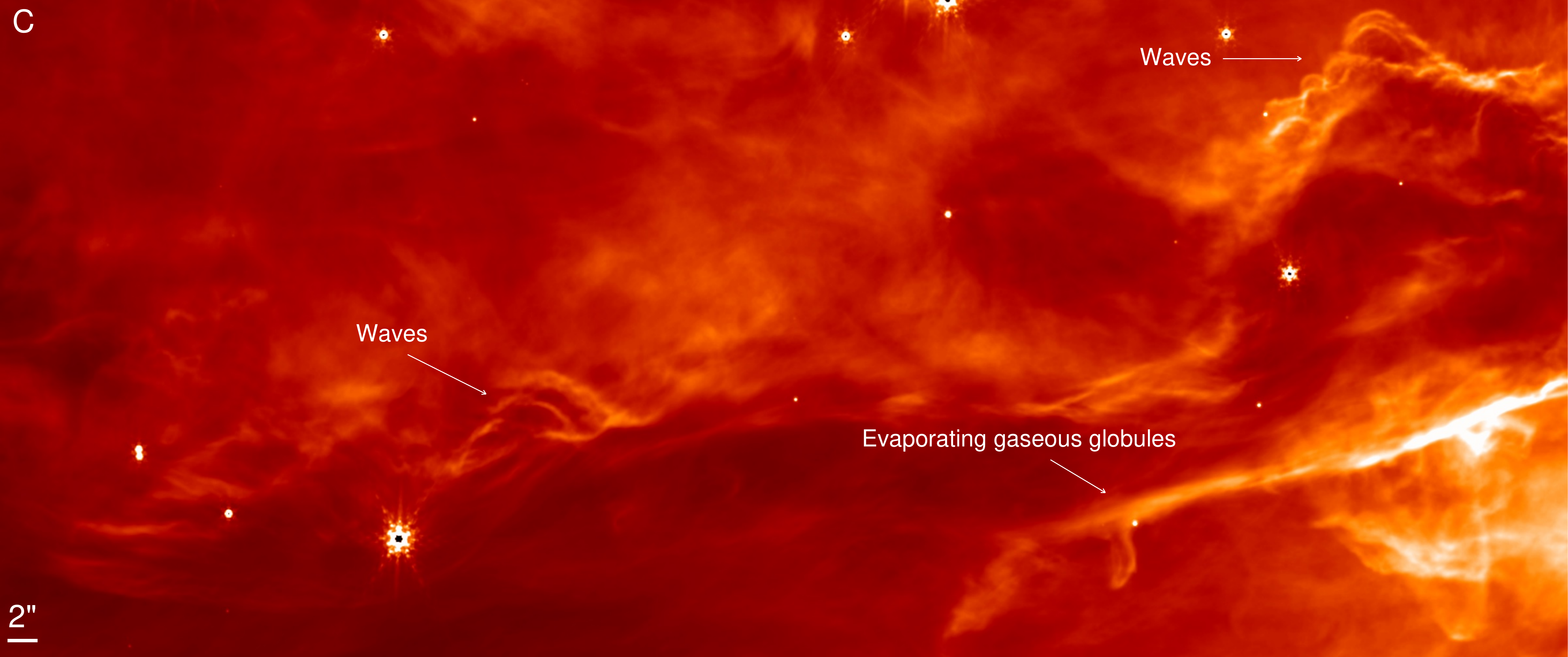}
\includegraphics[width=\textwidth]{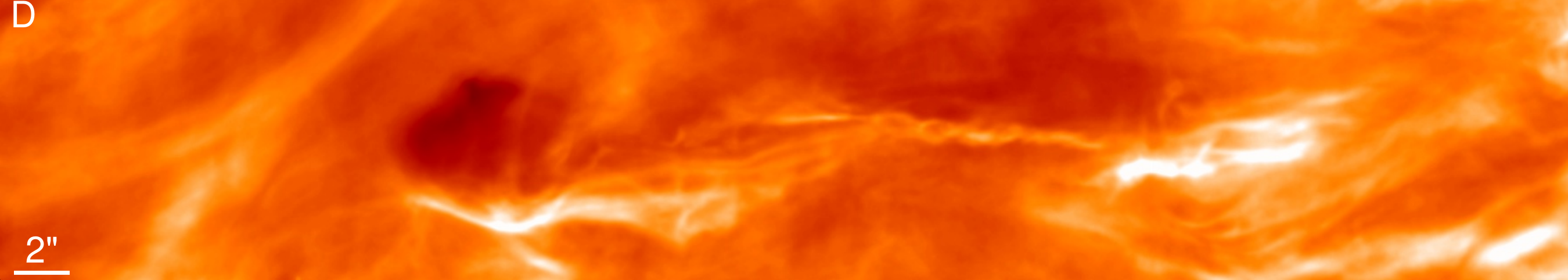}
\end{minipage}

\caption{Zoom into the F335M image areas shown in Figs. \ref{fig:composite-image-North-Orion-nebula} (for panels A, B) and \ref{fig:composite-image-Center_South_M43} (for panels B, C). {\bf A} Irradiated molecular edges showing many patterns such as ridges and arches (which are most likely bow-shocks) detected in the southeast area of M42 North. {\bf B} Irradiated molecular edges detected in the north area of M42 North. {\bf C} Irradiated molecular edges showing many patterns such as ridges, waves and evaporating gaseous globule detected in the northwest area of the M43 South field.
{\bf D} Features driven by a large outflow detected in the south area of the M43 South field.}
\label{fig:zoom-image-North-Orion-nebula-M43}
\end{center}
\end{figure*}

Here, we present  images within the fields covered the north of M42 obtained in the NIRCam parallel mode. 
As shown by the ionized gas line map in Fig. \ref{fig:composite-image-Center_South_M43}, the north of M42 contains one main ionized gas region in the southwest part of the map. This region is irradiated by the Trapezium  cluster. 
At the edge of the ionized region are the neutral zones, the PDRs.
Particularly interesting PDRs are the ones in panel C in Figs. \ref{fig:composite-image-Center_South_M43} and \ref{fig:zoom-image-North-Orion-nebula-M43} which are irradiated from the South by star $\theta ^1$ Ori C and from the north by the B0.5V star NU Ori in the center of M43 (see Sect. \ref{sec:M43-North}).
As shown in Fig. \ref{fig:composite-image-Center_South_M43}, a remarkable spatial correlation is observed between the F335M filter (AIBs) and the F212N filter (vibrationally excited line H$_2$ 1-0 S(1)).
 Most likely, these PDRs are in a lower $G_0/n_H$ regime than in the   Bar. In fact, the incident FUV flux from the south on that PDR can be estimated to be about 10 times lower than in the   Bar considering the projected distance from the star $\theta ^1$ Ori 
 C. 
 For lower $G_0/n_H$, the H$^0$/H$_2$ transition is not driven by dust opacity but  by H$_2$ self shielding  \cite[e.g,][]{Hollenbach99}, and the H$_2$ emission peak is expected to spatially coincide with the emission peak of the AIBs. Spatial coincidence between H$_2$ and AIB emission peaks has been observed in several lower excited PDRs such as the $\rho$ Oph-W and Horsehead nebula PDR \citep[e.g.,][]{habart_h_2003,habart05}.
In high $G_0/n_H$ regime (as at the main edge of the   Bar), where the H$^0$/H$_2$ is driven by dust opacity and H$_2$ is mostly photo-dissociated in the outer PDR layers ($A_V<1$), H$_2$ emission appears after the emission peak of the AIBs.
In the panel C of Fig. \ref{fig:zoom-image-North-Orion-nebula-M43}, a thin structure (width of 1$\arcsec$), probably an Evaporating Gaseous Globule, is detected at the PDR edge.

\subsection{M43}
\label{sec:M43-North}

\begin{figure*}[h!]
\begin{center}
\vspace*{0cm}
\includegraphics[width=0.8\textwidth]{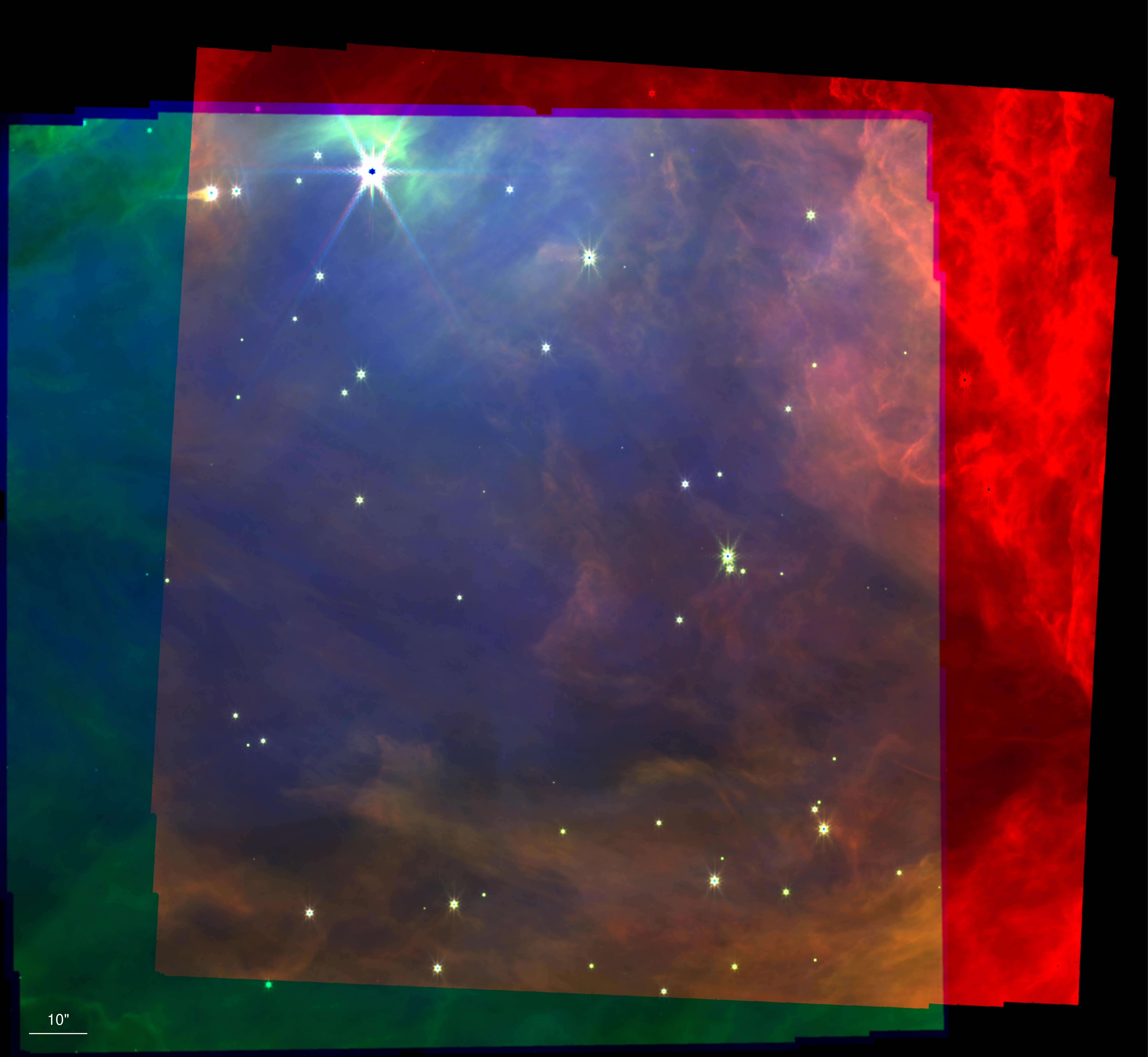}
\includegraphics[width=0.3\textwidth]{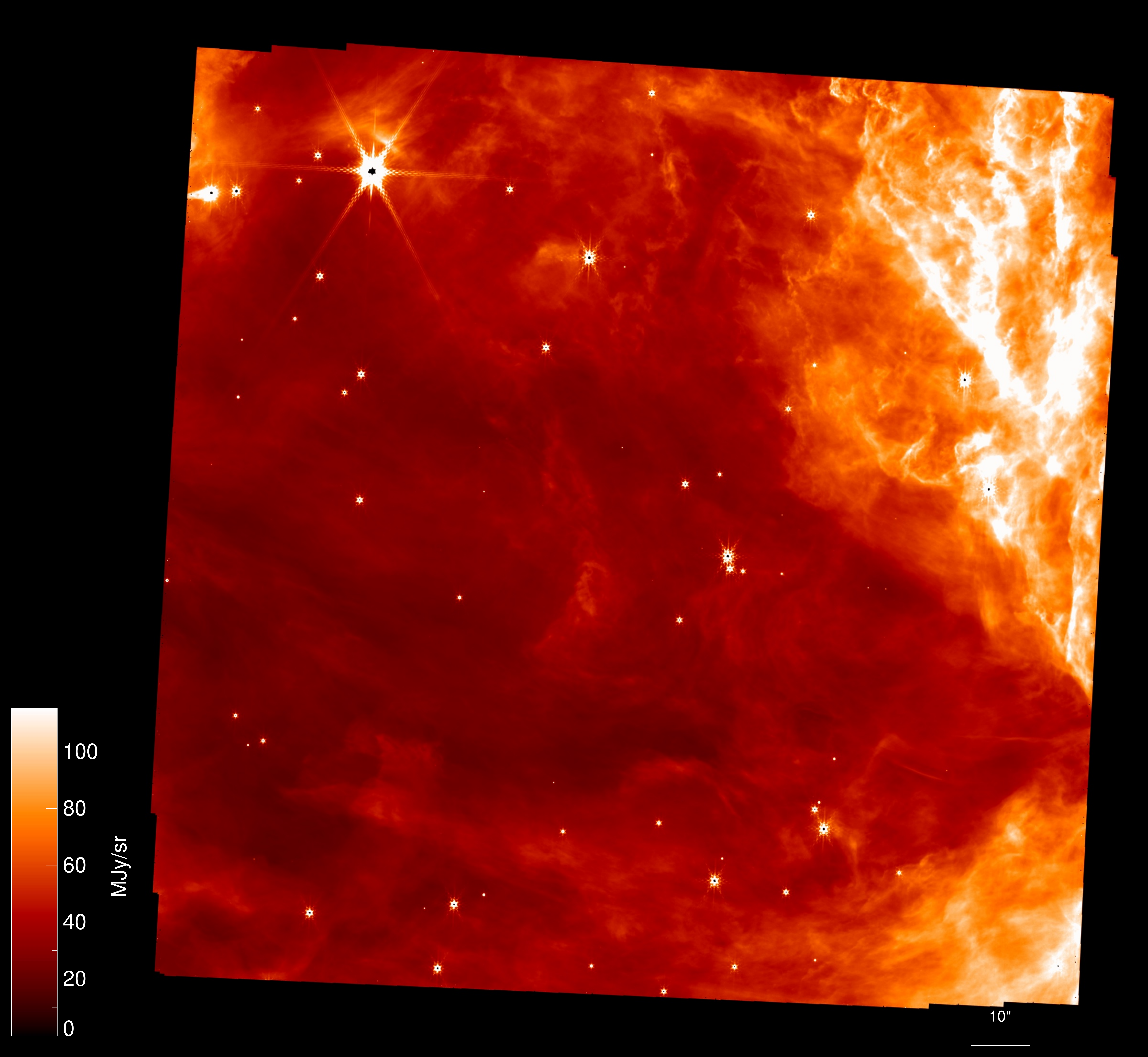}
\includegraphics[width=0.3\textwidth]{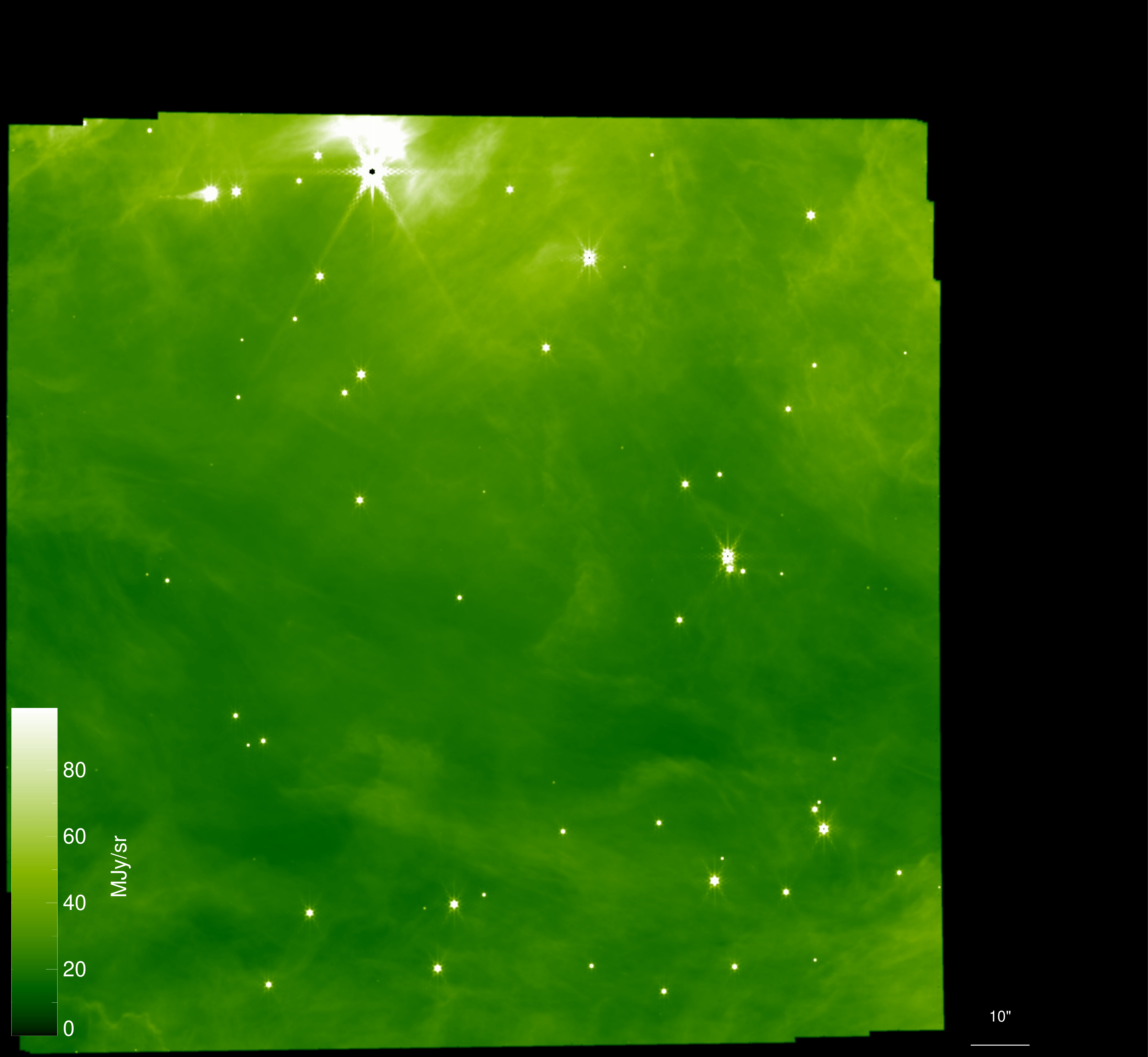}
\includegraphics[width=0.3\textwidth]{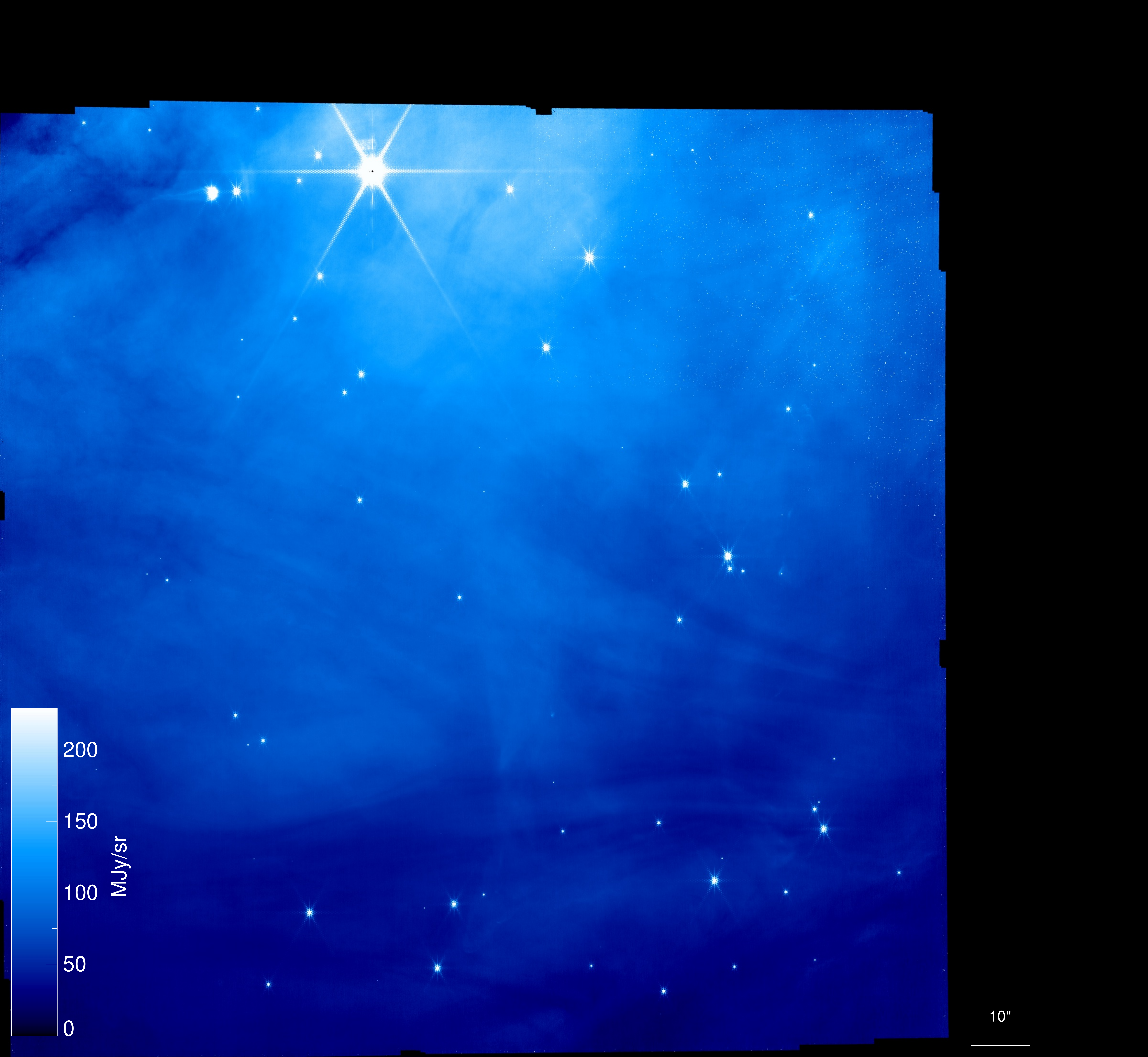}

\caption{The M43 region as seen by the JWST’s NIRCam instrument with north up and east left. Several images in different filters were combined to produce an RGB composite image : F335M (red), F212N (green) and F187N (blue) that traces respectively emission from hydrocarbons (AIBs),  molecular gas (H$_2$ 1-0 S(1) line) and ionized gas (Pa $\alpha$ line), respectively. 
The individual images used to make the RGB composite one are shown below. No continuum was subtracted. The size of the images is $\sim$158$\arcsec \times 158 \arcsec$  and it is centered on RA=05$^{\rm h}$35$^{\rm m}$28$\fs$92, DEC=-05$\degr$17$\arcmin$04.31$\arcsec$.  }
\label{fig:composite-image-Center_M43}
\end{center}
\end{figure*}

\begin{figure*}[h!]
\begin{center}
\vspace*{0cm}
\includegraphics[width=1\textwidth]{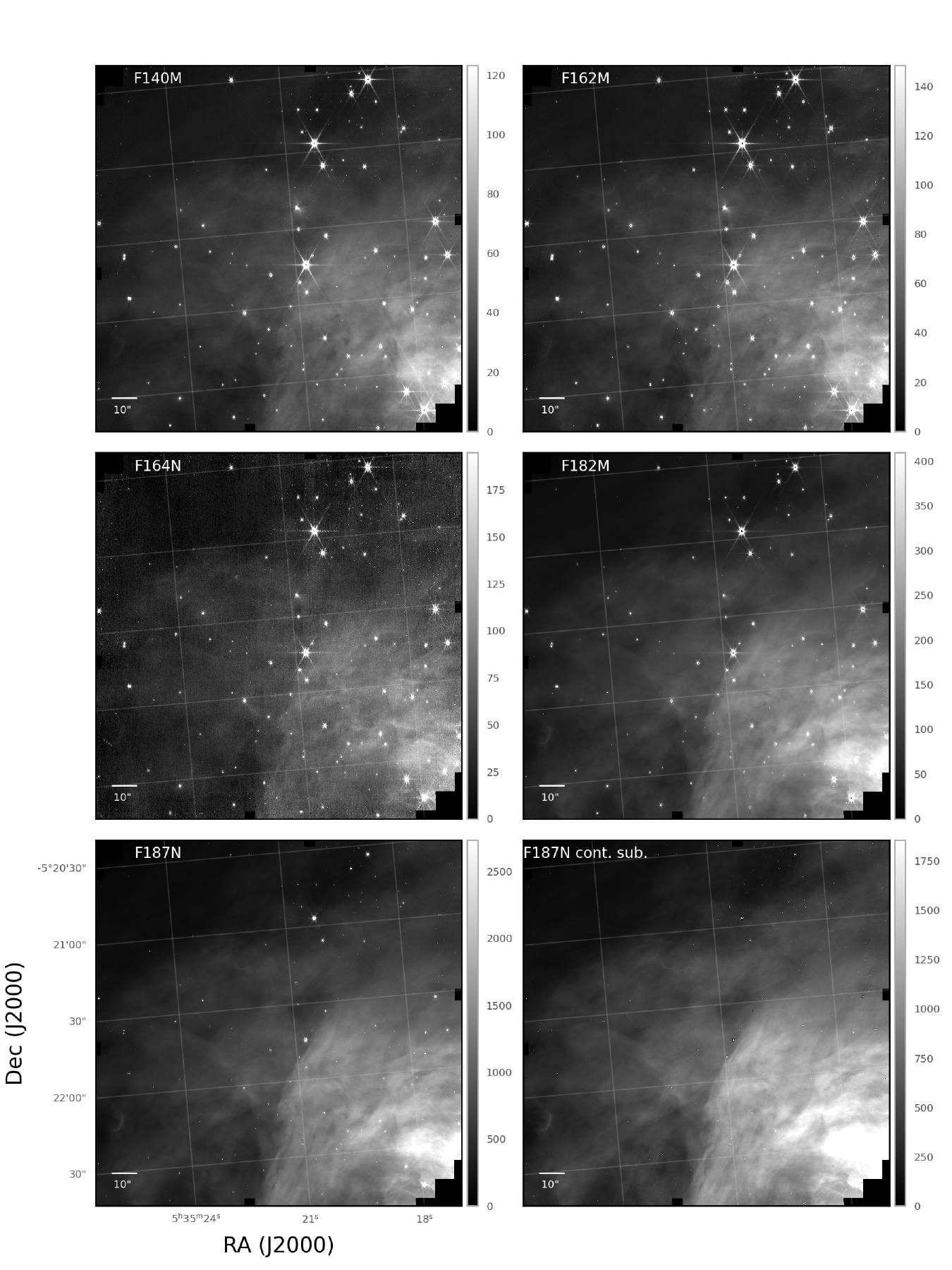}
\caption{NIRCam maps of the North region of  the Orion Nebula in different filters. Units are in MJy/sr.}
\label{fig:Annexe_Orion_north_all_filters_1}
\end{center}
\end{figure*}

\begin{figure*}[h!]
\begin{center}
\vspace*{0cm}
\includegraphics[width=1\textwidth]{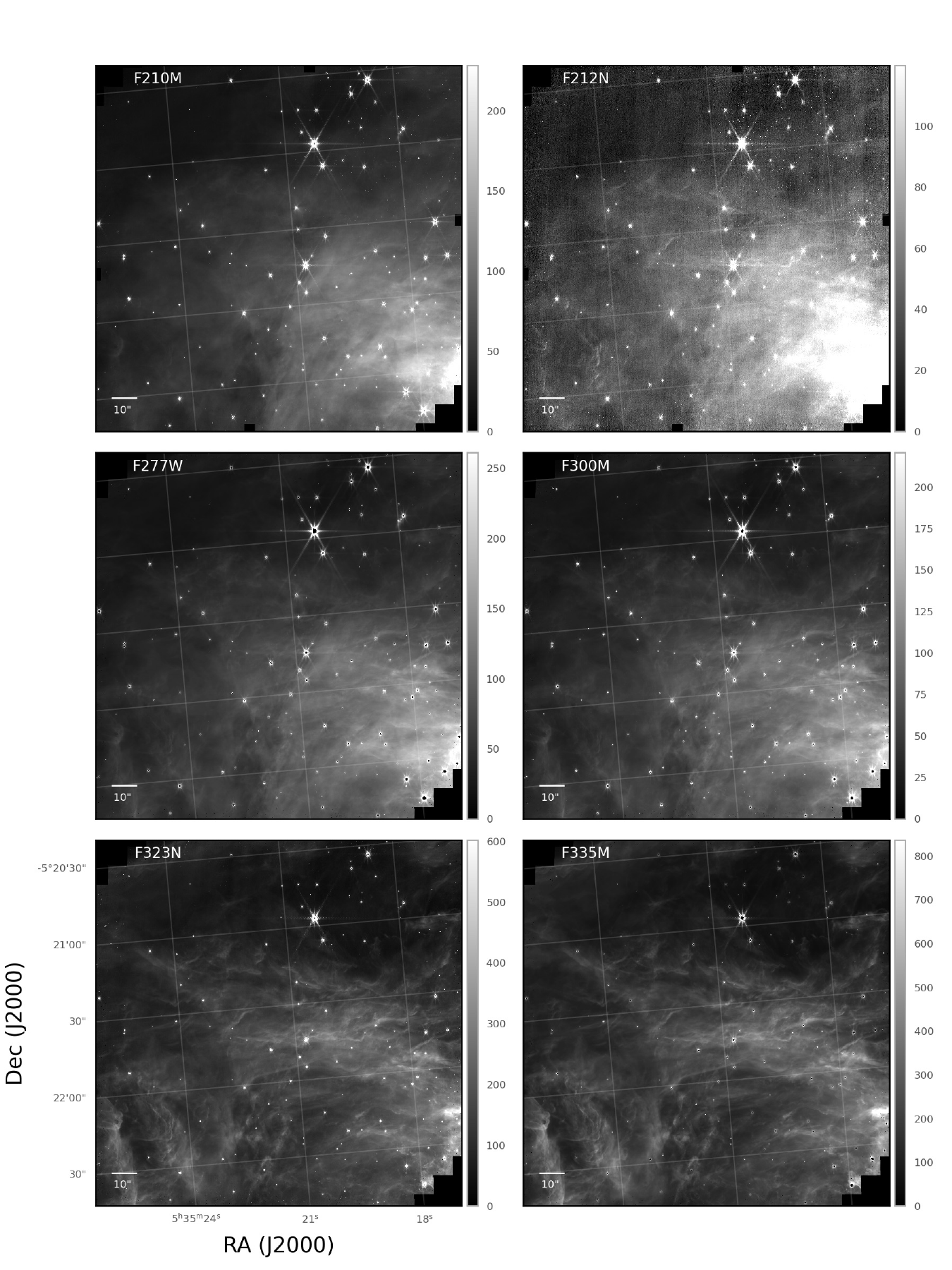}
\caption{Same as in Fig.~\ref{fig:Annexe_Orion_north_all_filters_1}, but for other filters.}
\label{fig:Annexe_Orion_north_all_filters_2}
\end{center}
\end{figure*}

\begin{figure*}[h!]
\begin{center}
\vspace*{0cm}
\includegraphics[width=1\textwidth]{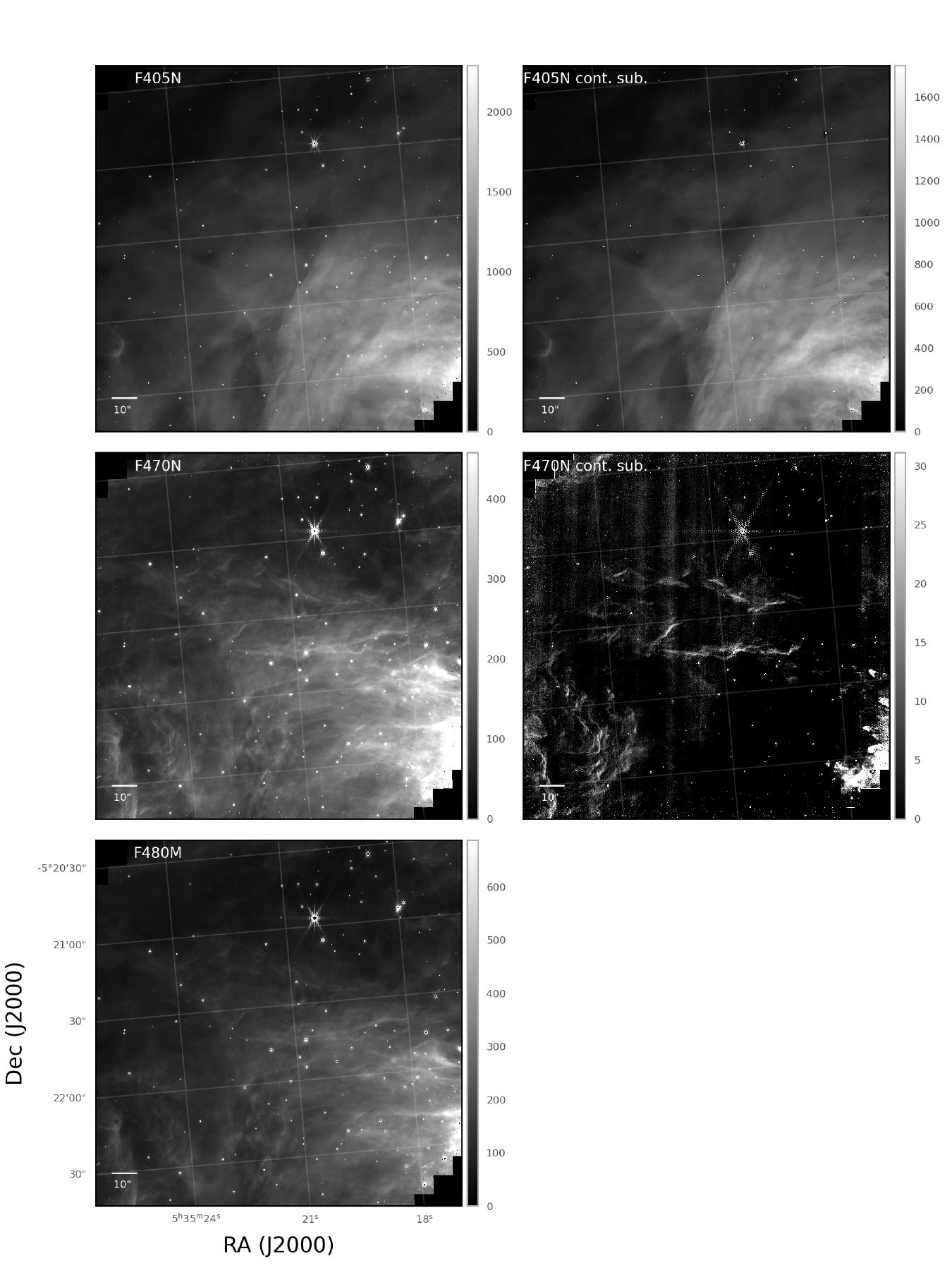}
\caption{Same as in Fig.~\ref{fig:Annexe_Orion_north_all_filters_1}, but for other filters.}
\label{fig:Annexe_Orion_north_all_filters_3}
\end{center}
\end{figure*}

\begin{figure*}[h!]
\begin{center}
\vspace*{0cm}
\includegraphics[width=1\textwidth]{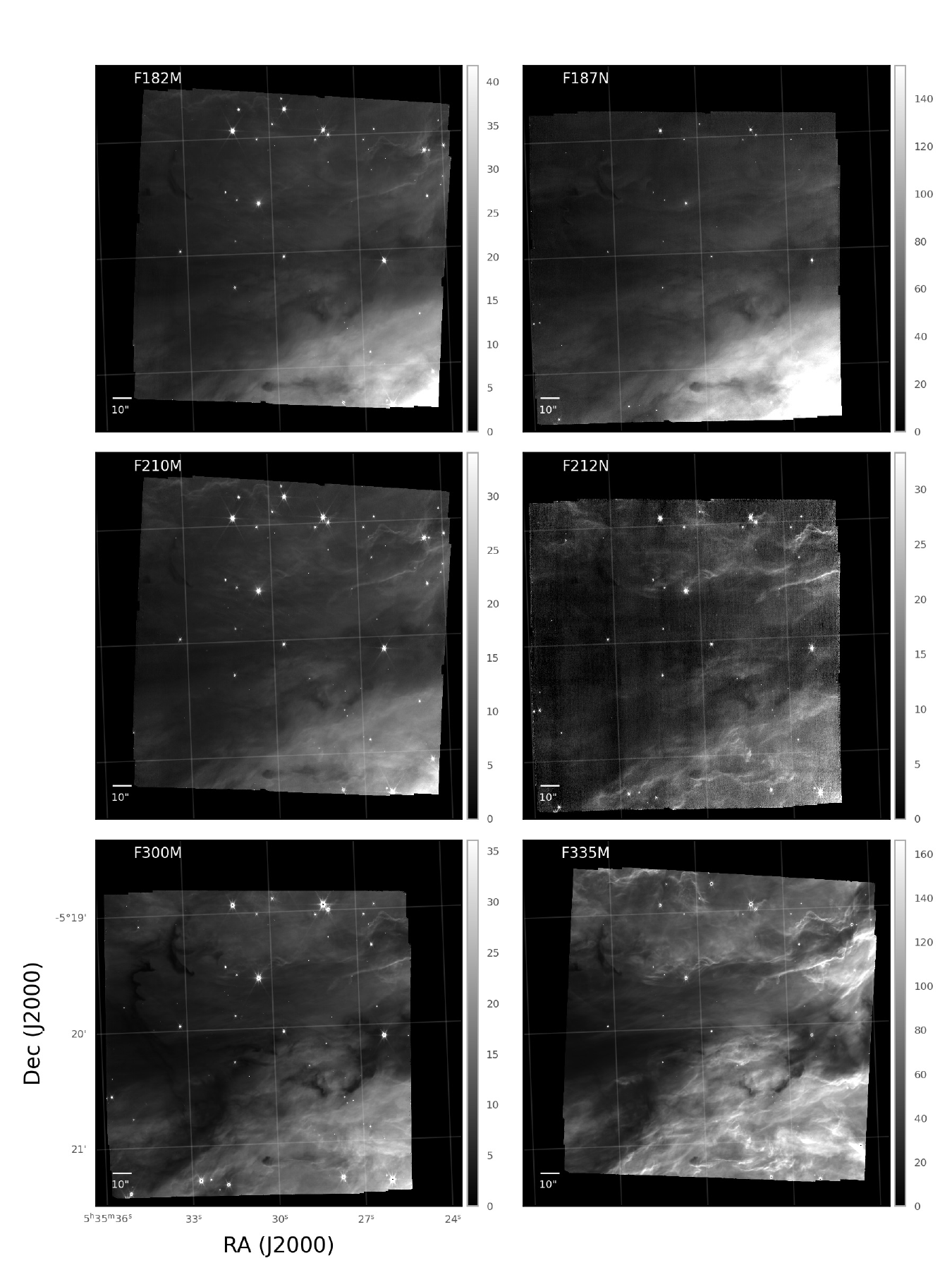}
\caption{NIRCam maps of the M43 South region in different filters. Units are in MJy/sr.}
\label{fig:Annexe_Center_south_M43_all_filters_1}
\label{fig}
\end{center}
\end{figure*}

\begin{figure*}[h!]
\begin{center}
\vspace*{0cm}
\includegraphics[width=1\textwidth]{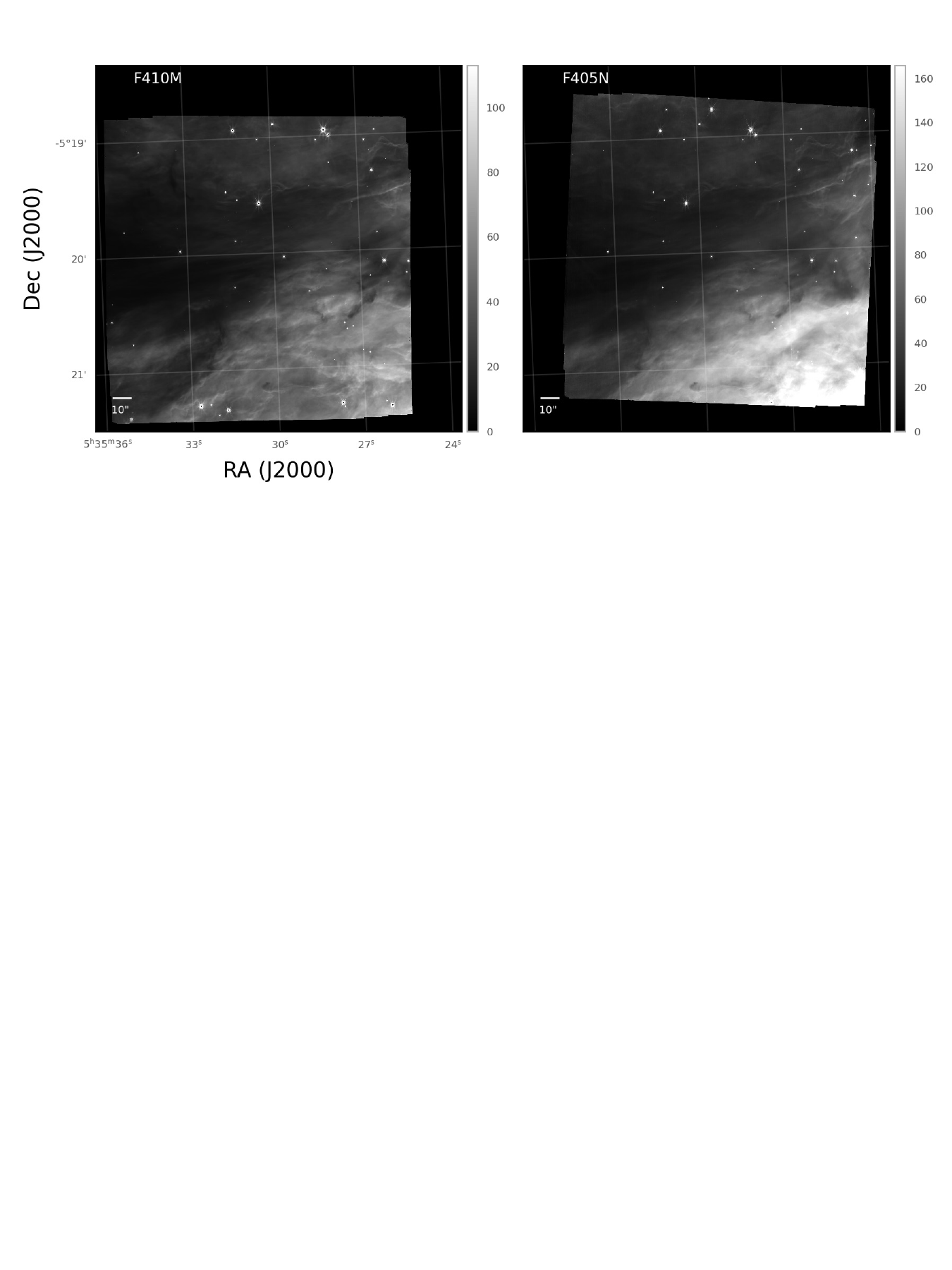}
\caption{Same as in Fig.~\ref{fig:Annexe_Center_south_M43_all_filters_1}, but for other filters.}
\label{fig:Annexe_Center_south_M43_all_filters_2}
\end{center}
\end{figure*}

\begin{figure*}[h!]
\begin{center}
\vspace*{0cm}
\includegraphics[width=1\textwidth]{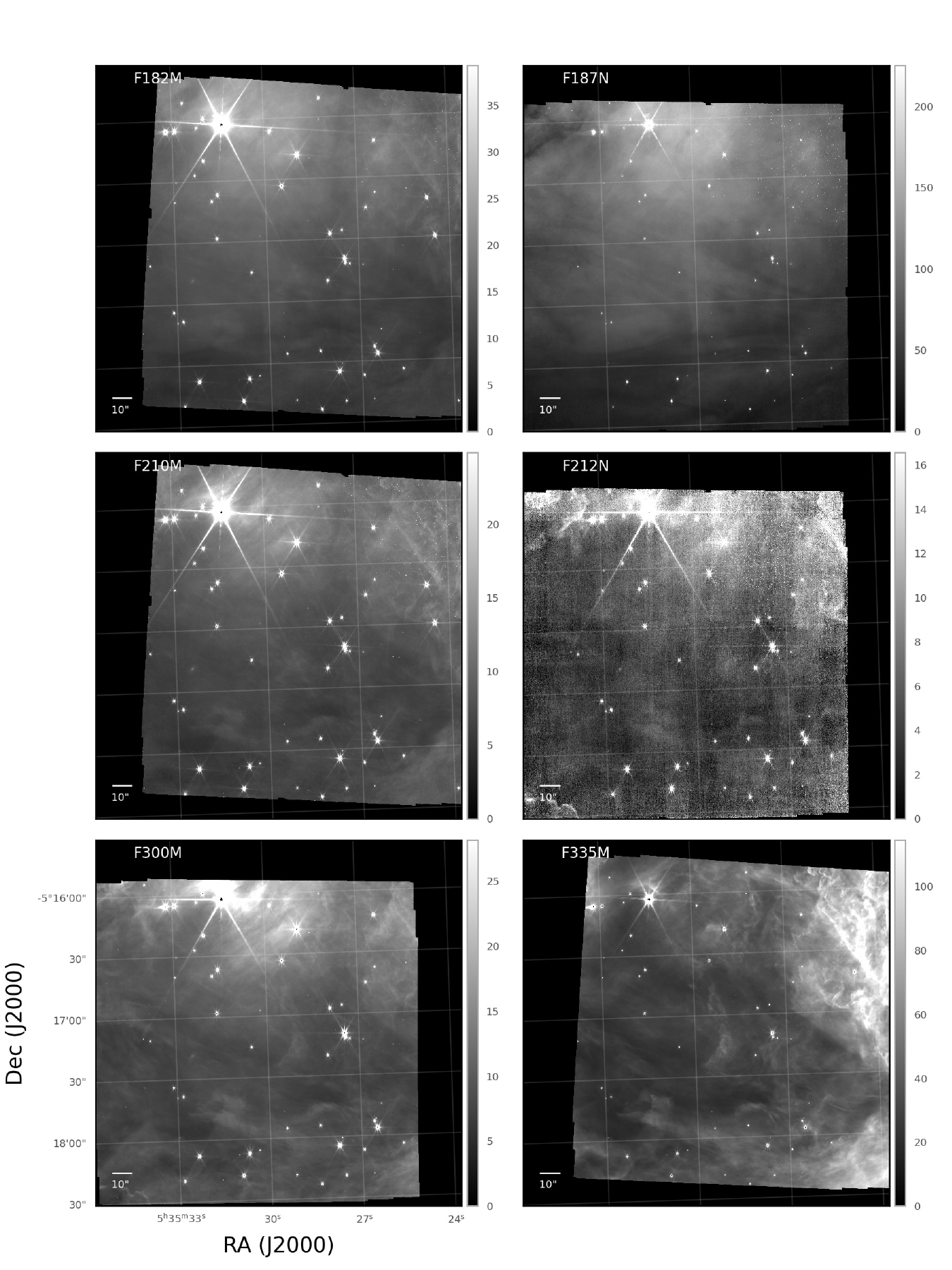}
\caption{NIRCam maps of the M43 center region in different filters. Units are in MJy/sr.}
\label{fig:Annexe_Center_M43_all_filters_1}
\end{center}
\end{figure*}

\begin{figure*}[h!]
\begin{center}
\vspace*{0cm}
\includegraphics[width=1\textwidth]{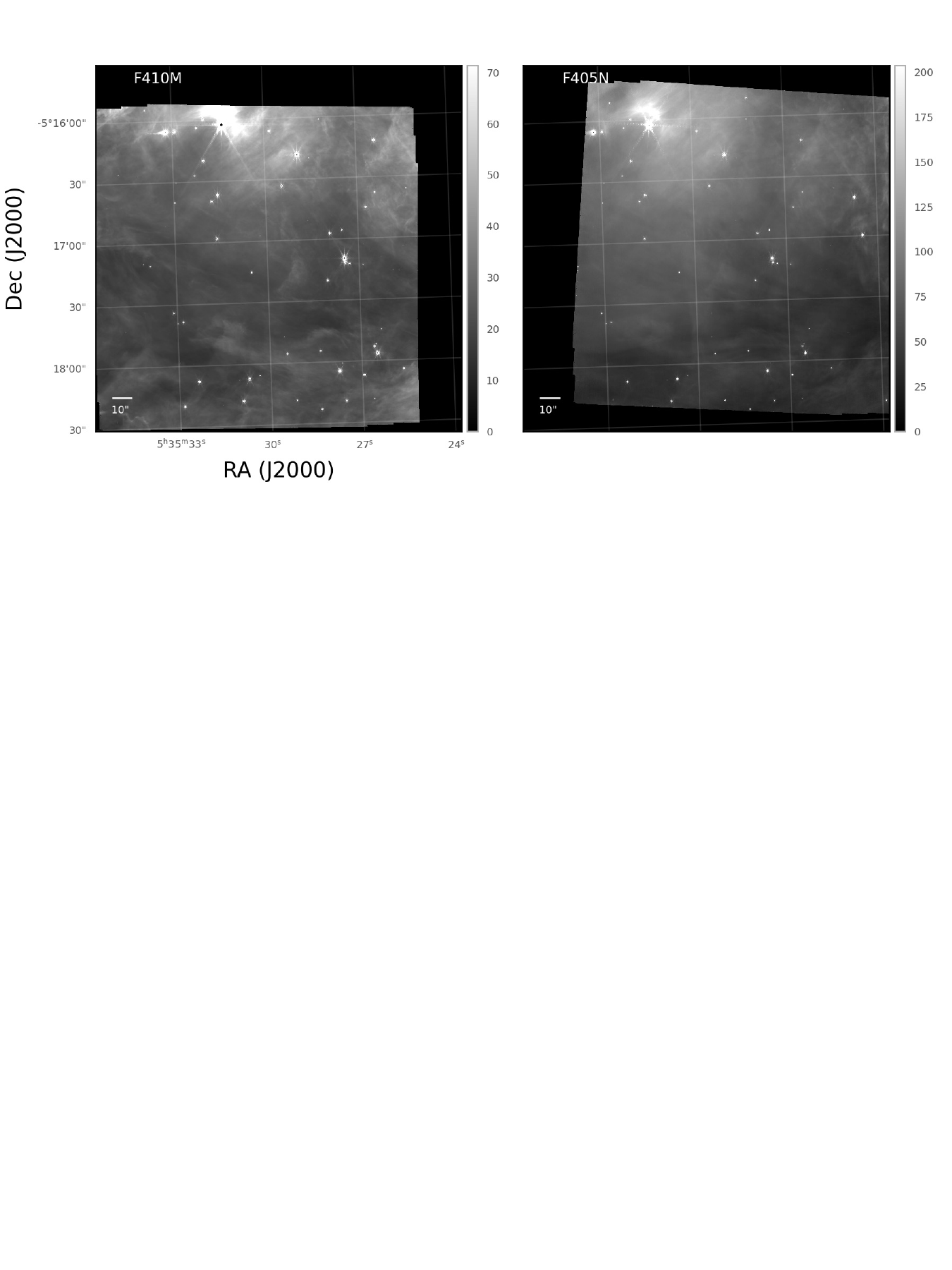}
\caption{Same as in Fig.~\ref{fig:Annexe_Center_M43_all_filters_1}, but for other filters.}
\label{fig:Annexe_Center_M43_all_filters_2}
\end{center}
\end{figure*}

\begin{figure*}[h!]
\begin{center}
\vspace*{0cm}
\includegraphics[width=1\textwidth]{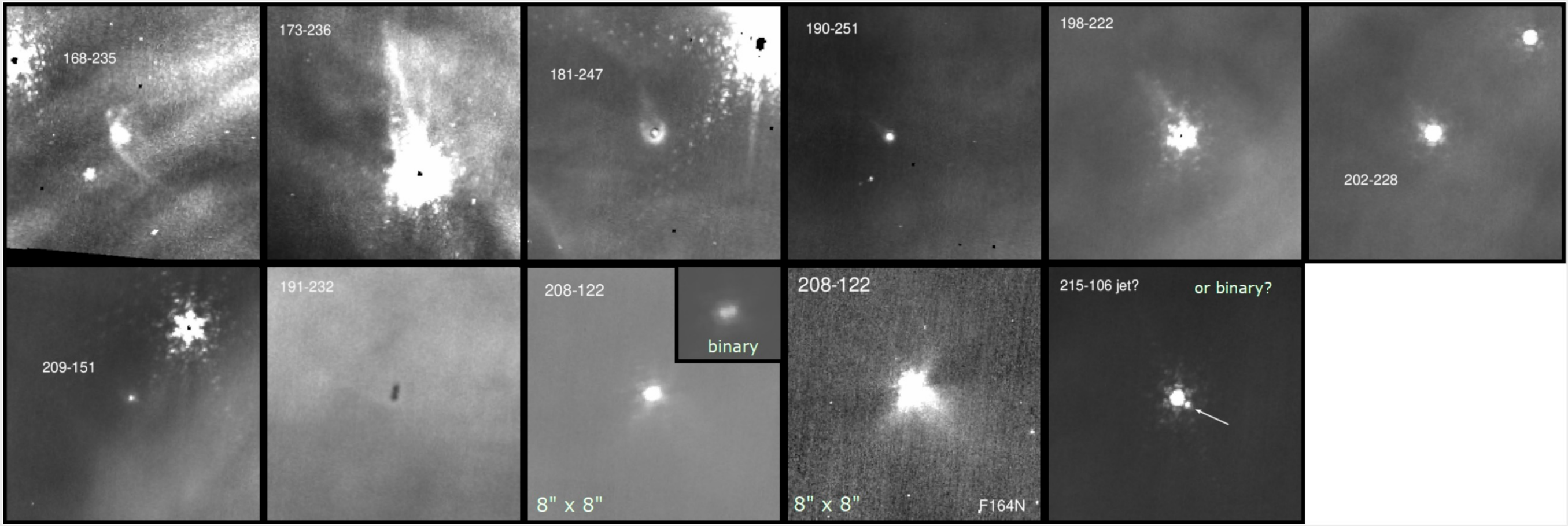}
\caption{Zoom on the proplyds with extended structure identified in the F187N NIRCam image of the M42 north-east region (module A). The tiles are $5\arcsec \times 5\arcsec$, with exception of the object 208-122 for which tiles are $8\arcsec \times 8\arcsec$, with north is up and east to the left.
Some images suffer from instrumental effects such as the diffraction pattern from bright stars and uncorrected cosmic ray events, affecting particularly the edge of the images because of lack of redundancy. }
\label{fig:proplyds_M42A}
\end{center}
\end{figure*}

\begin{figure*}[h!]
\begin{center}
\vspace*{0cm}
\includegraphics[width=1\textwidth]{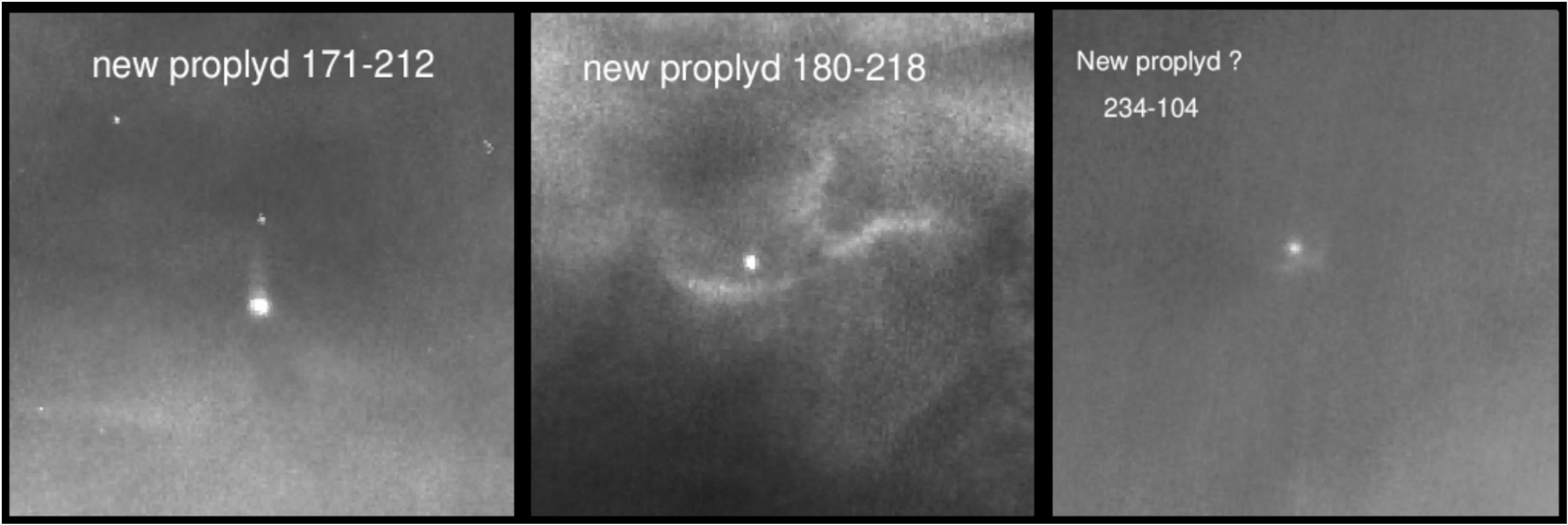}
\caption{Zoom on the 3 new bright proplyds identified in the F187N NIRCam image north-east of the Trapezium stars (module A): 171-212, 180-218 and 234-104. They have faint cusps and only one shows a tail in Pa$\alpha$, that is proplyd 171-212. The tiles are $5\arcsec \times 5\arcsec$ with north up and east to the left. }
\label{fig:new_proplyds_M42A}
\end{center}
\end{figure*}

\begin{figure*}[h!]
\begin{center}
\vspace*{0cm}
\includegraphics[width=1\textwidth]{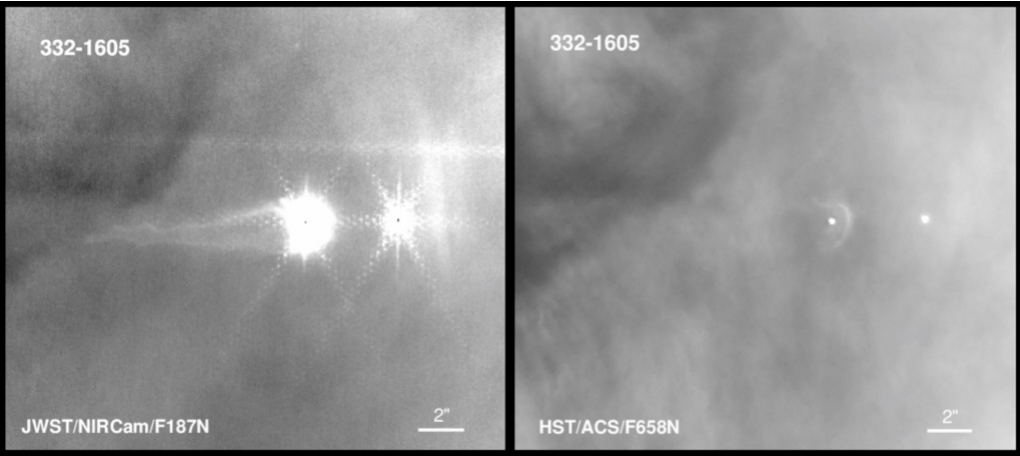}
\caption{Zoom on the giant proplyd 332-1605 found in M43 (module A) and pointing directly to the ionizing star NU\,Ori, just $27\arcsec$ to the west and not visible in the images. This object shows a long tail in Pa$\alpha$ that is nearly 10 times larger than proplyd HST10. The H$\alpha$ HST/ACS image of this object does not show any tail but only a faint cusp. The NIRCam F187N (left) and HST F1658N (right) images have similar spatial resolution allowing for a direct comparison of proplyd morphology in the optical and near-IR. North is up and east to the left.}
\label{fig:giant_proplyd_M43}
\end{center}
\end{figure*}

\begin{figure*}[h!]
\begin{center}
\vspace*{0cm}
\includegraphics[width=0.98\textwidth]{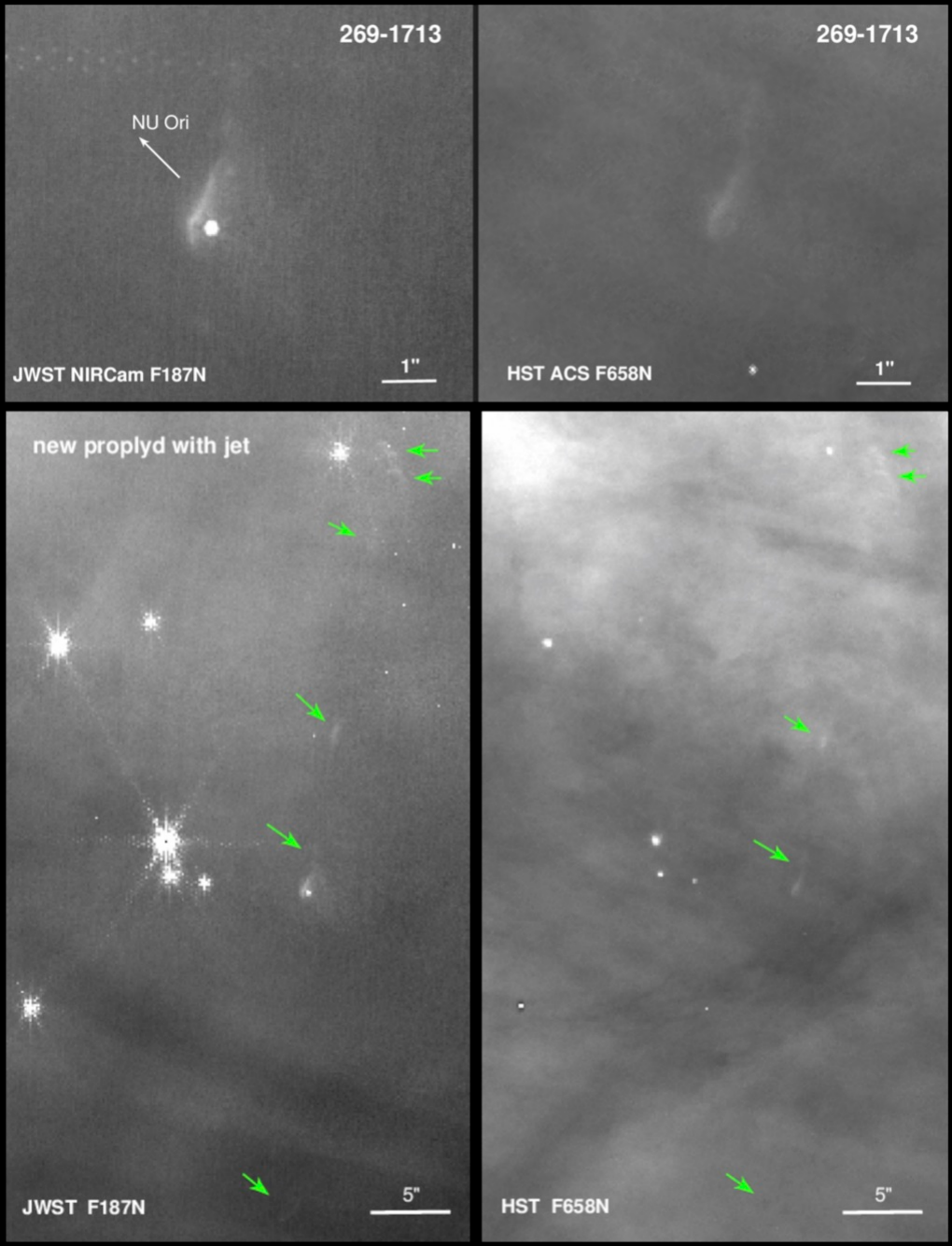}
\caption{Zoom on the new proplyd found in the M43 north region (detector A) at $97\arcsec$ southwest of the ionizing star NU\,Ori (top figure). This object also shows a prominent jet identified by a chain of knots visible in both Pa$\alpha$ and H$\alpha$ images (bottom figure). However, the central star is only visible in the near-IR, reflecting the youth of this object. The JWST/NIRCam F187N images in the left and the HST/ACS F658N images in the right have similar spatial resolution. North is up and east to the left.} 
\label{fig:new_proplyd_M43}
\end{center}
\end{figure*}

As shown in Fig. \ref{fig:composite-image-Center_M43},
M43 contains one main ionized gas region. 
M43 exhibits an expanding bubble structure, with an expansion velocity of 6~km~s$^{-1}$ \cite[]{Pabst20}. 
 Looking at the NIRCam images, the UV flux appears   significantly lower than in the inner Orion Nebula. The Pa $\alpha$ line is in fact much less bright that in the inner region of the nebula. The emission in the F335M filter is also significantly weaker.  
This is expected due to the much cooler and less luminous exciting star NU Ori.
\cite{Pabst20,Pabst22} estimated the local PDR physical conditions in M43, the incident radiation field $G_0$ ($\sim 10^3$) and the gas density $n$
from the SED results and their
[C II] and CO data.
In the rim of M43, \cite{Pabst20} found $n=10^4$~cm$^{-3}$ from the [C II] and CO peak separation. 
Gas temperature estimated from [CII] excitation temperature is about 100-120 K. Thus, the thermal pressure must be around few $10^6$~K~cm$^{-3}$, 10 to 100 times lower than in the Bar.
 The PDRs around M43 are bright in aromatic emission but quite faint in H$_2$. This is consistent with the fact that the density should not be too high ($n \le 10^4$~cm$^{-3}$).
The conditions of the local PDR are compatible with the conditions of pressure equilibrium with the ionized region using the results of \cite{Odell2010}. From their M43 Samples A in Table 10, the averages electronic temperature and density are $T_e=7950 \pm60$~K and $n_e=510 \pm 40~$cm$^{-3}$.

\begin{table*}[hbt!]
    \caption{New proplyd candidates identified in the NIRCam F187N images of M42 north and M43 (module A).}
    \label{tab:new_proplyds}
    \begin{center}
    \begin{tabular}{lccclc}
    \hline\hline
    \noalign{\smallskip}
    Proplyd{\boldmath $^{\mathrm{a}}$} & $\alpha_{J2000}$ & $\delta_{J2000}$ & Region & Comments & 2MASS \\
    \noalign{\smallskip}
     \hline 
     \noalign{\smallskip}
    $171-212$ & 05 35 17.14 & $-05$ 22 11.9 & M42 & bright cusp with tail & no \\
    $180-218$ & 05 35 18.04 & $-05$ 22 18.2 & M42 & near to filament, no tail & no \\
    $234-104$ & 05 35 23.37 & $-05$ 21 03.7 & M42 & faint cusp, no tail & no \\
    $269-1713$ & 05 35 26.87 & $-05$ 17 12.9 & M43 & bipolar jet, HH objects & yes, star\\
    \noalign{\smallskip}
    \hline
    \end{tabular}
    \end{center}
    \begin{list} {}{}
    \item[$^{\mathrm{a}}$]{\small Proplyd name following \citet{O'Dell1994} coordinate-based convention. }
    \end{list}
    \end{table*}


\end{appendix}

\end{document}